\documentclass{aa}  
\usepackage{graphicx}
\usepackage{subfigure}
\usepackage{txfonts}
\usepackage{verbatim}
\usepackage{soul}
\usepackage{float}
\usepackage{pifont}
\usepackage{tabularx,adjustbox,booktabs}

\begin{document}

\title{Pulsating hydrogen-deficient white dwarfs and pre-white dwarfs observed with {\it TESS}}
\subtitle{I. Asteroseismology of the GW Vir stars RX~J2117+3412,
  HS~2324+3944, NGC~6905, NGC~1501, NGC~2371, and K~1$-$16}

\author{Alejandro H. C\'orsico\inst{1,2}, 
        Murat Uzundag\inst{3,4},
        S. O. Kepler\inst{5}, 
        Leandro G. Althaus\inst{1,2}, 
        Roberto Silvotti\inst{6},
        Andrzej S. Baran\inst{7,8,9},
        Maja Vu\v{c}kovi\'{c}\inst{3},
        Klaus Werner\inst{10},
        Keaton J. Bell\inst{11,12},
        and Michael Higgins\inst{13}
        }
\institute{Grupo de Evoluci\'on Estelar y Pulsaciones. 
           Facultad de Ciencias Astron\'omicas y Geof\'{\i}sicas, 
           Universidad Nacional de La Plata, 
           Paseo del Bosque s/n, 1900 
           La Plata, 
           Argentina
           \and
           IALP - CONICET
           \and
           Instituto de F\'isica y Astronom\'ia, Universidad de Valpara\'iso, Gran Breta\~na 1111, Playa Ancha, Valpara\'iso 2360102, Chile
           \and
           European Southern Observatory, Alonso de Cordova 3107, Santiago, Chile
           \and
           Instituto de F\'{i}sica, Universidade Federal do Rio Grande do Sul, 91501-970, Porto-Alegre, RS, Brazil
           \and
           INAF-Osservatorio Astrofisico di Torino, strada dell'Osservatorio 20, 10025 Pino Torinese, Italy
           \and 
           Uniwersytet Pedagogiczny, Obserwatorium na Suhorze, ul. Podchor\c{a}\.zych 2, 30-084 Krak\'ow, Polska
           \and
           Embry-Riddle Aeronautical University, Department of Physical Science, Daytona Beach, FL\,32114, USA
           \and
           Department of Physics, Astronomy, and Materials Science, Missouri State University, Springfield, MO\,5897, USA
           \and
           Institute for Astronomy and Astrophysics, Kepler Center for Astro and Particle Physics, Eberhard Karls University, Sand 1,72076 T\"ubingen, Germany
           \and
           DIRAC Institute, Department of Astronomy, University of Washington, Seattle, WA-98195, USA
           \and 
           NSF Astronomy and Astrophysics Postdoctoral Fellow
           \and
           Department of Physics, Duke University, Durham, NC-27708, USA
           \\
\email{acorsico@fcaglp.unlp.edu.ar}
           }
\date{Received ; accepted }
\abstract{
The recent arrival of continuous photometric observations of
unprecedented quality from  space missions has largely fueled the
study of pulsating stars, bringing the area to an unprecedented
interest in stellar astrophysics. In the particular case of pulsating
white dwarfs, the {\it TESS}  mission is taking asteroseismology of
these compact stars to a higher level,  emulating or even surpassing
the  performance of its predecessor, the {\it Kepler} mission.}
{In this paper,  we present a detailed asteroseismological analysis of
  six GW Vir stars including the
  observations collected by the {\it TESS} mission.}
{ We processed and analyzed {\it TESS} observations of RX~J2117+3412
  (TIC~117070953),  HS~2324+3944 (TIC~352444061), NGC~6905
  (TIC~402913811),  NGC~1501 (TIC~084306468), NGC~2371
  (TIC~446005482),  and K~1$-$16 (TIC 233689607). We carried out a
  detailed asteroseismological analysis of these stars on the basis of 
  PG~1159 evolutionary models that take into account the  complete evolution
  of the progenitor stars. We constrained  the  stellar  mass  of
  these  stars  by comparing  the  observed period spacing with the
  average of the computed period spacings, and, when possible, 
  we employed the individual observed periods to search for a representative seismological model.} 
{In total, we extracted 58 periodicities from the {\it TESS} light
  curves of these GW Vir stars using a standard  pre-whitening
  procedure to derive the potential pulsation frequencies. All the
  oscillation frequencies that we found  are associated with $g$-mode
  pulsations with periods spanning from $\sim 817$ s to $\sim 2682$ s. 
  We find constant period spacings for
  all but one star (K~1$-$16), which allowed us to infer their
  stellar masses and constrain the  harmonic degree $\ell$ of the
  modes. Based on rotational frequency splittings, we derive the 
  rotation period of RX~J2117+3412, obtaining a value in agreement with previous 
  determinations. We performed period-to-period fit analyses on  five of the
  six analyzed stars. For four stars (RX~J2117+3412,  HS~2324+3944,
  NGC~1501, and NGC~2371), we were able to find an asteroseismological
  model with masses  in agreement with the stellar-mass values
  inferred from the period spacings,  and generally compatible with
  the spectroscopic masses. Obtaining seismological models
  allowed us to estimate the seismological distance and compare  it
  with the precise astrometric  distance measured with {\it
    GAIA}. Finally, we find that the period spectrum of K~1$-$16
  exhibits dramatic changes  in frequency and amplitude which, together
  with the scarcity of modes, prevented us from making a meaningful
  seismological modeling.} 
{The high-quality data collected by the {\it TESS}
  space mission,  considered simultaneously with  ground-based
  observations, are able to provide a very valuable input to the
  asteroseismology of GW Vir stars, similar  to the case of other
  classes of pulsating white-dwarf stars. The {\it TESS} mission, in
  conjunction with future space  missions and upcoming surveys, will
  make impressive progress in white-dwarf asteroseismology.}

\authorrunning{C\'orsico et al.}
\titlerunning{Asteroseismology of GW Vir stars with {\it TESS}}
\maketitle


\section{Introduction}

Pulsating white dwarfs (WD) and pre-WDs constitute a well established
class of variable stars that exhibit pulsation periods in the range
$100-7\,000$ s, associated to low-order ($\ell \leq 3$) nonradial $g$
(gravity) modes.  At present, there are about 350 known pulsating
WDs, spread among several subclasses such as ZZ Ceti stars or DAVs
(pulsating hydrogen-rich WDs), V777 Her stars or DBVs (pulsating
helium-rich WDs), and GW Vir stars or pulsating PG~1159 stars, among
others \citep[see the reviews
  by][]{2008ARA&A..46..157W,2008PASP..120.1043F,
  2010A&ARv..18..471A,2019A&ARv..27....7C}.  White-dwarf
asteroseismology has undergone substantial progress, thanks to
ground-based observations, mainly with the time-series photometric
observations of the ``Whole Earth  Telescope''
\citep[WET;][]{1990ApJ...361..309N}, followed by the spectral
observations of the Sloan Digital Sky Survey
\citep[SDSS,][]{2000AJ....120.1579Y}, and in recent years by the
availability of space missions that provide unprecedented high-quality
data. Indeed, the {\it Kepler} satellite, both the main mission
\citep{2010Sci...327..977B} and the {\it K2} mode \citep{2014PASP..126..398H},
allowed the study of 32 ZZ Ceti stars and two V777 Her stars
\citep{2011ApJ...736L..39O,2017ApJS..232...23H,2017ApJ...835..277H,
  2017ApJ...851...24B,2017PhDT........14C,2020FrASS...7...47C}, until
it was out of operation by October 2018.  The successor of {\it
Kepler} is the Transiting Exoplanet Survey Satellite \citep[{\it
TESS},][]{2015JATIS...1a4003R}.  {\it TESS} has provided extensive photometric observations of the $200\,000$ brightest stars in 85 \% 
of the sky in the first part of the mission, each observation with a 
time base of about 27 days per sector observed.

GW Vir stars are pulsating PG~1159 stars ---hot hydrogen (H)-deficient,
carbon (C)-,~ oxygen (O)-, helium (He)-rich pre-WD stars---  that
include PNNV stars --- still surrounded by a nebula ---  and  DOV
stars -- that lack a nebula \citep{1991ApJ...378..326W}. The
classification of GW Vir stars includes also the pulsating Wolf-Rayet
central stars of planetary nebula ([WC])  and Early-[WC] = [WCE] stars,
 because they share the same pulsation properties of pulsating PG~1159 stars \citep{2007ApJS..171..219Q}.  GW Vir stars exhibit multiperiodic
luminosity variations with periods in the range 300--6000 sec, induced
by nonradial gravity ($g$)-mode pulsations driven by the
$\kappa$-mechanism due to partial ionization of C and O in the outer
layers \citep{1983ApJ...268L..27S,1984ApJ...281..800S,
  1991ApJ...383..766S,2005A&A...438.1013G,
  2006A&A...458..259C,2007ApJS..171..219Q}.  PG~1159 stars represent
the evolutionary connection between post-AGB stars and most of the
H-deficient WDs, including DO and DB WDs \citep{2006PASP..118..183W}.
These stars likely have their origin in a born-again episode induced
by a post-AGB He thermal pulse \citep[see][for
  references]{2001ApJ...554L..71H,2001Ap&SS.275....1B,
  2005A&A...435..631A,2006A&A...449..313M}.

In this work, we  determine the  internal  structure  and  evolutionary  
status of the pulsating pre-WD stars RX~J2117.1+3412, HS~2324+3944,   NGC~6905, 
NGC~1501, NGC~2371, and K~1$-$16 on the basis of the full 
PG~1159 evolutionary models of \cite{2005A&A...435..631A} and  \cite{2006A&A...454..845M}. 
These models were derived
from the complete evolution of progenitor stars through the thermally
pulsing AGB phase and born-again episode.  We  compute adiabatic
$g$-mode pulsation  periods  on  PG~1159  evolutionary models with
stellar masses ranging from 0.530 to $0.741\ M_{\odot}$. 

The paper is
organized as  follows. In  Sect. \ref{target} we provide an account of
the main characteristics of the studied  GW Vir stars.
Sect. \ref{observations} is focused on describing the methods we apply
to obtain the pulsation periods of each target
star. Sect. \ref{period-spacing} is devoted to search for a constant
period spacing in the list of periods of each star by applying three
significance tests. A brief summary of the stellar models of PG~1159
stars employed for the asteroseismological analysis of these stars is
provided  in Sect.  \ref{models}.   In Section \ref{spec_mass} we
derive the spectroscopic masses of the  target stars on the basis of
their published values of $T_{\rm eff}$ and $\log g$. In
Sect. \ref{astero} we carry out a detailed asteroseismological
analysis for each star, by assessing the stellar mass of each object
through the use of the period spacing when possible, and by performing
period-to-period fits with the aim of finding an asteroseismological
model for each pulsating PG~1159 star.  Finally, in
Sect. \ref{conclusions}, we  summarize  our  main  results  and  make
some  concluding remarks.

\section{The  targets}  
\label{target}  

\begin{figure*} 
\includegraphics[clip,width=1.9\columnwidth]{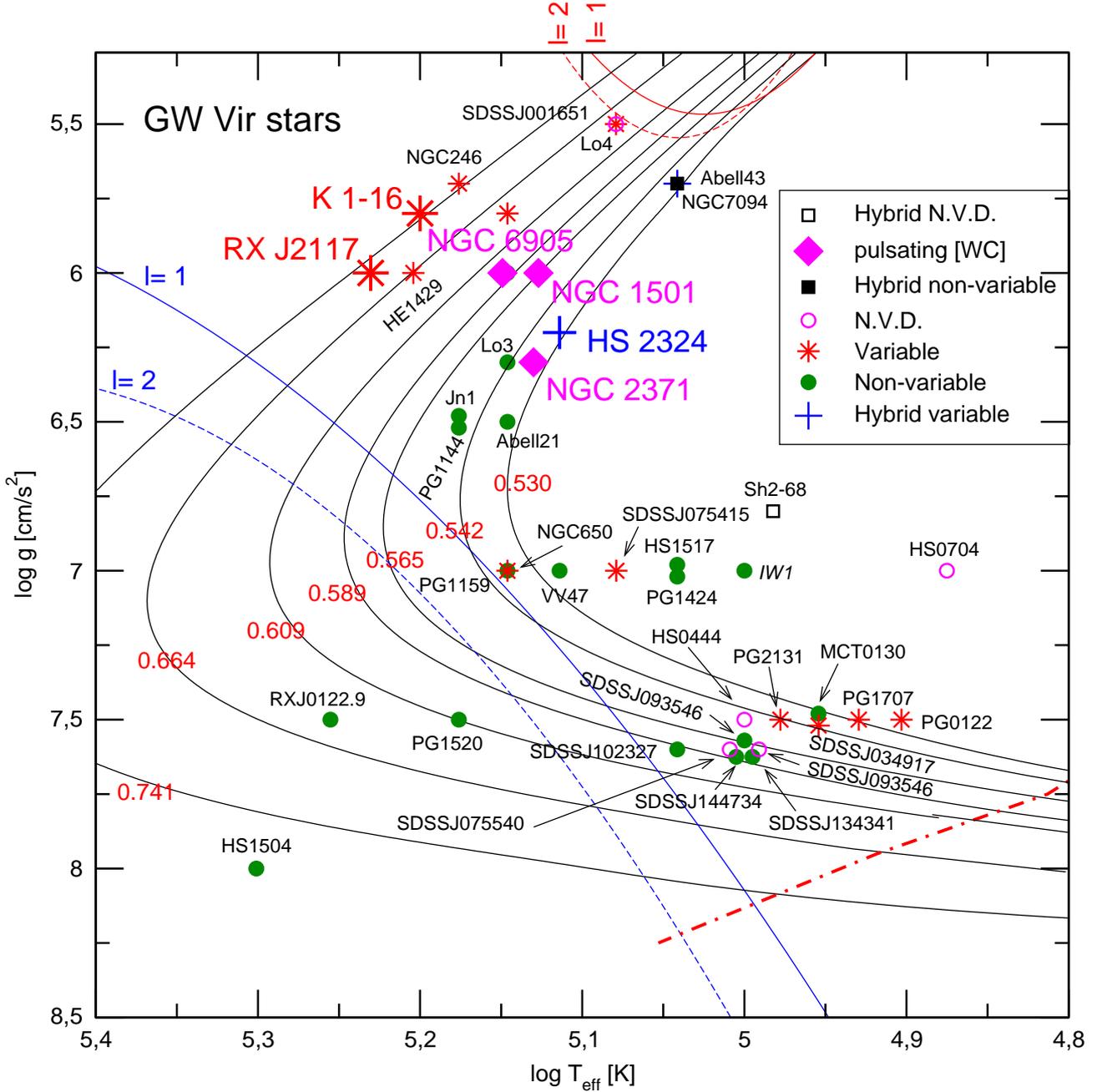}
\caption{Location of the spectroscopically calibrated  variable and
  non-variable PG~1159 stars and variable [WCE] stars 
  in the $\log T_{\rm eff}-\log g$ plane.
  PG~1159 stars with no variability data (N.V.D.) are depicted with
  hollow circles.  Thin solid  curves  show  the evolutionary tracks
  from \cite{2006A&A...454..845M} for stellar masses: 0.530, 0.542,
  0.565, 0.589, 0.609, 0.664, and $0.741 M_{\odot}$.  Parameterizations
  of the theoretical pulsational dipole (solid curves) and quadrupole
  (dashed curves) red  and  blue  edges  of  the  instability domain
  according to \cite{2006A&A...458..259C} are also displayed. The
  location of the GW Vir stars RX~J2117 (red star), K~1$-$16 (red
  star),  HS~2324 (blue plus), NGC~6905, NGC~1501, and NGC~2371
  (magenta diamonds) are emphasized with large symbols.}
\label{fig:1} 
\end{figure*}

Unlike the {\it Kepler} space telescope, characterized by a restricted sky
coverage,  the {\it TESS} mission  is providing unprecedented photometric 
observations of GW Vir stars, given its expansive sky 
coverage. In this work, we report on new TESS observations of the
already known pulsating pre-WD stars RX~J2117.1+3412, HS~2324+3944, NGC~6905,
NGC~1501, NGC~2371, and K~1$-$16. Note that NGC~6905 and NGC~1501 are
not formally spectroscopic PG~1159 stars, but instead  [WCE] stars,
and NGC~2371 is classified as a [WC]-PG~1159 transition object. The
essential difference between the PG~1159, [WC] and [WCE]
spectroscopic classes is the presence of much stronger winds and mass
losses  in the [WC] and [WCE] objects, leading to emission line
spectra  \citep[][]{2006PASP..118..183W}. Following \cite{2007ApJS..171..219Q}, 
we include in this paper the stars NGC~6905, NGC~1501, and  NGC~2371 in the 
category of GW Vir variable stars. The location of the six stars in the
$\log T_{\rm eff}$ vs. $\log g$ diagram is displayed in
Fig. \ref{fig:1}.  Note that, according to the surface gravity of each
star and the  evolutionary tracks of \cite{2006A&A...454..845M}, all
the stars to  be analyzed in this  study are evolving at stages prior
to their maximum possible temperatures,  before entering the cooling
branch of WDs. It is expected, then,  that all these stars are heating
and contracting, with the consequent  secular shortening of the
periods of pulsation of their $g$ modes (see Sect. \ref{astero}).

\begin{table*}
  \caption{The list of six pulsating pre-white dwarf stars studied in this work. Columns 1, 2, 3, 4, 5, 6, 7 and 8 correspond to
  the {\it TESS} input catalog number, name of the object, effective temperature, surface gravity, GW Vir class, spectral 
  classification, parallax, and distance, respectively. For details, see the text.}
  \begin{tabular}{cccccccc}
\hline
TIC        & Name             &  $T_{\rm eff}$ & $\log g$ & GW Vir  & Spectral type & $\pi$  &  $d$  \\
           &                  &    [K]         & [cgs]    & class    &               &[mas]   &   [pc]\\
\hline
 117070953 & RX~J2117.1+3412  & $170\,000\pm10\,000$ & $6.00\pm0.03$ & PNNV & PG~1159 & $1.991\pm0.050$ & $502\pm12$\\
 352444061 & HS~2324+3944     & $130\,000\pm10\,000$ & $6.20\pm0.20$ & DOV  & PG~1159 (hybrid) & $0.691\pm0.050$ & $1448\pm105$ \\
 402913811 & NGC~6905         & $141\,000\pm10\,000$ & $6.00\pm0.20$ & PNNV & [WCE] & $0.612\pm0.059$ & $1634\pm158$ \\
 084306468 & NGC~1501/[WO4]   & $134\,000\pm10\,000$ & $6.00\pm0.20$ & PNNV & [WCE] & $0.567\pm0.025$ & $1763\pm79$     \\
 446005482 & NGC~2371         & $135\,000\pm10\,000$ & $6.30\pm0.20$ & PNNV & [WC]-PG~1159 & $0.532\pm0.066$ & $1879\pm 232$ \\
 233689607 & K~1$-$16          & $160\,000\pm16\,000$ & $5.80\pm0.30$ & PNNV & PG~1159 & $0.466\pm0.053$  & $2146\pm244$  \\
\hline
\label{basic-parameters-targets}
\end{tabular}
\end{table*}

We describe the basic characteristics of these stars below:

\begin{itemize}

\item RX~J2117.1+3412 (hereafter RX~J2117 and TIC~117070953): a PNNV
  star characterized by an effective temperature of $T_{\rm eff}=
  170\,000\pm 10\,000$ K, a surface gravity of $\log g= 6.00\pm 0.3$,
  and a surface composition of $(X_{\rm He}, X_{\rm C}, X_{\rm O})=
  (0.39, 0.55, 0.06)$ \citep{2006PASP..118..183W}.  This is one of the
  PG~1159 stars in  which iron has been detected
  \citep{2010ApJ...719L..32W}.  At this effective temperature, the
  star is the hottest known GW~Vir star.  The location of this star in
  the $\log T_{\rm eff}-\log g$ diagram is displayed in
  Fig. \ref{fig:1}. The {\it Gaia} DR2 parallax and corresponding
  distance for this star are  $\pi= 1.991 \pm 0.050$~mas and $d= 502
  \pm 12$~pc. The variability of this star was independently
  discovered by \cite{1992IAUC.5603....1W} and
  \cite{1993A&A...267L..35V}. \cite{2002A&A...381..122V} analyzed the
  star in depth and published  a large  amount of  results  from  a
  multisite photometric campaign with the WET
  \citep{1990ApJ...361..309N}. A search for $p$ (pressure)  modes in this star was carried
  out by \cite{2013A&A...558A..63C}, with null results.
  RX~J2117 has been the focus of a
  detailed asteroseismological modelling by
  \cite{2007A&A...461.1095C}, yielding a variety of constraints; among
  them, a value of the asteroseismic stellar mass that is considerably
  lower than suggested by spectroscopy coupled to evolutionary
  tracks. \\

\item HS~2324+3944 (hereafter HS~2324 and TIC~352444061): one out of
  four peculiar  members of the PG~1159 spectral class that exhibit H
  at the surface, called ``hybrid PG~1159 stars''
  \citep{1991A&A...249L..16N}.  The star exhibits a surface chemical
  composition of  $(X_{\rm H}, X_{\rm He}, X_{\rm C}, X_{\rm O})=
  (0.17, 0.35, 0.42, 0.06)$ \citep{2006PASP..118..183W}. This object
  is characterized by an effective temperature $T_{\rm eff}= 130\,000
  \pm 10\,000$ K and a surface gravity $\log g= 6.2\pm 0.2$ (see
  Fig. \ref{fig:1}). The {\it Gaia} DR2 parallax and corresponding
  distance for this star are $\pi= 0.691 \pm 0.0502$~mas and $d= 1448
  \pm 105$~pc. HS~2324 was discovered to be pulsating by
  \cite{1996A&A...309L..23S} \citep[see,
    also,][]{1997A&A...326..692H}, and it was the target of a
  multisite photometric campaign carried out by
  \cite{1999A&A...342..745S}.  This object has not been  the subject
  of a published asteroseismological modeling. \\

\item NGC~6905 (also TIC~402913811): a [WCE]-type star  discovered by
  \cite{1996AJ....111.2332C}, characterized by $T_{\rm eff}=
  141\,000$~K and $\log g= 6$  (see Fig. \ref{fig:1}), and pulsation
  periods in the range $710-912$ s. \cite{1996AJ....111.2332C} found
  that, while the principal pulsation modes lie within a fairly
  restricted range of  frequencies, the individual modes that are
  actually observed change completely on a timescale of a few months
  (or less). The surface chemical composition of NGC~6905 is ($X_{\rm
    He}, X_{\rm C}, X_{\rm O})= (0.60, 0.25, 0.15)$
  \citep{2001Ap&SS.275...41K}.   The {\it Gaia} DR2 parallax and
  corresponding distance for this object are $\pi= 0.612 \pm0.059$ mas
  and $d= 1634\pm 158$~pc. This star has not been
  asteroseismologically modelled before.\\

\item NGC~1501 (also TIC~084306468): classified as a [WCE]-type
  star. The effective temperature and gravity of this star are $T_{\rm
    eff}= 134\,000$ K and $\log g=6.0$ (Fig. \ref{fig:1}), and its
  surface chemical composition is ($X_{\rm He}, X_{\rm C}, X_{\rm O})=
  (0.50, 0.35, 0.15)$ \citep{1997IAUS..180..114K,2006PASP..118..183W}.
  The parallax and corresponding  distance for this star extracted
  from {\it Gaia} DR2 are $\pi= 0.567  \pm 0.025$~mas and $d= 1763 \pm
  79$~pc.  The variability of NGC~1501 was detected by
  \cite{1996AJ....111.2332C}, who measured ten periodicities ranging
  from 5200~s down to 1154~s, with the largest-amplitude pulsations
  occurring between 1154~s and 2000~s. As for NGC~6905, in the case of
  NGC~1501 the observed pulsation  spectrum varies on timescales of
  months \cite{1996AJ....111.2332C}. Based on period-spacing data,
  \cite{1996AJ....112.2699B} inferred a stellar mass of $0.53 \pm 0.03
  M_{\odot}$ for NGC~1501.  The star was a target of a detailed
  asteroseismological analysis by \cite{2009A&A...499..257C}, who
  derived a stellar mass of $\sim 0.57 M_{\odot}$ based on the
  period-spacing data.  However, they were unable to find an
  unambiguous best-fit model with a period-fit procedure.\\

\item NGC~2371 (also TIC~446005482): a GW Vir variable classified  as
  a [WC]-PG~1159  transition object. It was discovered to be pulsating
  by \cite{1996AJ....111.2332C}.  The star shows periodicities in the
  range $923-1825$ s, and has atmospheric parameters $T_{\rm eff}=
  135\,000$ K and $\log g= 6.3$ (see Fig. \ref{fig:1}), and  ($X_{\rm
    He}, X_{\rm C}, X_{\rm O})= (0.54, 0.37, 0.08)$
  \citep{2004ApJ...609..378H}. Like NGC~6905 and NGC~1501, NGC~2371
  exhibits variations of the characteristics of the frequency
  spectrum on timescales of months or shorter
  \citep{1996AJ....111.2332C}.  From {\it Gaia} DR2, we known that the
  parallax and distance of this object are $\pi= 0.532 \pm0.066$ mas
  and $d= 1879\pm 232$~pc, respectively. This star has not been the
  focus of any detailed asteroseismological analysis.\\

\item K~1$-$16 (also Kohoutek~1$-$16 and TIC 233689607): a PNNV star 
characterized by  an
  effective temperature of $T_{\rm eff}= 140\,000\pm 10\,000$ K, a
  surface gravity of $\log g= 6.40\pm 0.3$, and a surface composition
  of $(X_{\rm He}, X_{\rm C}, X_{\rm O})= (0.33, 0.50, 0.17)$
  \citep{2006PASP..118..183W}. \cite{2007A&A...474..591W}  and
  \cite{2010ApJ...719L..32W} discovered high-ionization lines of neon
  (Ne VIII) and iron (Fe X) in K~1$-$16,  and re-determined its
  effective  temperature and gravity as $T_{\rm eff}= 160\,000\pm
  16\,000$ K and   $\log g= 5.8\pm 0.3$ (see Fig. \ref{fig:1}). Also,
  the chemical composition was re-determined: $(X_{\rm He}, X_{\rm C},
  X_{\rm O})= (0.58, 0.34, 0.06)$
  \citep{2010ApJ...719L..32W}. K~1$-$16 was the
  first PNNV star to be discovered \citep{1984ApJ...277..211G}. These
  authors detected periodicities  with a dominant period of $\sim
  1700$ s and a semi-amplitude of about 0.01 mag.
  \citet{1984ApJ...277..211G} found that several additional periods
  sometimes appear in power  spectra derived from light curves, and on
  two occasions a rapid drop into  ---or emergence from--- a state in
  which no detectable variations were present.
  \cite{1995PASP..107..914F} claimed spectral variability and
  speculated about a  connection of this phenomenon to the changes in
  the pulsating nature of this star. From {\it Gaia} DR2, the parallax
  and distance of K~1$-$16 are $\pi= 0.466 \pm0.053$ mas and $d=
  2146\pm 244$~pc, respectively. The asteroseismological potential of
  this interesting object has not been thoroughly studied yet.\\

\end{itemize}

We summarize the stellar properties of the six target stars in Table \ref{basic-parameters-targets}. 

\begin{table*}
  \caption{The list of  six GW Vir stars reported from {\it
      TESS} observations,  including the name of
    the targets, {\it TESS} magnitude, observed sectors,  date, \textit{CROWDSAP} keyword,  and length of the runs 
    (columns 1, 2, 3, 4, 5 and 6, respectively). From the Fourier Transform of the original and shuffled data, three different set of parameters: resolution, average
    noise level of amplitude spectra, and  detection threshold which
    we define as the amplitude at $0.1\%$ false alarm probability, FAP, are presented
    in columns 7, 8 and 9, respectively. The \textit{CROWDSAP} keyword shows the ratio of the target flux to the total flux in the optimal \textit{TESS} light curve aperture.}
  \begin{tabular}{ccccccccc}
\hline
Name &  $T_{\rm mag}$ & Obs.   & Start Time        &  \tt{CROWDSAP} & Length & Resolution & Average Noise & $0.1\%$\,FAP \\
     &                & Sector & (BJD-2\,457\,000) &   & [d]    & $\mu$Hz    &  Level [ppt]  &    [ppt]  \\
\hline
RX~J2117.1+3412 & 11.73 & 15     & 1711.3688 & 0.519 & 22.76  &  0.76  & 0.26 & 1.18 \\
HS~2324+3944    & 14.81 & 16-17  & 1738.6566 & 0.571 & 49.14  &  0.35  & 0.44 & 2.07 \\
NGC~6905        & 13.90 & 14     & 1683.3582 & 0.087 & 26.85  &  0.43  & 0.73 & 3.35 \\
NGC~1501/[WO4]  & 12.40 & 19     & 1816.0888 & 0.556 & 25.06  &  0.46  & 0.11 & 0.53 \\
NGC~2371        & 12.90 & 20     & 1842.5093 & 0.356 & 26.31  &  0.44  & 0.20 & 0.90 \\
K~1$-$16        & 14.36 & 14-17  & 1683.3542 & 0.570 & 298.93 &  0.038 & 0.21 & 0.86 \\
                &       & 19-20  &           &       &        &        &      &      \\        
                &       & 22-26  &           &       &        &        &      &      \\
\hline
\label{DOVlist}
\end{tabular}
\end{table*}

\begin{figure*} 
\includegraphics[clip,width=1.\linewidth]{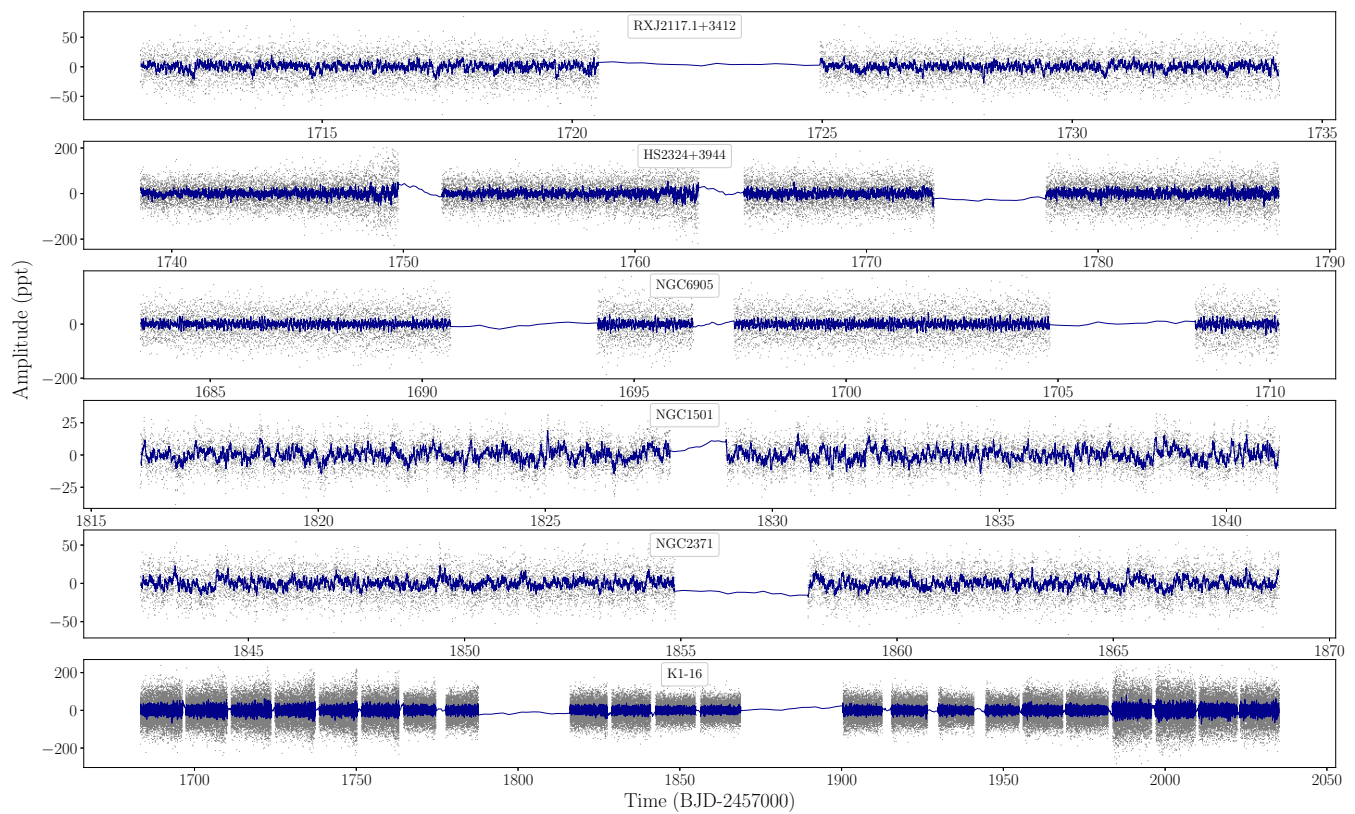}
\caption{Reduced \textit{TESS} light curves of the pulsating pre white-dwarf stars studied in this work. The black dots are the {\it TESS} data sampled every two minutes. The blue lines are binned light curves which are calculated by running mean every 20 points (corresponding to 38 minutes).
K~1$-$16 was observed during nine sectors (14, 15, 16, 17, 19, 20, 22, 23, 24, 25 and 26),
HS~2324 was observed during two consecutive sectors (16 and 17), and the other 
stars were observed during a single sector each. }
\label{fig:lightcurves} 
\end{figure*} 

\begin{figure*} 
\includegraphics[clip,width=1\linewidth]{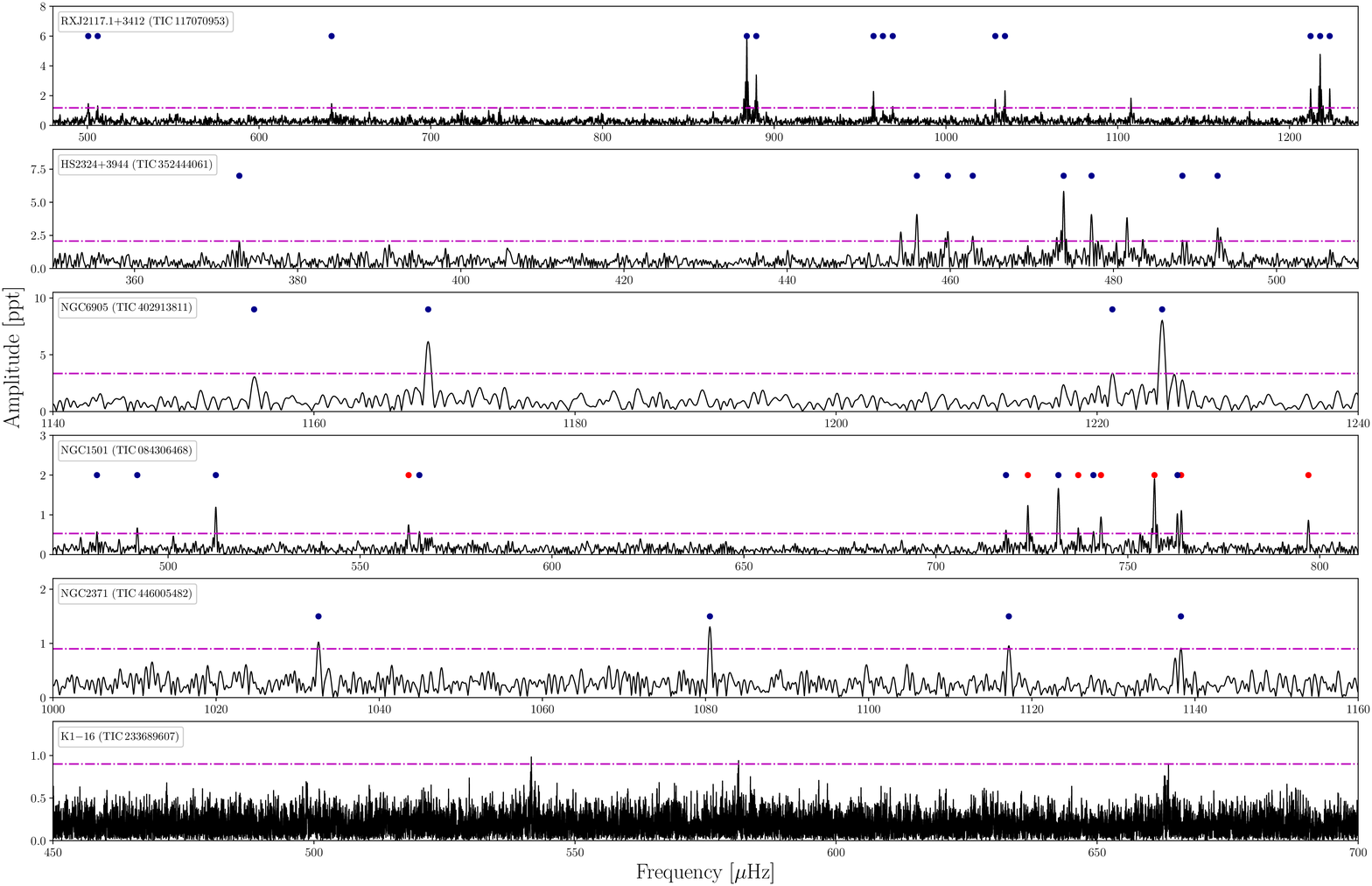}
\caption{The FTs of the GW Vir stars studied in this work, showing 
  identified dipole ($\ell= 1$, blue dots) and quadrupole ($\ell= 2$, red dots) modes. The horizontal magenta dashed line shows
  the detection threshold, which is defined as $0.1\%$ FAP.}
\label{fig:MODE-ID} 
\end{figure*}

\section{Observations and data reduction}  
\label{observations}  

The primary mission of {\it TESS} is to search for exoplanets around bright
target stars. It was  launched successfully on 18 April 2018
\citep{2014SPIE.9143E..20R}. The mission-data products of  {\it TESS} 
also allow us to study 
stellar variability, including pulsations. Thanks to nearly continuous, 
stable photometry, as well as its expansive 
sky coverage, {\it TESS} has made a significant contribution to the study of 
stellar pulsations in evolved compact objects \citep[][]{2019A&A...632A..42B,2020ApJ...888...49W,2020A&A...633A..20A,
2020A&A...638A..82B}.

In this work, we study six known  pulsating pre-white dwarf stars by investigating their
pulsational  characteristics with the high-precision photometry of
{\it TESS}  (see Table \ref{DOVlist}).  The transformations between
the photometric systems, from the visual magnitude ($V_{\rm mag}$) to
{\it TESS} $T_{\rm mag}$ magnitudes, are calculated using publicly
available code of {\tt ticgen}\footnote{\tt
  https://github.com/tessgi/ticgen}. The stars were observed in 2-min
short-cadence mode of {\it TESS}, 
corresponding to a Nyquist frequency of 4200~$\mu$Hz. 
We downloaded the light curves from The Mikulski Archive for Space Telescopes,
which is hosted by the Space Telescope Science Institute
(STScI)\footnote{\tt http://archive.stsci.edu/} as FITS format.  The
data contained in the FITS files are processed based on the Pre-Search Data
Conditioning Pipeline \citep{jenkins2016}. We extracted times and fluxes 
(PDCSAP FLUX) from the FITS files. The times are given in barycentric 
corrected dynamical Julian days \citep[BJD - 2457000, corrected for leap seconds,
see][]{2010PASP..122..935E}. The fluxes are converted to fractional variations
from the mean, i.e. differential intensity $\Delta I/I$, and transformed to
amplitudes in parts-per-thousand (ppt). The ppt unit corresponds
to the milli-modulation amplitude (mma) unit\footnote{1 mma= 1/1.086 mmag= 0.1 \% = 1 ppt; see, e.g., \cite{2016IBVS.6184....1B}.} used in the past.
We sigma-clipped the data at 5~$\sigma$ to remove the outliers which appear 
above 5 times the median of intensities --- i.e. departed from the (local) 
mean by 5 $\sigma$.

The final light curves of the target stars are  shown in
Fig. \ref{fig:lightcurves}. After detrending the light curves, we
calculated their Fourier transforms (FTs) and examined them for
pulsations and binary signatures.  For pre-whitening, 
we employed our customized tool in which, using a nonlinear
least square (NLLS) method, we simultaneously fitted each pulsation
frequency in a waveform $A_i \sin(\omega_i\ t + \phi_i)$,
with $\omega=2\pi/P$, and $P$ the period.
This iterative process has been done starting with the highest peak, 
until there is no peak that appears above $0.1 \%$ FAP 
significance threshold. We analyzed the concatenated light curve 
from different sectors, if observed.  The false alarm probability was 
calculated randomizing the timings, i.e., shuffling the observations 
one thousand times and recalculating the  FTs. We calculated the amplitude 
at which there is a 0.1\%= 1/1000 probability of any peak being due to noise
\citep[e.g.,][]{1993BaltA...2..515K}. 

Before describing the observations of each individual target star, we would like to emphasize that most pulsating WDs and pre-WDs, including GW Vir stars, show amplitude and frequency variations. A similar phenomenon is observed in pulsating sdB stars. 
An example of this phenomenon is documented in this paper for 
K~1$-$16 (see Sect. \ref{obser-k1-16} below). In general terms, some modulations could be probably due to simple photon-count noise caused by contamination of the background light in the aperture.  Other modulations could come from the rotation-to-pulsation energy interchange, as the amplitude of the components of the multiplets changing with time have been detected over the years. In the case of GW Vir stars, however, considering that the theoretical $e$-folding times are much longer than the observations, we do not expect real mode energy interchanges, as we do observe for the DBV star GD 358, for example \citep{2009ApJ...693..564P}.

\subsection{RX~J2117}
\label{obser-rxj2117}

RX~J2117 (TIC~117070953, $T_{\rm mag}= 11.73$, $m_{\rm V}= 12.33$) was observed by
{\it TESS} on Sector 15 between 2019  Aug 15 and 2019 Sep 11. 
The temporal resolution is $1.5/T = 0.76\ \mu$Hz ($T$ is data span of
22.8 days).  The average noise level of the amplitude spectra is of
0.29 ppt. The  upper panel of Fig. \ref{fig:lightcurves} 
depicts the light curve,  and 
the upper panel of Fig. \ref{fig:MODE-ID} shows the FT  
corresponding to  RX~J2117. In Table \ref{table:RXJ2117} we
show the list of periods of RX~J2117 detected with {\it TESS}. The period
at $1038$~s corresponds to a peak with S/N lower than 4. However, we have included 
it because it corresponds to the central component of a triplet and has been
observed in ground-based data \citep{2002A&A...381..122V}.

\begin{table}
\centering
\caption{Independent frequencies, periods, and 
amplitudes, their uncertainties, and the 
signal-to-noise ratio in the  data of RX~J2117. Frequencies in boldface 
correspond to components of rotational triplets (see Fig. \ref{fig:RM-RXJ}).
Errors are given in parenthesis to 2 significant digits.}
\begin{tabular}{ccccr}
\hline
\noalign{\smallskip}
Peak &$\nu$      & $\Pi$  &  $A$   & S/N \\
& ($\mu$Hz) &  (s)   & (ppt)  &   \\
\noalign{\smallskip}
\hline
\noalign{\smallskip}
f$_{\rm 1}$ &  {\bf 500.559(39)} &   1997.760(16) &   1.45(23) &   5.5  \\
f$_{\rm 2}$ &  {\bf 506.057(43)} &   1976.060(17) &   1.33(23) &   5.1  \\
         &&&\\
f$_{\rm 3}$ &  642.255(40) &   1557.010(10) &   1.44(23) &   5.5  \\
f$_{\rm 4}$ &  740.266(49) &   1350.870(9)  &   1.17(23) &   4.5  \\
         &&&\\
f$_{\rm 5}$ &  {\bf 884.017(9)}  &   1131.200(12) &   6.15(23) &  23.6  \\
f$_{\rm 6}$ &  {\bf 889.556(17)} &   1124.156(21) &   3.37(23) &  12.9  \\
         &&&\\
f$_{\rm 7}$ &  {\bf 957.817(25)} &   1044.041(27) &   2.30(23) &   8.8  \\
f$_{\rm 8}$ &  {\bf 963.28(6)}   &     1038.120(6) &   1.03(23) &   3.9  \\
f$_{\rm 9}$ &  {\bf 969.013(44)} &    1031.978(47) &   1.29(23) &   4.9  \\
         &&&\\
f$_{\rm 10}$ & {\bf 1028.729(33)} &    972.073(31) &   1.72(23) &   6.6  \\
f$_{\rm 11}$ & {\bf 1034.356(25)} &    966.785(23) &   2.31(23) &   8.8  \\
         &&&\\
f$_{\rm 12}$ & 1107.713(31) &    902.761(25) &   1.84(23) &   7.0  \\
         &&&\\
f$_{\rm 13}$ & {\bf 1212.297(23)} &    824.880(16) &   2.48(23) &   9.5  \\
f$_{\rm 14}$ & {\bf 1217.872(12)} &    821.105(8) &   4.78(23) &  18.3  \\
f$_{\rm 15}$ & {\bf 1223.429(24)} &    817.375(16) &   2.41(23) &   9.2  \\
\noalign{\smallskip}
\hline
\end{tabular}
\label{table:RXJ2117}
\end{table}

\begin{figure} 
\includegraphics[clip,width=1.\linewidth]{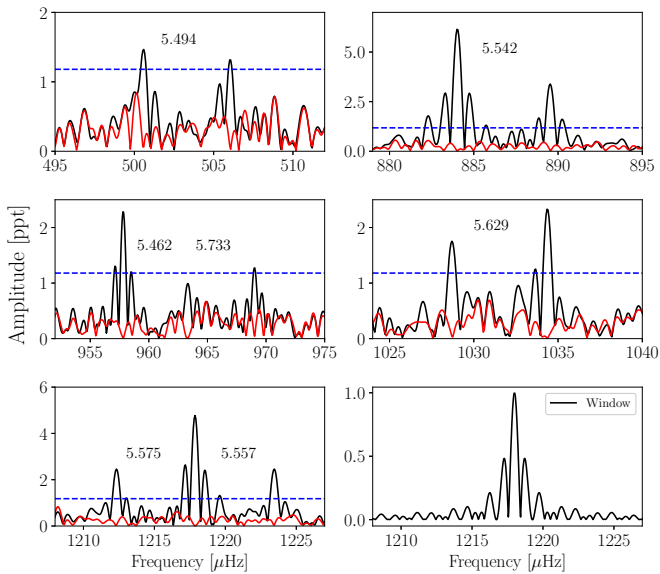}
\caption{The amplitude spectra of RX~J2117 showing rotational multiplets. 
The subfigures present rotational splittings in different region in 
the amplitude spectra with an average of splitting $\delta\nu = 
5.570\ \mu$Hz. The last subfigure (bottom right) presents the window 
function of RX~J2117 for the frequency $1217.872\ \mu$Hz.
The horizontal blue dashed line shows the confidence level of $0.1\%$ FAP. 
The red lines are residuals after extraction of the signals.}
\label{fig:RM-RXJ} 
\end{figure}

The frequencies emphasized with boldface in Table \ref{table:RXJ2117}
can be interpreted  as components of rotational triplets ($\ell= 1$),
with an average frequency separation of $\delta \nu \sim 5.57\ \mu$Hz. This frequency separation is in agreement with  the value
found by \cite{2002A&A...381..122V}. The rotational multiplets are
depicted in Fig. \ref{fig:RM-RXJ}. Let us briefly describe the effect
of rotation on the pulsation frequencies of a star. In the presence of
stellar rotation, non-radial modes of degree $\ell$ split into
$2{\ell}+1$ components differing in azimuthal ($m$) number. In the
case of slow and solid rotation, the frequency splitting can be
obtained as:  $\delta \nu_{\ell, k, m}= m\ (1-C_{\ell,k})\ \Omega_{\rm
  R}$, $\Omega_{\rm R}$ being the angular rotational frequency  of the
pulsating star, and $m= 0, \pm 1, \pm 2, \cdots, \pm \ell$. The
condition of slow rotation translates in $\Omega_{\rm R} \ll
\nu_{\ell, k}$.  The coefficients $C_{\ell, k}$, called Ledoux
coefficients \citep{1958HDP....51..353L}, adopt a simple form in the
asymptotic limit of high radial-order $g$ modes ($k \gg \ell$):
$C_{\ell,k} \sim [\ell(\ell+1)]^{-1}$.  For dipole ($\ell= 1$) and
quadrupole ($\ell= 2$) modes, we have  $C_{1,k}\sim  0.5$ and $C_{2,k}
\sim 0.17$, respectively. The presence of multiplets in the frequency
spectrum of a pulsating WD can be very useful to identify the harmonic
degree of the pulsations. This method to infer the rotation period has
been successfully applied to several pulsating WD stars
\citep[see][for the case of ZZ Ceti stars observed during the {\it
    Kepler} and {\it K2} missions]{2017ApJS..232...23H}. In the case
of RX~J2117, since the multiplets exhibited by the star are triplets
(two complete triplets and three triplets with one undetected component
each), this means that $\ell= 1$, and so, $C_{\ell,k}\sim 0.5$.  We
assume that the central peak of the multiplets ($m = 0$ components)
are 1997.75, 1124.15, 1038.12, 972.07 and 889.557~$\mu$Hz.  Therefore,
we can obtain an estimate of the rotational period of RX~J2117 of
$P_{\rm rot}=1/\Omega_{\rm R} \sim 1.04$ days. 

An interesting feature of the FT of RX~J2117 is a series of 
at least seven low-frequency harmonics of a $9.509\pm 0.004\ \mu$Hz 
signal. These are suggestive of variability from a photometric binary with an 
orbital period of $1.2172\pm0.0005$\,days. Extracting and analyzing 
light curves from every individual pixel in the TESS target pixel file 
for TIC\,117070953 reveals a clear binary eclipse signature. This is 
strongest outside the aperture used by the TESS pipeline to extract 
the light curve for RX~J2117, but near enough to contribute some light 
to the aperture. We conclude that the eclipses do not originate 
from the PNNV target. A public software tool to aid the 
identification of contaminating variable stars in the 
large 21\arcsec\ TESS pixels is under development (Higgins \& Bell, 
in prep.).

\subsection{HS~2324}
\label{obser-hs2324}

\begin{figure}[h!] 
\includegraphics[clip,width=1.0\linewidth]{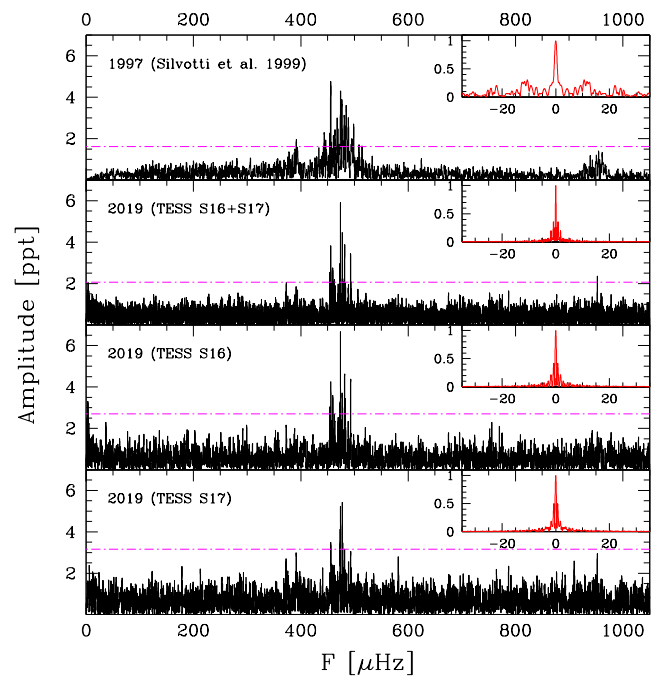}
\caption{Comparison between the amplitude spectrum of HS~2324 in 1997 
\citep{1999A&A...342..745S} and 2019 ({\it TESS} data).
The detection threshold (magenta dash-dotted line) in 2019 is slightly
higher (2.07 ppt with sector 16 and 17 together vs 1.62 ppt in 1997),
but the spectral window (red, upper-right panels) is much cleaner. 
This partially explains the higher density of peaks in 1997. 
The low level of noise below $\sim$100~ $\mu$Hz in 1997 is due
to artificial filtering.}
\label{fig:HS2324_1997_vs_2019_dfts} 
\end{figure}

\begin{figure}
\begin{minipage}[c]{8.5cm}
\hspace{4mm}
\includegraphics[clip,width=1.0\linewidth]{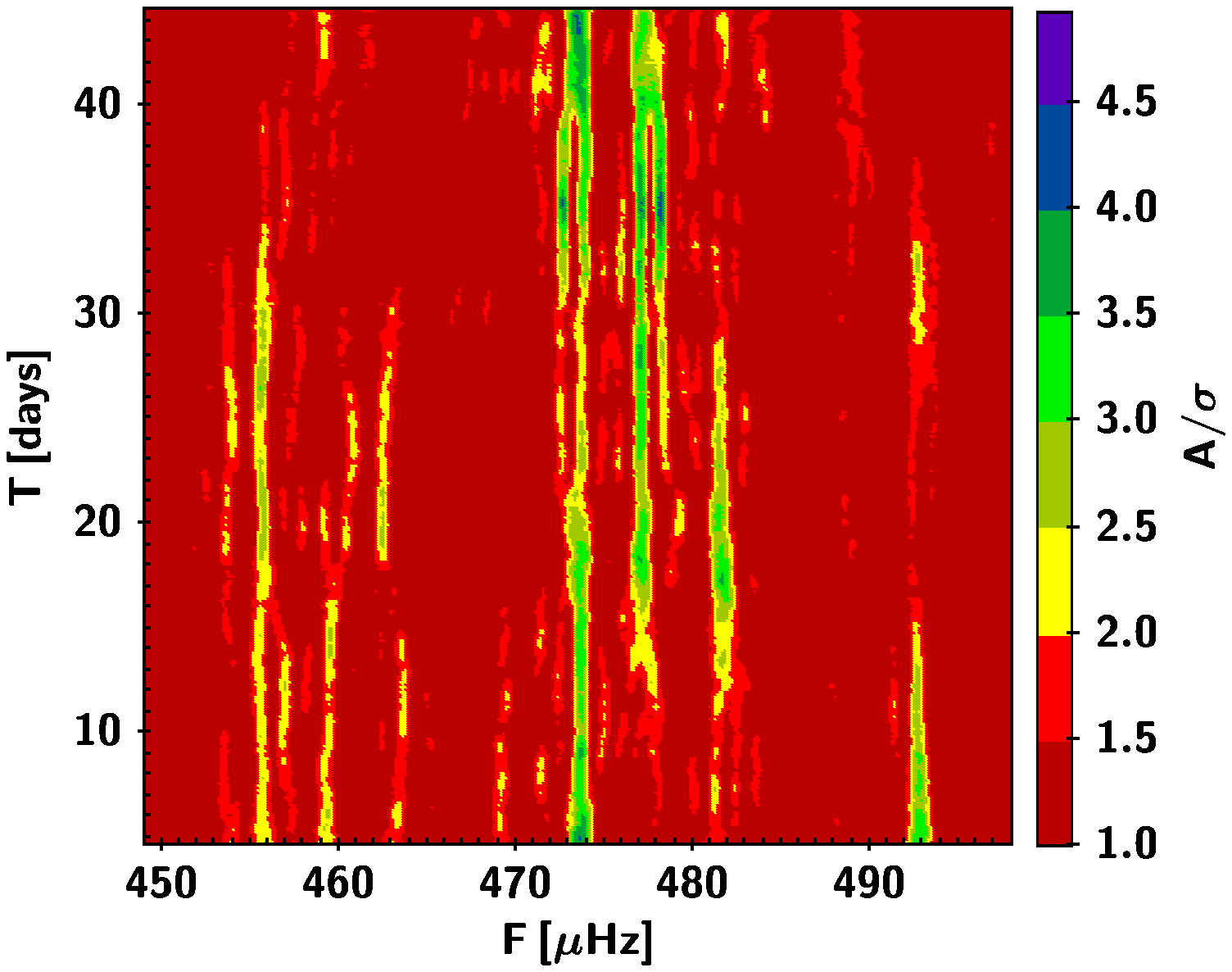}
\end{minipage}
\begin{minipage}[c]{8.5cm}
\hspace{4.8mm}
\includegraphics[clip,width=0.97\linewidth]{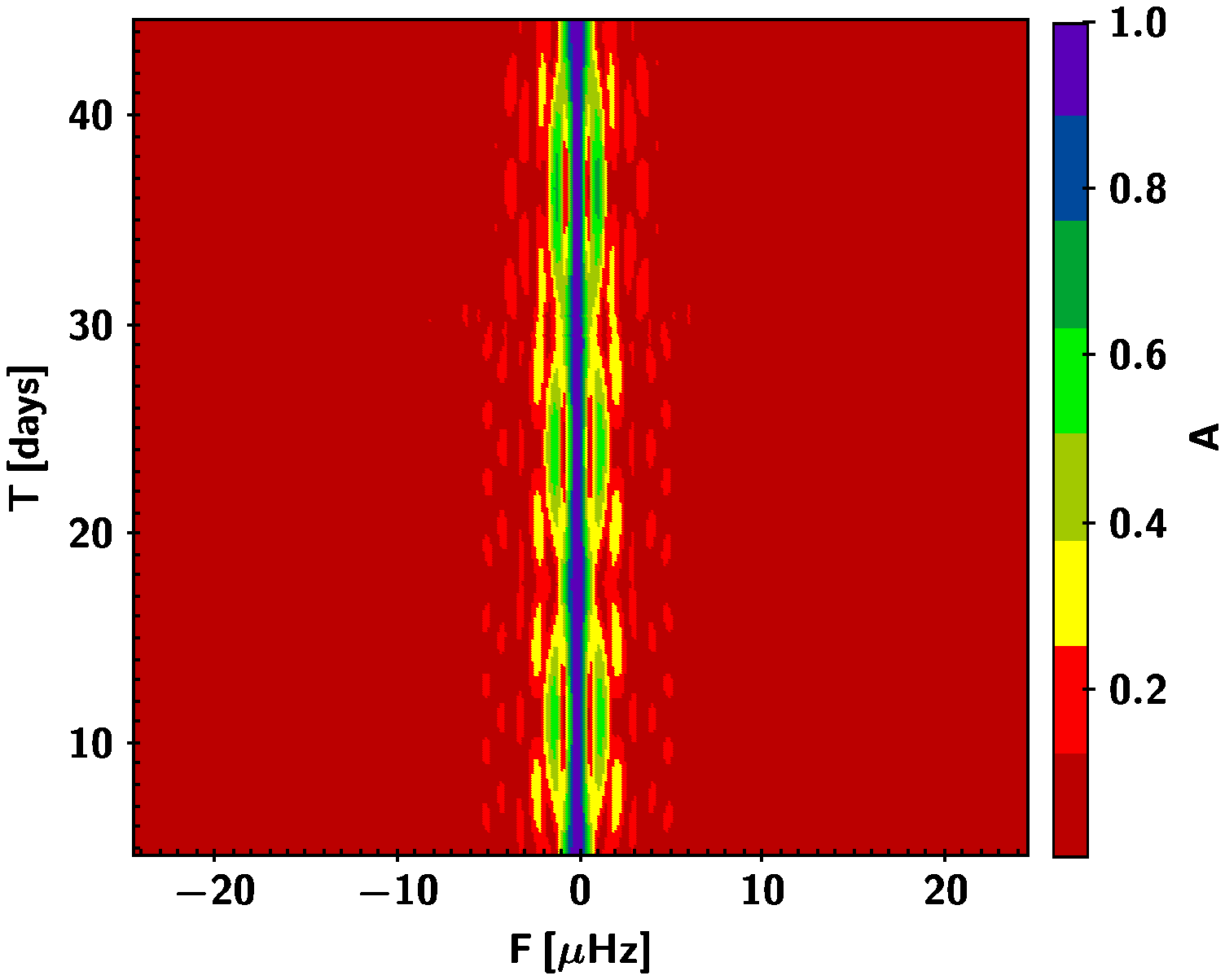}
\end{minipage}
\caption{Top: sliding Fourier Transform (sFT) of HS~2324 in the region between 449 and 498~$\mu$Hz. 
The color-coded amplitude is given in $\sigma$ units, i.e. amplitude divided by the mean noise of each FT.
Since each FT is computed on a short subrun of 9.3 days, the FT mean noise is high and this is why the color-coded S/N appears much lower than in Table \ref{table:HS2324}.
Bottom: sliding Window Function (sWF) obtained using a single sinusoid of
constant frequency and constant amplitude of one, without any noise.
The frequency scale is the same in both panels.
See text for more details on sFT/sWF computations.}
\label{fig:HS2324-sFT} 
\end{figure} 

\begin{table}
\centering
\caption{Independent frequencies, periods, and 
amplitudes (and their uncertainties) and the 
signal-to-noise ratio in the  data of HS~2324.}
\begin{tabular}{ccccr}
\hline
\noalign{\smallskip}
Peak & $\nu$     & $\Pi$  &  $A$   & S/N \\
  & ($\mu$Hz)  &  (s)   &  (ppt) &   \\
\noalign{\smallskip}
\hline
\noalign{\smallskip}
f$_{\rm 1}$ &  372.849(23) &   2682.05(17) &   2.04(38) &   4.6  \\
f$_{\rm 2}$ &  453.928(19) &   2202.99(9) &   2.52(38) &   5.7  \\
f$_{\rm 3}$ &  455.909(12) &   2193.42(6) &   4.00(38) &   9.0  \\
f$_{\rm 4}$ &  459.709(16) &   2175.29(8) &   2.96(38) &   6.7  \\
f$_{\rm 5}$ &  462.755(19) &   2160.97(9) &   2.50(38) &   5.6  \\
f$_{\rm 6}$ &  473.900(8)  &   2110.152(35) &   6.01(38) &  13.6  \\
f$_{\rm 7}$ &  477.316(11) &   2095.046(49) &   4.26(38) &   9.6  \\
f$_{\rm 8}$ &  481.670(13) &   2076.11(5) &   3.73(38) &   8.4  \\
f$_{\rm 9}$ &  488.458(22)$^*$ &   2047.26(9) &   2.10(38) &   4.7  \\
f$_{\rm 10}$ &  492.769(16) &   2029.35(7) &   3.12(39) &   7.0  \\
f$_{\rm 11}$ &  493.215(22) &   2027.52(9) &   2.32(39) &   5.2  \\
f$_{\rm 12}$ &  952.492(22) &   1049.877(24) &   2.16(38) &   4.9 \\
\noalign{\smallskip}
\hline
\noalign{\smallskip}
\multicolumn{5}{l}{$^*$Alternatively, it could be 488.984~$\mu$Hz.
The two close peaks}\\
\multicolumn{5}{l}{ \hspace{0.5mm} have similar amplitude
and when one of the two is selected,}\\
\multicolumn{5}{l}{\hspace{0.5mm} the other 
one gets down below the detection threshold.}
\end{tabular}
\label{table:HS2324}
\end{table}

HS~2324 (TIC~352444061, $T_{\rm mag}= 14.81$, $m_{\rm V}= 15.41$), was
observed for 49.15 days on Sectors 16 and 17 with a duty cycle of
91.02\%. The light curve is shown in Figure
\ref{fig:lightcurves}.  In Fig. \ref{fig:MODE-ID} we show the FT.   
In Table \ref{table:HS2324} we list the periods above 0.1\% FAP level of 2.07 ppt (corresponding to a S/N of 3.23) detected in the {\it TESS} light curve.  While 11
frequencies are populating a small region of the power spectrum
between $\sim 450\ \mu$Hz and $\sim 500\ \mu$Hz, one frequency is
located in the low frequency region ($\sim 372.8\ \mu$Hz) and a peak is
found in the high frequency region ($\sim 952.5\ \mu$Hz). 
The signal with $952.5\ \mu$Hz is close to twice the
main frequency group and thus we can not exclude that it is
a linear combination of the main peaks. However, it does not
correspond exactly to any combination, the closest combination
being at $951.2\ (= 473.9+477.3)\ \mu$Hz.
A direct comparison between the amplitude spectrum in 1997 
\citep{1999A&A...342..745S} and the {\it TESS} amplitude spectrum of 2019 is shown in 
Fig. \ref{fig:HS2324_1997_vs_2019_dfts}.

\cite{1999A&A...342..745S}, with a frequency resolution of $1.4\
\mu$Hz, suggested a  possible rotation period of $\sim$2.3 days for
HS~2324.  They also proposed that the detected peaks were part of
dipole and quadrupole  asymptotic sequences, elaborating on period
spacings of 18.8 s and 10.4 s for  $\ell = 1$ and $\ell = 2$,
respectively. Here, we find that all the periodicities detected in {\it TESS} data can be associated to dipole sequences (see
Fig. \ref{fig:MODE-ID} and Sect. \ref{period-spacing-hs2324}).  
With a frequency resolution of $0.35\ \mu$Hz, we
find no clear indications of rotational multiplets, suggesting that the 
star may have a rotation period longer than $\sim$49 days (the duration 
of the run) and/or that the star is seen pole-on, or also that
the modes with $m=\pm 1$ are not excited during the observations
\citep[see, e.g.,][]{2015A&A...573A..52B,1991ApJ...378..326W}.

Notwithstanding the above, we have also examined the possibility that the two pairs of frequencies at $(481.67-477.316)\ \mu$Hz and $(492.769-488.458)\ \mu$Hz
are two ($\ell =1$) incomplete rotational triplets
with frequency separations of $4.354\ \mu$Hz and $4.310\ \mu$Hz, 
respectively. As a matter of fact, a rotational splitting of $\sim 4.3\ \mu$Hz 
can not be totally excluded, and it is very close to the $\sim 4.2\ \mu$Hz 
splitting found by \cite{1999A&A...342..745S}. 
If these two pairs of frequencies were actually part of two rotational triplets, then only one component of each pair should be used in the
assessment of the dipole period spacing in Sect. \ref{period-spacing-hs2324}. 
As we shall see in that Section, the assessment of the period-spacing 
of HS~2324 remains almost unaltered even if we remove two frequencies 
from the analysis. 

In Fig. \ref{fig:HS2324-sFT} we show the sliding FT (sFT) of HS~2324 
in the main region of the spectrum (top panel), that can be compared 
with the sliding Window Function (sWF, bottom panel). We see that most frequencies are 
relatively stable, with some variations in amplitude.
This figure clearly exclude the damping oscillator hypothesis
that was considered by \citet{1999A&A...342..745S}.
Some additional details on sFT/sWF computations are given at the end of Section \ref{obser-k1-16}. For HS~2324 each data subset includes 6715 data points 
(corresponding to $\sim$9.3 days for continuous data) and the step between 
one subset and the next corresponds to 51 data points  (0.07 days for continuous 
data).

To continue a photometric monitoring of this star would be useful in order to 
detect new pulsation frequencies, verify the presence of rotational multiplets, and confirm the identification of some $\ell= 2$ pulsation modes 
(see Sect. \ref{modelling-hs2324} and Table \ref{table:HS2324-asteroseismic-model}).

\subsection{NGC~6905}
\label{obser-ngc6905}

\begin{table}
\centering
\caption{Independent frequencies, periods, and 
amplitudes (and their uncertainties) and the 
signal-to-noise ratio in the  data of NGC~6905.}
\begin{tabular}{ccccr}
\hline
\noalign{\smallskip}
Peak & $\nu$    &  $\Pi$  &  $A$   &  S/N \\
 & ($\mu$Hz)      &  (s)   & (ppt)   &   \\
\noalign{\smallskip}
\hline
\noalign{\smallskip}
f$_{\rm 1}$ & 1155.415(43) &    865.490(32) &   3.1(6) &   4.2  \\
f$_{\rm 2}$ & 1168.752(22) &    855.613(16) &   6.1(6) &   8.4  \\
f$_{\rm 3}$ & 1221.170(38) &    818.887(26) &   3.5(6) &   4.8  \\
f$_{\rm 4}$ & 1224.984(17) &    816.337(11) &   8.0(6) &  11.1  \\
\noalign{\smallskip}
\hline
\end{tabular}
\label{table:NGC6905}
\end{table}

NGC~6905 (TIC~402913811, $T_{\rm mag}= 13.90$, $G=14.5545 \pm 0.0027
$), was  observed for 26.85 days on Sector 14. {\it TESS} observations
of NGC~6905 started on 18 July 2019 and lasted on 15 August
2019. These observations yield $13\,615$ data points after we removed
the outliers. The final light curve after detrending is shown in
Figure \ref{fig:lightcurves}.  NGC~6905 is a good illustration of 
cross contamination in {\it TESS}
photometry.  Each {\it TESS} pixel comprises 21 arcsec.  NGC~6905's planetary nebula
extends to 32 arcsec, and there is an additional target ({\it Gaia} magnitude 
of 14.5) a mere 10 arcsec from NGC~6905.  This contamination reflects itself as additional
noise in the FT of NGC~6905.  Indeed, NGC~6905 has the highest average noise level 
(0.73 ppt) of our sample. All the oscillation amplitudes  that are extracted
from the light curve are reduced due to this  contamination from
the nearby stars and nebula.  The FT of NGC~6905 showed
a few signals within S/N of 4.6. We calculated the threshold at 0.1\%
false alarm probability as 3.35 ppt corresponding to S/N of
4.6.  The average noise level for NGC6905 is 0.73 ppt. The peaks that 
are extracted from the light curve appear at 4.14, 8.14, 4.67 and 10.68 $\sigma$ 
($\sigma$ is the median of noise level of amplitude).  
We included the peak of 1155.415 $\mu$Hz 
as a possible oscillation. Additionally, we extracted three more periodicities from the light
curve. These three frequencies, 1168.752, 1221.170 and 1224.984~$\mu$Hz,
are robust detections with S/N of 8.4, 4.8 and 11.1,
respectively. After the extraction of these three peaks from the light
curve, we calculated the average noise level of amplitude spectra and
defined the new detection threshold. The average noise level of the FT
excluding these three modes is remained almost the same as 0.73 ppt
(the difference is negligible with 0.001 ppt). The 0.1\% FAP was also
not affected drastically and found as 3.34 ppt (including all the 
peaks 0.1\% FAP = 3.35).   If we consider the detection threshold 4
$\sigma$ of 3 ppt, then 1155.415~$\mu$Hz (f$_{\rm 1}$) can
conveniently be evaluated as true pulsational peak.  These four
extracted peaks are shown in Fig. \ref{fig:MODE-ID} and listed in
Table \ref{table:NGC6905}.

%

\subsection{NGC~1501}
\label{obser-ngc1501}

NGC~1501 (TIC~084306468, $T_{\rm mag} = 12.40$, $m_{\rm V}= 13.0$) was
observed on sector 19, which spanned about 25 days from 27 November
2019 to 24 December 2019.  We have performed the  FT 
from the light curve shown in Fig. \ref{fig:lightcurves}, consisting of
16893 measurements with a duty cycle of 92.8\%. The frequency spectra
shows a rich content of peaks with 16 pulsational signals above the
detection limit of 0.1\%\, FAP= 0.53 ppt. In Table~\ref{table:NGC1501}
we show the list of periods of NGC~1501. The period spectrum 
detected with {\it TESS} is markedly different from that found by 
\cite{1996AJ....112.2699B}. Indeed, the 11 periods derived by \cite{1996AJ....112.2699B} 
are more evenly distributed over a  wider range of periods. A graphic 
comparison between the ground- and space-based period 
spectra of NGC~1501 is made in Sect. \ref{period-spacing-ngc1501}.

 In Fig. \ref{fig:MODE-ID} we display
the  FT of NGC~1501 from {\it TESS}. 
The pulsation power concentrates on a frequency (period)
interval from $481\ \mu$Hz (2077 s) to $797\ \mu$Hz (1254.6 s).  The
highest-amplitude ($A= 1.938$ ppt) peak is located at a frequency of
$756.954\ \mu$Hz in the high-frequency region of the spectrum.  The
other bunch of peaks are in the high-frequency region between $718\
\mu$Hz and $797\ \mu$Hz.
Concerning the peak at $\sim 68\ \mu$Hz, this is probably due to contamination,
since it is not the result of a difference of other frequencies. Also, 
the corresponding period at $\sim 14\,800$ s is too long for GW Vir stars, 
so that it cannot represent an eigenmode of the star. Since this mode is
not included in the seismic analysis of this study, we have not considered it 
further in the paper.

\begin{table}
\centering
\caption{Independent frequencies, periods and 
amplitudes (and their uncertainties) and the 
signal-to-noise ratio in the  data of NGC~1501.}
\begin{tabular}{ccccr}
\hline
\noalign{\smallskip}
Peak & $\nu$      &  $\Pi$  &  $A$   &  S/N \\
& ($\mu$Hz) &   (s)   & (ppt)  &    \\
\noalign{\smallskip}
\hline
\noalign{\smallskip}
f$_{\rm 1}$ &   67.617(13) &  14789.2(2.9) &   1.77(9) &  15.4  \\
f$_{\rm 2}$ &  481.465(43) &   2077.00(19) &   0.55(9) &   4.7  \\
f$_{\rm 3}$ &  491.971(34) &   2032.64(14) &   0.69(9) &   6.0  \\
f$_{\rm 4}$ &  512.428(20) &   1951.49(8) &   1.18(9) &  10.2  \\
f$_{\rm 5}$ &  562.655(31) &   1777.29(10) &   0.77(9) &   6.7  \\
f$_{\rm 6}$ &  565.448(39) &   1768.51(12) &   0.60(9) &   5.2  \\
f$_{\rm 7}$ &  718.266(35) &   1392.24(7) &   0.67(9) &   5.8  \\
f$_{\rm 8}$ &  723.958(19) &   1381.295(36) &   1.23(9) &  10.7  \\
f$_{\rm 9}$ &  731.915(14) &   1366.279(27) &   1.65(9) &  14.3  \\
f$_{\rm 10}$ &  737.094(34) &   1356.68(6) &   0.69(9) &   6.0  \\
f$_{\rm 11}$ &  741.039(42) &   1349.46(8) &   0.57(9) &   4.9  \\
f$_{\rm 12}$ &  743.012(25) &   1345.872(44) &   0.96(9) &   8.3  \\
f$_{\rm 13}$ &  756.948(13) &   1321.095(23) &   1.93(9) &  16.8  \\
f$_{\rm 14}$ &  762.954(23) &   1310.696(40) &   1.07(10) &   9.3  \\
f$_{\rm 15}$ &  763.892(22) &   1309.086(37) &   1.14(10) &   9.9  \\
f$_{\rm 16}$ &  797.047(27) &   1254.632(43) &   0.87(9) &   7.5  \\

\noalign{\smallskip}
\hline
\end{tabular}
\label{table:NGC1501}
\end{table}

\subsection{NGC~2371}
\label{obser-ngc2371}

NGC~2371 (TIC\,446005482, $T_{\rm mag}= 12.90$, $m_{\rm V}= 13.5$) 
 was observed in sector 20 of {\it TESS} (Table \ref{DOVlist},
Fig. \ref{fig:lightcurves}). We extracted 4 frequencies above 
the detection threshold of 0.9 ppt, corresponding to a 0.1 \% FAP 
(Table \ref{table:NGC2371}). The periodicities that we extracted
from the light curve  populate a small region of the power spectrum
with periods (frequencies) between 878.524 s (1138.272~$\mu$Hz) and 1032.558 s 
(968.467~$\mu$Hz). In Fig. \ref{fig:MODE-ID} we show the  FT of NGC~2371.
The peak at $\sim 106\ \mu$Hz is probably due to contamination.
This mode is not included in the seismic analysis and it is not
considered further in the paper.

Given the frequency resolution of $0.44\ \mu$Hz and the average noise
level of $0.22$~ppt, we did not detect any rotational multiplets in
the frequency spectra. However, there is a hint regarding a possible
constant period-spacing pattern (see Sect. \ref{fig:MODE-ID}). 

\begin{table}
\centering
\caption{Independent frequencies, periods, and 
amplitudes (and their uncertainties) and the 
signal-to-noise ratio in the  data of NGC~2371.}
\begin{tabular}{ccccr}
\hline
\noalign{\smallskip}
Peak & $\nu$      &  $\Pi$  &  $A$   &  S/N \\
 & ($\mu$Hz) &   (s)   & (ppt)  &    \\
\noalign{\smallskip}
\hline
\noalign{\smallskip}
f$_{\rm 1}$ &  105.657(25) &   9464.6(2.2) &   1.60(17) &    8.0  \\
f$_{\rm 2}$ & 1032.553(38) &    968.474(36) &   1.03(17) &   5.1  \\
f$_{\rm 3}$ & 1080.538(30) &    925.465(26) &   1.31(17) &   6.5  \\
f$_{\rm 4}$ & 1117.180(42) &    895.111(34) &   0.93(17) &   4.6  \\
f$_{\rm 5}$ & 1138.274(43) &    878.523(33) &   0.91(17) &   4.5  \\
\noalign{\smallskip}
\hline
\end{tabular}
\label{table:NGC2371}
\end{table}


\subsection{K~1$-$16}
\label{obser-k1-16}

\begin{table}
\centering
\caption{Independent frequencies, periods, and amplitudes and the 
signal-to-noise ratio in the data of K~1$-$16.}
\begin{tabular}{ccccccc}
\hline
\noalign{\smallskip}
Peak & $\nu$     & $\Pi$ & $A$   & S/N & $A_{max}$ & S/N\\
     & ($\mu$Hz) &  (s)  & (ppt) &     & (ppt)     &\\
\noalign{\smallskip}
\hline
\noalign{\smallskip}
(f$_{\rm 1}$ & 229.75     & 4352.6 & 1.70$^1$ & 4.2 & 2.43$^3$ & 4.2)\\
 f$_{\rm 2}$ & 471.74     & 2119.8 & 1.78$^1$ & 4.4 & 2.30$^4$ & 4.0\\
 f$_{\rm 3}$ & 498.73     & 2005.1 & 0.69$^2$ & 3.3 & 4.09$^3$ & 7.1\\
 f$_{\rm 4}$ & 541.63$^6$ & 1846.3 & 1.08$^2$ & 5.1 & 5.50$^3$ & 9.5\\
 f$_{\rm 5}$ & 568.97     & 1757.6 & 2.00$^1$ & 4.9 & 2.93$^3$ & 5.1\\
 f$_{\rm 4}$ & 581.33$^6$ & 1720.2 & 0.90$^2$ & 4.3 & 3.00$^5$ & 4.6\\
 f$_{\rm 6}$ & 663.04$^7$ & 1508.2 & 0.80$^2$ & 3.8 & 3.27$^3$ & 5.7\\
\noalign{\smallskip}
\hline
\noalign{\medskip}
\multicolumn{7}{l}{\small $^1$ Frequency and amplitude from sectors 22+23.}\\
\multicolumn{7}{l}{\small $^2$ Frequency and amplitude from all sectors.}\\
\multicolumn{7}{l}{\small $^3$ Maximum amplitude in sector 23.}\\
\multicolumn{7}{l}{\small $^4$ Maximum amplitude in sector 22.}\\
\multicolumn{7}{l}{\small $^5$ Maximum amplitude in sector 17.}\\
\multicolumn{7}{l}{\small $^6$ Could be aliases of super-Nyquist counterparts
at 7791.7 and}\\ 
\multicolumn{7}{l}{\small \hspace{1.5mm} 7752.0~$\mu$Hz respectively 
(cf. text). }\\
\multicolumn{7}{l}{\small $^7$ A close peak at 663.67~$\mu$Hz could be either 
an independent}\\ 
\multicolumn{7}{l}{\small \hspace{1.5mm} frequency or the product of frequency 
variations (cf. Fig. \ref{fig:sft1}).}\\
\end{tabular}
\label{table:K1-16}
\end{table}

\begin{figure*}
\begin{center}
\begin{minipage}[c]{8.8cm}
\hspace{8mm}
\includegraphics[height=5.6cm]{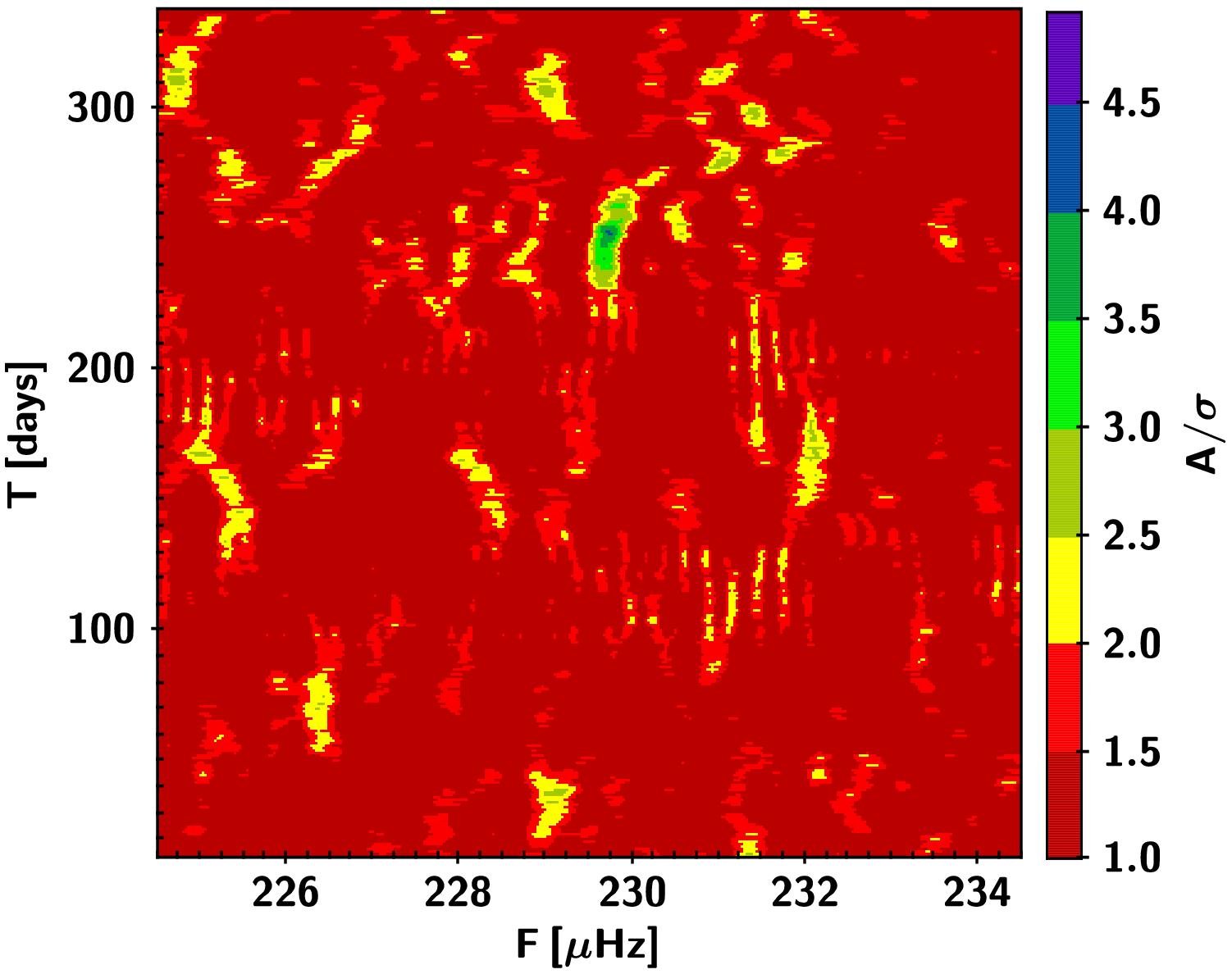}
\end{minipage}
\begin{minipage}[c]{8.8cm}
\hspace{8mm}
\includegraphics[height=5.6cm]{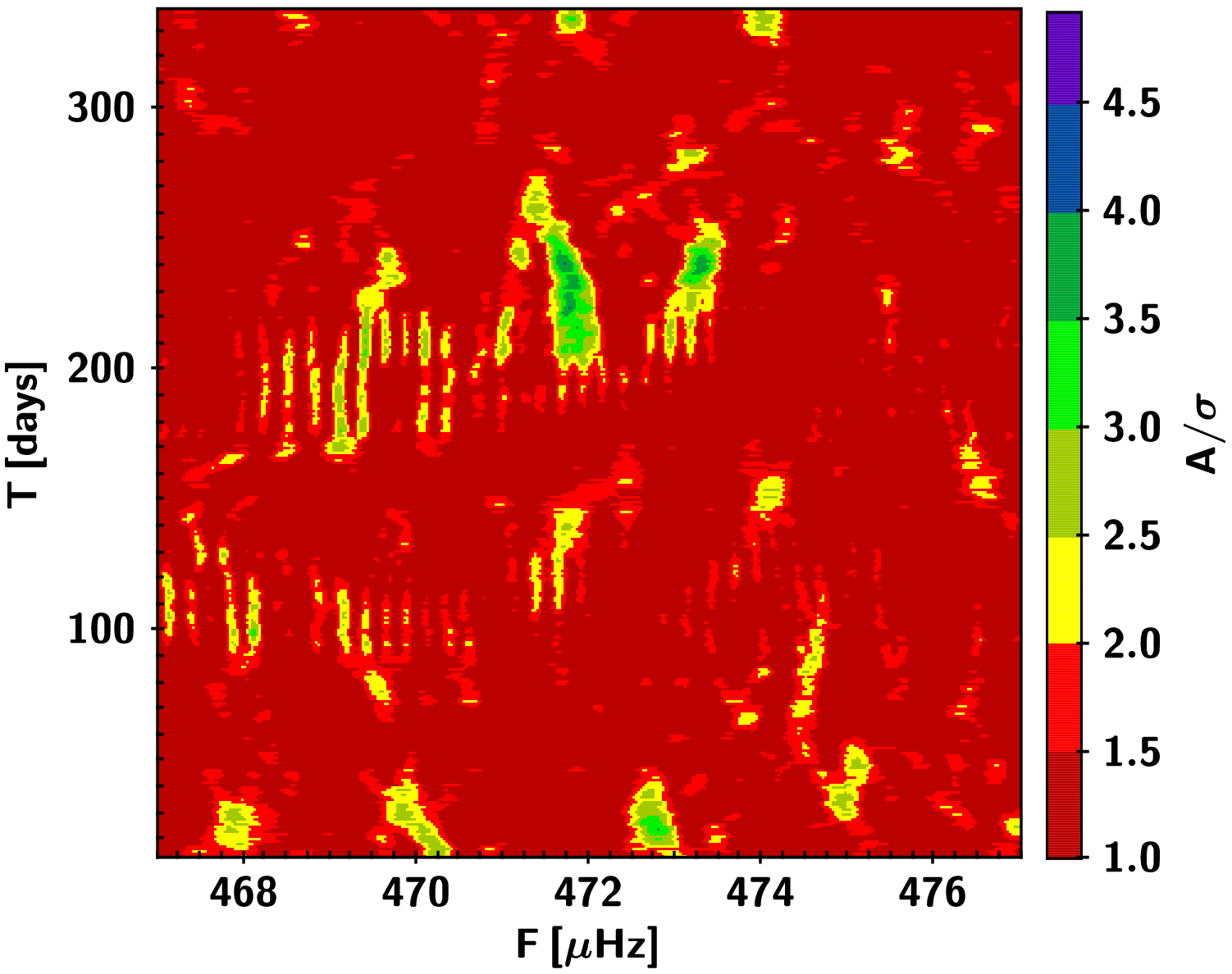}
\end{minipage}
\begin{minipage}[c]{7.8cm}
\vspace{1.5mm}
\hspace{-2.2mm}
\includegraphics[clip,height=5.6cm]{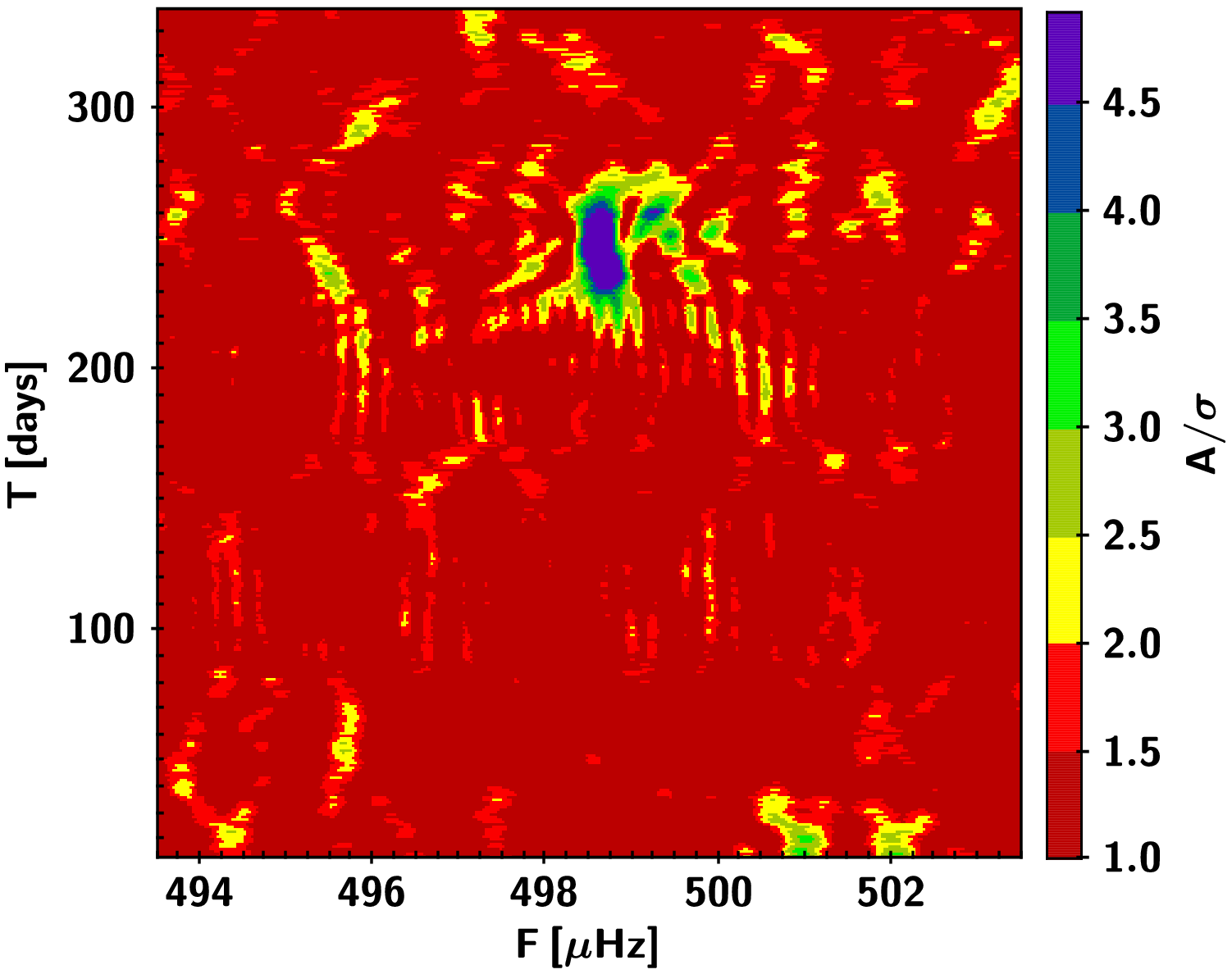}
\end{minipage}
\begin{minipage}[c]{7.8cm}
\vspace{1.5mm}
\hspace{8mm}
\includegraphics[clip,height=5.6cm]{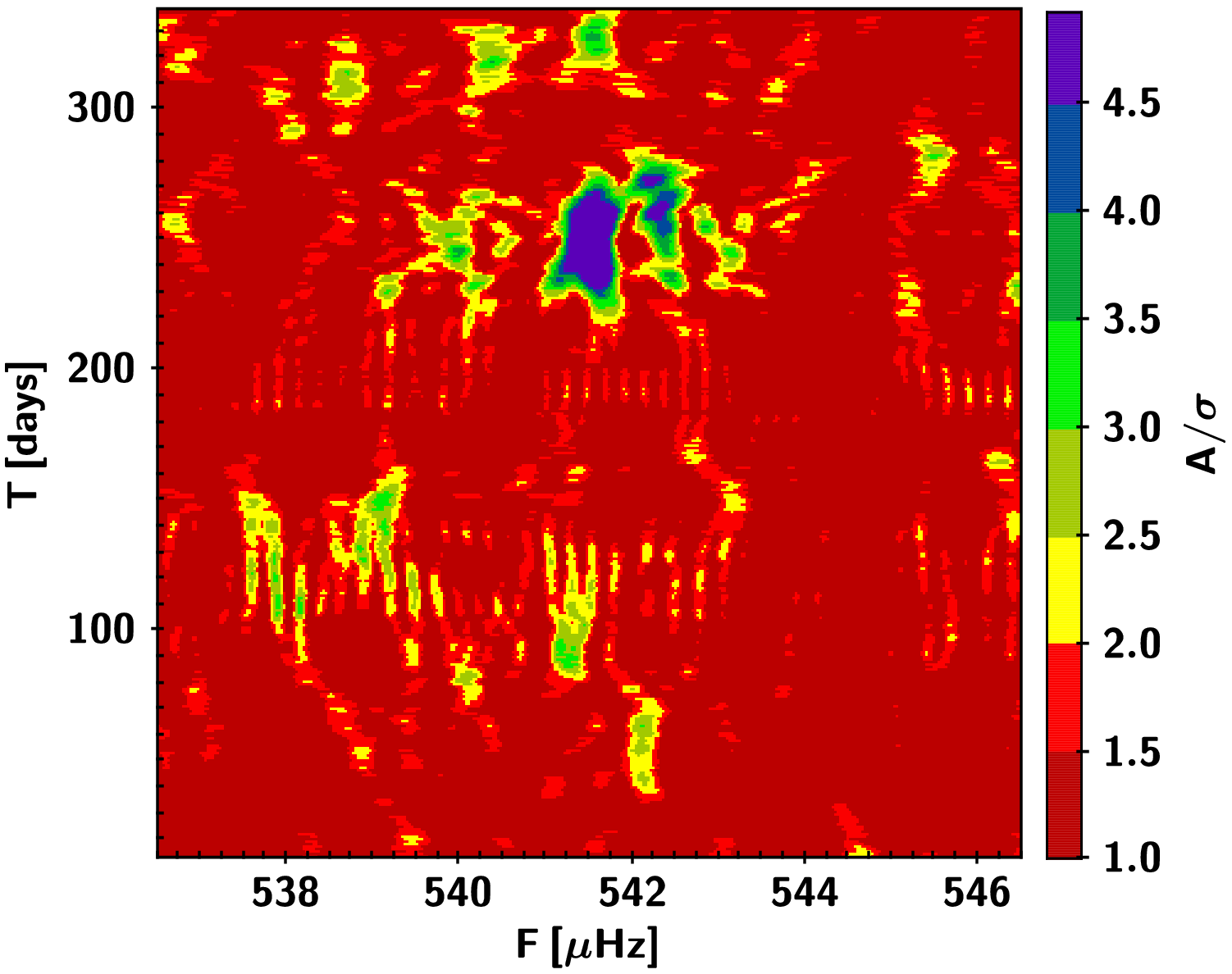}
\end{minipage}
\begin{minipage}[l]{7.8cm}
\vspace{1.5mm}
\hspace{-2.2mm}
\includegraphics[clip,height=5.6cm]{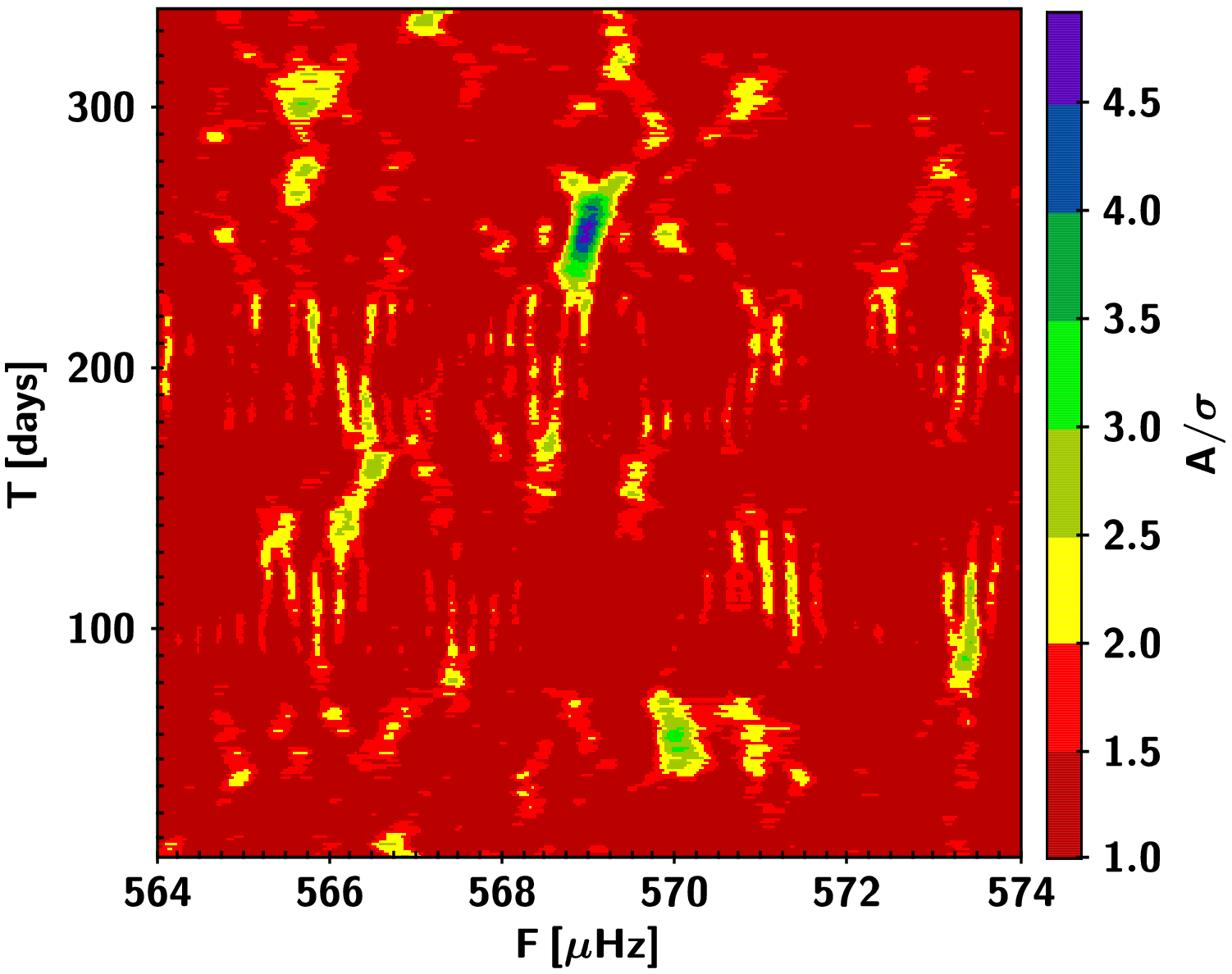}
\end{minipage}
\begin{minipage}[r]{7.8cm}
\vspace{1.5mm}
\hspace{8mm}
\includegraphics[clip,height=5.6cm]{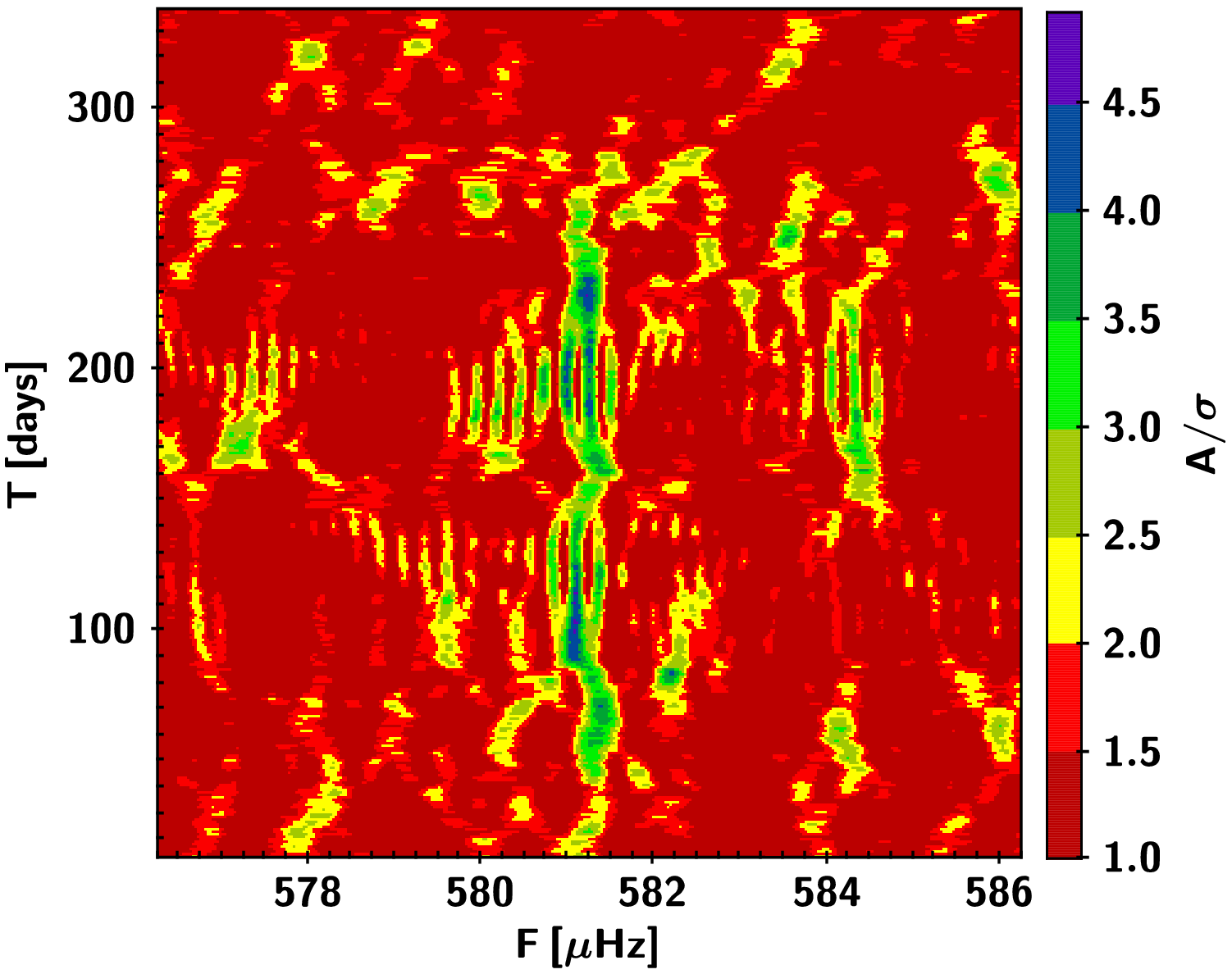}
\end{minipage}
\begin{minipage}[l]{7.8cm}
\vspace{1.5mm}
\hspace{-2.2mm}
\includegraphics[clip,height=5.6cm]{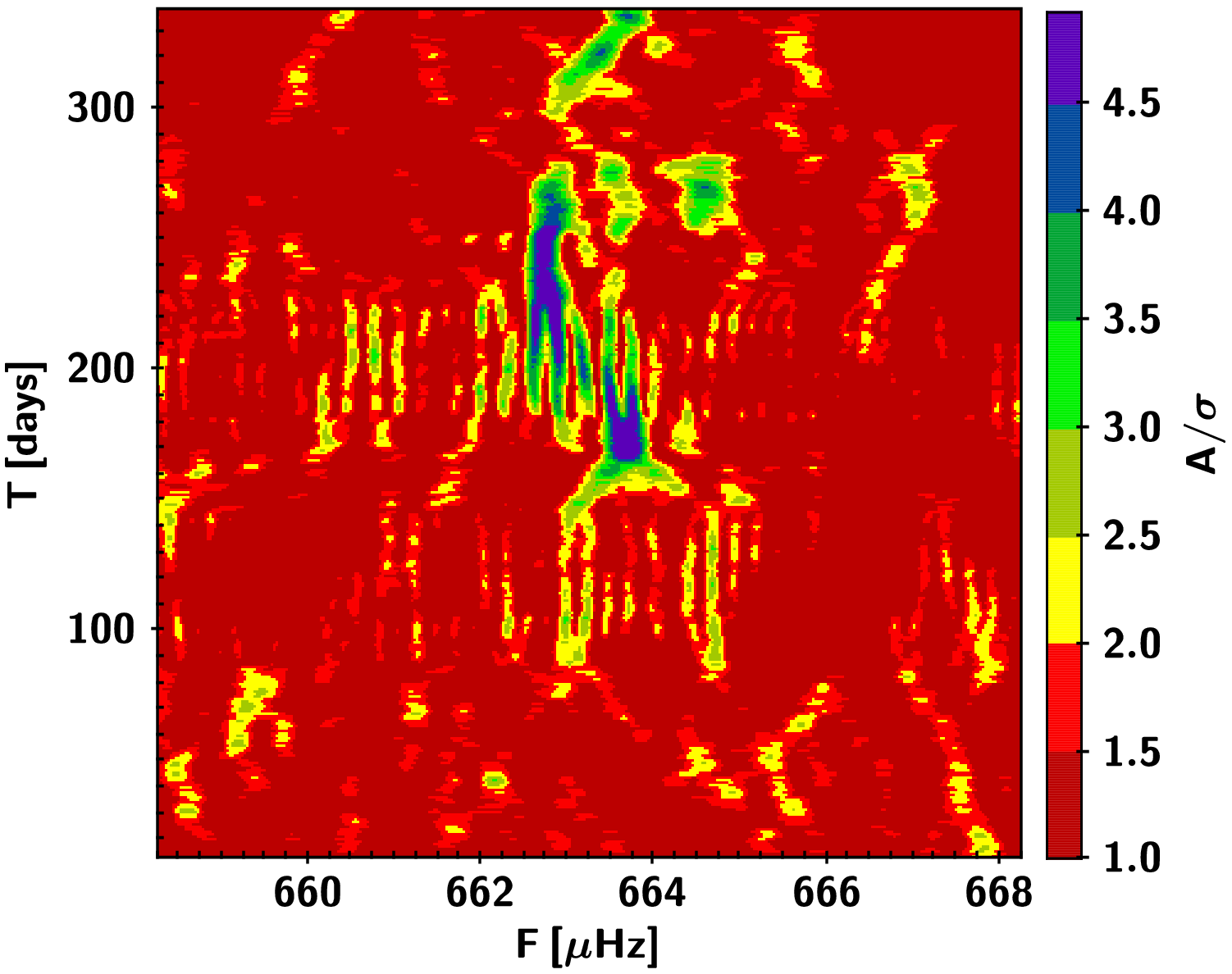}
\end{minipage}
\begin{minipage}[r]{7.8cm}
\vspace{2.2mm}
\hspace{8mm}
\includegraphics[clip,height=5.52cm]{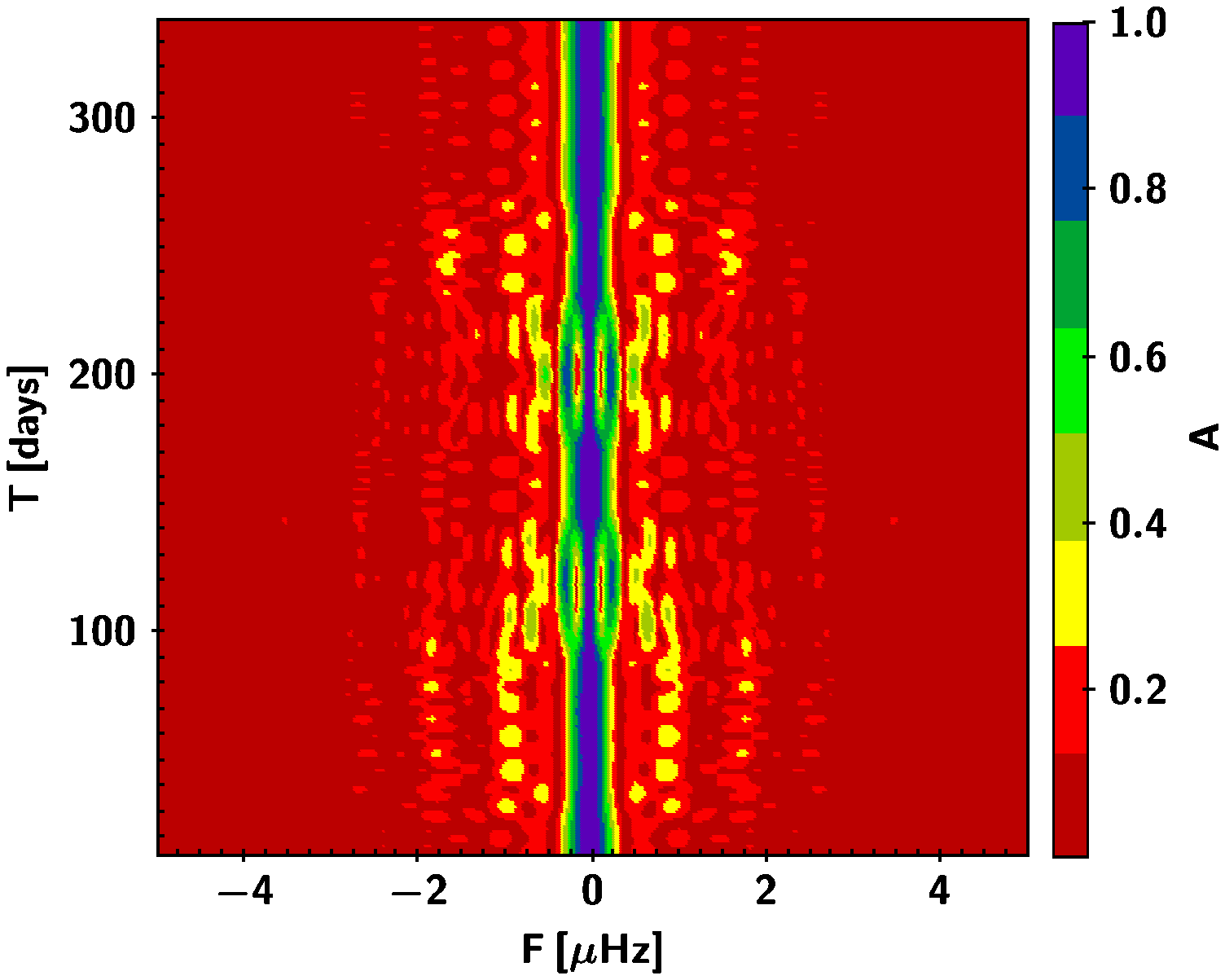}
\end{minipage}
\caption{Panels 1-7: sliding Fourier Transform (sFT) of K~1$-$16 centered
on the peaks at $\sim$229.7, $\sim$471.7, $\sim$498.7,  $\sim$541.6, 
$\sim$569.0, $\sim$581.3 and $\sim$663.0 $\mu$Hz.
The color-coded amplitudes are given in $\sigma$ units, i.e. amplitude
divided by the mean noise of each FT.
The oscillations are not stable over the course of the \textit{TESS} 
observations and in the first $\approx$200 days (up to sector 22) some of
the modes are absent.
Panel 8: sliding Window Function (sWF) obtained using a single sinusoid 
of constant frequency and constant amplitude of one, without any noise.
The frequency scale is the same in all panels.
See text for more details on sFT/sWF computations.}
\label{fig:sft1} 
\end{center}
\end{figure*} 

K~1$-$16 (TIC\,233689607, $T_{\rm mag}= 14.36$, $m_{\rm V}= 14.96$)
was observed on sectors $14-17$, $19-20$, and $22-26$ of {\it TESS},
between 18 July 2019 and 4 July 2020. The light curve of K~1$-$16 is
presented in Fig. \ref{fig:lightcurves}. If we look at individual
sectors, the pulsations come and go, which means that the amplitudes
and frequencies change significantly. When combining all data
together, we obtain a threshold of 0.93\,ppt, which implies that only
two peaks should be considered real. However, given that amplitudes
(and frequencies) vary significantly on timescales of weeks/months like
in other PG~1159 stars, a few more peaks have amplitudes well
above the threshold in single sectors. Adding also these peaks, we
obtain a list of five frequencies at S/N higher than 4.5
(Table~\ref{table:K1-16}).
Two more frequencies at 229.75 and 471.74~$\mu$Hz are included in
Table~\ref{table:K1-16} as they reach a S/N of 4.4 and 4.5 respectively,
when we consider only a section of the light curve in which they
reach the maximum amplitude, from BJD 2458904.35 to 2458954.87 and from
2458900.38 to 2458940.35, respectively (or days 221-271.5 and 
217-257 in Fig. \ref{fig:sft1}).
Moreover, we investigated the possibility that some of the frequencies
in Table~\ref{table:K1-16} may be aliases of super-Nyquist peaks.
The possibility of detecting super-Nyquist pulsation frequencies is 
well described by \citet{2015MNRAS.453.2569M}, who showed that 
introducing a time-offset between one sector and the next may greatly 
help. Even though (unfortunately) this suggestion was not applied 
to the \textit{TESS} observing strategy, however it happens that by chance 
a small offset of $\sim$24 seconds, i.e. 20\% of the 2-min sampling 
time, exists between sector 17 and 19. In another two cases, between 
sector 19-20 and 20-22, time is off by $\sim$4\% of the sampling time.
This allows us to compare the amplitude of each peak with its 
super-Nyquist counterpart. Without offsets, e.g. using only sectors 
22+23 in which the star shows the highest number of significant peaks,
these amplitudes are almost identical. But when we add the sectors 
with the time offsets, the amplitudes differ and we see a larger 
amplitude for the peak at 7752.0~$\mu$Hz (or 129.0\,s) respect to its 
sub-Nyquist counterpart at 581.3~$\mu$Hz.
We see a similar behavior also for the peak at 541.6~$\mu$Hz and
its super-Nyquist counterpart at 7791.7~$\mu$Hz (or 128.3\,s), although 
in this case the effect is less pronounced.
Therefore we can not totally exclude that the peak at 581.3~$\mu$Hz is
actually the alias of a super-Nyquist counterpart at 7752.0~$\mu$Hz
(and to less extent the same might be true for the sub/super-Nyquist
pair at 541.63/7791.7~$\mu$Hz).
However, we decided to favor the sub-Nyquist solution based on
findings of \citet{1984ApJ...277..211G}. Despite their poor frequency
resolution that could not really resolve the region between
$\approx$530 and $\approx 620\ \mu$Hz (cf. their Fig.2b), the main peak 
they found was at $\sim 590\ \mu$Hz, quite close to the 581.3~$\mu$Hz
frequency detected by {\it TESS}.
Moreover, their 50\,s binned integration (and sampling) time, with original 
integrations of 5 \,s, should have allowed them to detect periods
near 130\,s.
And finally we note that the peak at 7752.0~$\mu$Hz, if true, would have
a much larger amplitude due to smearing (see e.g. 
\citealt{2017ApJ...851...24B}), an order of magnitude larger 
than the amplitude detected in the \textit{TESS} light curve of 1.35 ppt 
(and up to 2.54 ppt in sector 22).


Since the profiles of the peaks associated to unstable modes in the
amplitude spectrum may be complex and a pre-whitening process gives
only rough results, we do not report frequency and amplitude 
uncertainties in Table~\ref{table:K1-16}. Frequencies and amplitudes 
are those obtained from the complete data set or from individual sectors
if the peak is detected only in some sectors. Moreover, we report in
Table~\ref{table:K1-16} also the maximum amplitude registered in a
single sector. 
The long photometric measurements of K~1$-$16 allow us
to construct sFTs to examine the temporal evolution of
the pulsation modes over the course of the \textit{TESS} observations
and highlight the dramatic changes in frequency and amplitude of this
star. First we divide the \textit{TESS} light curve into 500 subsets
of 17943 data each (corresponding to $\sim$24.9 days for continuous
data) and step them by 324 data (0.45~d for continuous data).
Afterward, we calculate the  FT of each subset and stack
them on top of each other. 
The sFTs of the seven detected frequencies are shown in 
Fig. \ref{fig:sft1} and can be compared with the sliding Window Function
(sWF) in the last panel of Fig. \ref{fig:sft1}.
The color-coded amplitudes of the sFTs are given in $\sigma$ units, i.e. 
amplitude divided by the mean noise of each FT.
In most cases the frequencies are totally absent (no pulsation) in the
first \textit{TESS} observation sectors, and acquire measurable
amplitudes only in sectors 22 and 23. This is particularly evident for
the peaks near 498.7, 541.6 and 569.0~$\mu$Hz.


\section{Period spacing}
\label{period-spacing}

The $g$ modes responsible for the brightness variations of WDs and
pre-WDs can be excited in a sequence of consecutive radial orders,
$k$, for each value of $\ell$. In the asymptotic limit ($k \gg \ell$),
$g$ modes of consecutive radial overtone are approximately uniformly
spaced in period \citep{1990ApJS...72..335T}. The period spacing
is given by 

\begin{equation}\label{eq:1}
\Delta \Pi_{\ell}^{\rm a} = \frac{{\Pi}_{0}}{\sqrt{\ell(\ell+1)}},
\end{equation} 
\noindent where $\Delta \Pi_{\ell}^{\rm a}$ is the asymptotic period spacing,
and $\Pi_{0}$ is a constant defined as:
\begin{equation}\label{eq:2}
\Pi_{0}=  \frac{2 \pi^2}{\left[ \int_{r_1}^{r_2}
\frac{N}{r} dr \right]}, 
\end{equation} 
\noindent where $N$ is the Brunt-V\"ais\"al\"a frequency. 

While for chemically homogeneous stellar models the asymptotic formula
(\ref{eq:1})  constitutes a very precise description of their
pulsational properties, the  $g$-mode period spacings in chemically
stratified PG~1159  stars show  appreciable departures from uniformity
caused by the mechanical resonance  called "mode trapping".   The
presence of one  or more narrow regions in which the abundances of
nuclear species  vary rapidly strongly modifies the character of
the resonant cavity in which modes should propagate as standing
waves. Specifically, chemical interfaces act like  reflecting  walls
that partially  trap  certain modes, forcing them to oscillate with
larger amplitudes in specific regions --- bounded either by two
interfaces or by one interface and the stellar center  or surface ---
and with  smaller  amplitudes  outside  of  those regions. The
requirement for a mode to be trapped is that the wavelength of its
radial eigenfunction matches the spatial separation between two
interfaces or between one interface and the stellar center or
surface. Mode trapping  has been the subject of intense study in the
context of stratified DA and DB WD pulsations \citep[see,
  e.g.,][]{1992ApJS...80..369B,1993ApJ...406..661B,2002A&A...387..531C}. In
the case of PG~1159 stars, mode trapping has been extensively explored
by \cite{1994ApJ...427..415K} and \cite{2006A&A...454..863C}; we refer
the reader to those works for details.  

\subsection{Period spectrum of RX~J2117} 
\label{period-spacing-rxj2117}

\begin{figure} 
\includegraphics[clip,width=1.0\columnwidth]{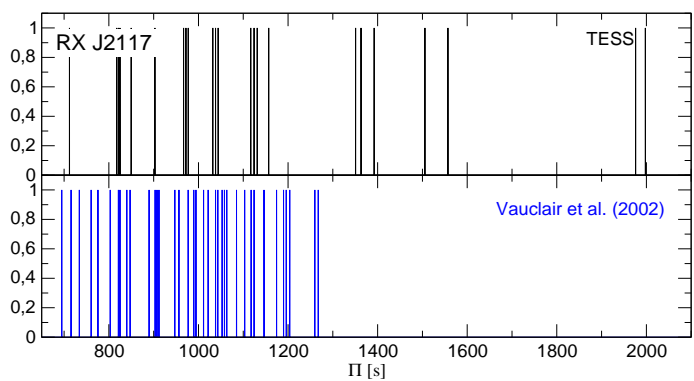}
\caption{Schematic distribution of the pulsation periods of RX~J2117
  according  to {\it TESS} (15 periods, black lines, upper panel), and
  according to \cite{2002A&A...381..122V}  (37 periods, blue lines,
  lower panel). The amplitudes have been arbitrarily  set to one for
  clarity.}
\label{fig:compara-tess-vea02} 
\end{figure}

When we compare the pulsation spectrum of RX~J2117 detected by {\it
  TESS} and that  observed through the ground-based monitoring by
\cite{2002A&A...381..122V}, we realize  that the pulsation spectra are
quite different. To begin with, the number of  pulsation periods
detected by {\it TESS}  and reported here (15 periods; see  Table
\ref{table:RXJ2117}) is substantially smaller than the number of
periods measured by \cite{2002A&A...381..122V}, who identified 37 ones
(see their Table 9). But the differences are not limited only to the
number of periods detected, which can be mainly attributed to the
small size of the {\it TESS} telescope, but also the {\it
  distribution} of the periods is completely different. In order to
envisage this, in Fig. \ref{fig:compara-tess-vea02} we have
schematically plotted the periods detected with {\it TESS} (upper
panel), and the periods detected by \cite{2002A&A...381..122V} (lower
panel), with arbitrary amplitudes set to one to facilitate
visualization. Clearly, the periods of the star detected by {\it TESS}
are distributed over a much wider range of periods, compared to the
periods measured by \cite{2002A&A...381..122V}. Only  six periods,
around 821, 825, 903, 1038, 1044 and 1124~s  are almost identical
between both sets of data.  The reason for the discrepancy  between
ground-based and space observations is that the {\it TESS} mission can
only detect large-amplitude modes, both because of its small telescope
(15~cm diameter), and because it only observes redder than
6000~$\textup{\r{A}}$, where the pulsation amplitudes are small.  The
WET ground-based telescopes  on which the data of
\cite{2002A&A...381..122V}  are based are much larger.

In order to extract as much information as possible with the tools of
asteroseismology, it is crucial to exploit all the pulsation data
available --- which should represent eigenvalues for the star.
Therefore, to identify the pulsation modes and determine the period
spacing  of RX~J2117, which is essential to estimate the stellar mass
(see Sect. \ref{modelling-rxj2117}), we decided to expand the list of
periods by adding the 20 dipole $m=0$ periods found by
\cite{2002A&A...381..122V} to the list of periods collected by  {\it
  TESS} (Table \ref{table:RXJ2117}).  Note that only half of these 20 $m=
0$ periods were directly detected, the others  were deduced through a
detailed analysis from the presence of the $m= -1$ and/or $m= +1$
components of each $\ell=1$ triplet.  For the periods close to 821~s,
972~s, 1038~s, and 1124~s, detected (or inferred) in both data sets,
we adopted the  periods measured by {\it TESS} because  they are in
general more accurate, due to the continuous long dataset.   The
extended list of periods to be used in our analysis contains 31
periods and is shown in Table \ref{table:RXJ2117-extended}.

\begin{table}
\centering
\caption{Enlarged list of periods of RX~J2117. Column 1 corresponds to
  20 $\ell= 1$ $m= 0$ periods measured by \cite{2002A&A...381..122V}
  (VEA02), and column 2 corresponds to the 15 periods detected by {\it TESS}   
  (Table \ref{table:RXJ2117}). The
  periods with an asterisk are the ones used in the linear least
  square fit (Fig. \ref{fig:fit-RXJ2117}).}
\begin{tabular}{cc|ccc}
\hline
\noalign{\smallskip}
$\Pi_i^{\rm O}$ (s) & $\Pi_{i}^{\rm O}$ (s) & $\Pi_{\rm fit}$ (s) & 
$\delta\Pi$ (s) & $\ell^{\rm O}$\\
  VEA02 & {\it TESS} & & & \\
\noalign{\smallskip}
\hline
\noalign{\smallskip}      
692.267*  &	      &  691.298 &  0.969 & 1 \\ 
712.975*  &	      &  712.967 &  0.008 & 1 \\
733.948*  &	      &  734.636 & $-0.688$ & 1 \\
757.354*  &	      &  756.305 &  1.049 & 1 \\
778.921*  &	      &  777.974 &  0.947 & 1 \\
799.495*  &	      &  799.643 & $-0.148$ & 1 \\
          & 817.375   &          &        & 1 \\ 
821.145   & 821.105*  &  821.312 & $-0.207$ & 1 \\
          & 824.880   &          &        & 1 \\
843.692*  &	      &  842.981 &  0.711 & 1 \\ 
885.736*  &	      &  886.319 & $-0.583$ & 1 \\
          & 902.761   &          &        & ? \\ 
907.489*  &	      &  907.988 & $-0.499$ & 1 \\
951.750*  &	      &  951.326 &  0.424 & 1 \\
          & 966.785   &          &        & 1 \\
972.247   & 972.073*  &  972.995 & $-0.922$ & 1 \\
 994.387* &	      &  994.664 & $-0.277$ & 1 \\
1016.467* &	      & 1016.333 &  0.134 & 1 \\
          & 1031.978  &          &        & 1 \\
1038.118  & 1038.120* & 1038.002 &  0.118 & 1 \\ 
          & 1044.041  &          &	  & 1 \\ 
1058.026* &	      & 1059.671 & $-1.645$ & 1 \\
1103.292* &	      & 1103.009 &  0.283 & 1 \\
1124.117  & 1124.156* & 1124.678 & $-0.522$ & 1 \\
          & 1131.200  &  	 &	  & 1 \\ 
1146.346* &	      & 1146.347 & $-0.001$ & 1 \\
1189.956* &	      & 1189.685 &  0.271 & 1 \\
          & 1350.870  &          &        & ? \\
          & 1557.010  & 1558.058 & $-1.048$ & 1 \\
          & 1976.060  &          &        & ? \\
          & 1997.760  &          &        & ? \\
\noalign{\smallskip}
\hline
\end{tabular}
\label{table:RXJ2117-extended}
\end{table}

\begin{figure} 
\includegraphics[clip,width=1.0\columnwidth]{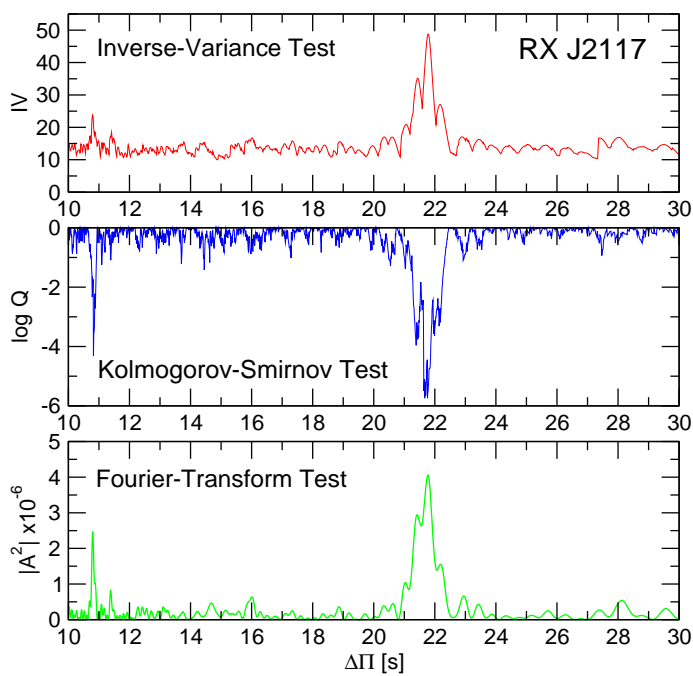}
\caption{I-V  (upper panel),  K-S  (middle panel), and  F-T  (bottom
  panel)  significance  tests  to  search  for  a constant  period
  spacing  in RX~J2117. The tests are applied to the set of 31
  pulsation periods of Table \ref{table:RXJ2117-extended}. A clear
  signal of a constant period spacing at  $\sim 21.8$~s is evident.
  See text for details.}
\label{fig:tests-RXJ2117} 
\end{figure}

\begin{figure} 
\includegraphics[clip,width=1.0\columnwidth]{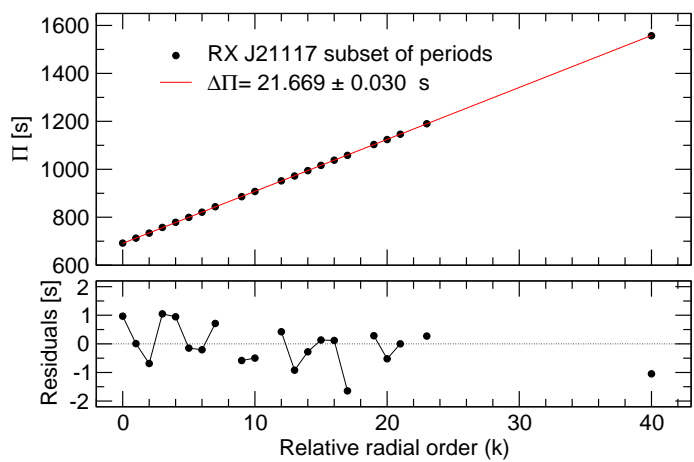}
\caption{Upper panel: linear least-squares fit to the 20 periods of
  RX~J2117 marked with asterisks in Table
  \ref{table:RXJ2117-extended}. The derived period spacing from this
  fit is $\Delta \Pi= 21.669\pm 0.030$~s.  Lower panel: the residuals
  of the period distribution relative to the mean period spacing,
  revealing signals of mode trapping in the period spectrum of
  RX~J2117. Modes with consecutive radial order are connected with a
  thin black line.}
\label{fig:fit-RXJ2117} 
\end{figure} 

We  searched for a  constant  period  spacing  in  the  data of
RX~J2117  using the  Kolmogorov-Smirnov
\citep[K-S;][]{1988IAUS..123..329K}, the inverse  variance
\citep[I-V;][]{1994MNRAS.270..222O} and the Fourier Transform
\citep[F-T;][]{1997MNRAS.286..303H} significance tests.  In the K-S
test, the quantity $Q$ is defined as the probability that the observed
periods are randomly distributed. Thus, any uniform or at least
systematically non-random period spacing in the period spectrum of the
star will appear as a minimum in $Q$. In the I-V test, a maximum of
the inverse variance will indicate a constant period spacing. Finally,
in the F-T test, we calculate the FT of a Dirac comb
function (created from a set of observed periods), and then we plot
the square of the amplitude of the resulting function in terms of the
inverse of the frequency. A maximum in the square of
the amplitude will indicate a constant period spacing.

Fig.~\ref{fig:tests-RXJ2117} displays the results of applying the K-S,
I-V, and F-T significance tests to the set of 31 periods of Table
\ref{table:RXJ2117-extended}. The three tests point to the existence
of a pattern of $\ell= 1$ constant period spacing of $\Delta \Pi\sim
22$~s\footnote{Note also the presence of a peak at $\sim 11$ s, that is
simply the sub-harmonic of this $\ell= 1$ period spacing ($\Delta \Pi/2$).}. 
To derive a refined value of the period  spacing, we have
carried out a linear least-squares fit to the 20 periods marked  with
an asterisk in Table \ref{table:RXJ2117-extended},  excluding those
with $m \neq 0$ and those for which mode identification is uncertain.
We obtain a period spacing  of $\Delta \Pi= 21.669\pm 0.030$~s (see
upper panel of Fig. \ref{fig:fit-RXJ2117}).  This value is very close
to the period spacing derived by \cite{2002A&A...381..122V} on the
basis of ground-based observations alone ($\Delta \Pi= 21.618 \pm
0.008$~s). With the derived value of the  mean period spacing we
determine one more period compatible with the $\ell =1$ string, at
about 1557~s.  The remaining periods can be associated with $\ell= 1$
modes (which, due to mode trapping effects, deviate from the derived
sequence of almost equally-spaced periods) or  with modes with $\ell=
2$ (or possibly higher). The relevance of finding  a constant period
spacing  is twofold: on the one hand it allows the identification of
the harmonic degree of the modes (that is, the assignment of the
harmonic degree $\ell$; see Table \ref{table:RXJ2117-extended}),  and
on the other hand, it enables us to estimate the stellar mass.  This
will be addressed for  RX~J2117 in Sect. \ref{modelling-rxj2117}.

In the lower panel of Fig. \ref{fig:fit-RXJ2117} we show the residuals
($\delta \Pi$) between the observed periods ($\Pi_i^{\rm O}$) and the
periods derived from the mean period spacing ($\Pi_{\rm fit}$).  The
presence of  several minima in the distribution of residuals strongly
suggests the mode-trapping  effects inflicted by the presence of
internal chemical transition regions.

\begin{figure} 
\includegraphics[clip,width=1.0\columnwidth]{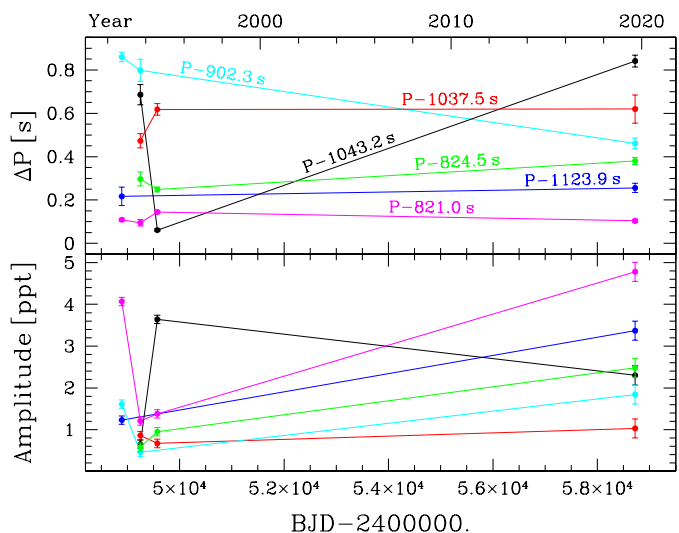}
\caption{Period variations (upper panels) and amplitudes 
(lower panels) of the six periods of RX~J2117
that are common to the \cite{2002A&A...381..122V} and {\it TESS}  data 
sets.  In the upper panel we have subtracted to each period a constant value in order to fit all periods within a narrow range.}
\label{fig:RXJ2117-period-stability} 
\end{figure}

Since six periods of RXJ~2117 that are common to both
\cite{2002A&A...381..122V}  and {\it TESS} data sets were measured at
different epochs spanning 26 or 27  years, it is interesting to test
their stability in time. 
However, with only a few measurements and a 
$\sim$25-years gap, we can not know if the periods that seem relatively 
stable in the upper panel of Fig. \ref{fig:RXJ2117-period-stability} 
are {\it really} stable.
On the other hand, we see at least two periods, those near 902
and 1043~s, that show significant period variations.
In particular, the period at $\sim$1043~s show strong variations
that are clearly anticorrelated with amplitude variations
(lower panel of Fig. \ref{fig:RXJ2117-period-stability}),
and we know that correlated or anticorrelated variations of period and
amplitude are typical of nonlinear interactions between different
pulsation  modes \citep[see, e.g.,][]{2018ApJ...853...98Z}.
The lower panel of Fig. \ref{fig:RXJ2117-period-stability} shows
that, except for the period at $\sim 1037$~s, all the others vary in amplitude.
In addition to the period at $\sim$1043~s, which we have already mentioned, another period at 821.1~s, rather stable over the period, shows strong variations in amplitude, up to a factor $\sim$4 in about one year.


\subsection{Period spectrum  of HS~2324}
\label{period-spacing-hs2324}

\begin{figure} 
\includegraphics[clip,width=1.0\columnwidth]{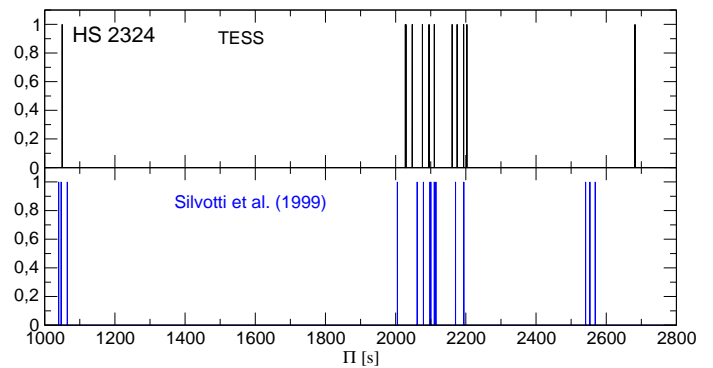}
\caption{Schematic distribution of the pulsation periods of HS~2324 according 
  to {\it TESS} data (12 periods, black lines, upper panel), and according to
  \cite{1999A&A...342..745S} (19 periods, blue lines, lower panel). The
  amplitudes have been arbitrarily set to one for clarity.}
\label{fig:compara-tess-sea99} 
\end{figure}

At variance with the case of RX~J2117, for HS~2324 we find a {\it
  TESS} pulsation spectrum quite similar to that resulting from the
ground-based observations. In Fig.~\ref{fig:compara-tess-sea99} we
schematically show the 12 periods detected with {\it TESS} (upper
panel), and the 19  periods detected by \cite{1999A&A...342..745S}
(lower panel). Only two periods, around 2110~s and  2194 s, are nearly
identical between both sets of data. In order to find a period spacing
in HS~2324,  we enlarged the list of periods by adding some of the 19
periods  observed by \cite{1999A&A...342..745S}.

In that article, the authors note that the solution reported  in their
Table~2 was not univocal, probably because of insufficient frequency
resolution and/or amplitude variations with time.  We re-analyzed
those data with different criteria for the pre-whitening  procedure,
in order to verify which frequencies are found in different
solutions, and we selected 11 periods that were added   to the list of
periods collected by {\it TESS} (Table \ref{table:HS2324}).  The
periods are slightly different from those in Table~2 of
\cite{1999A&A...342..745S} because we use the mean values of the
different solutions. For the two periods found in both data sets at
$\sim 2110$~s and $\sim 2194$~s, we adopt the periods measured by {\it
  TESS} because  they have  smaller uncertainties.  The extended list
of periods to be used in  our analysis contains 21 periods and is
presented in Table~\ref{table:HS2324-extended}.

\begin{table}
\centering
\caption{Enlarged list of periods of HS~2324. Column 1 corresponds to
  a subset of 11 periods estimated by \cite{1999A&A...342..745S} (SEA99)
  with small differences (see text),
  and column 2 corresponds to the 12 periods detected with {\it TESS}
  (Table~\ref{table:HS2324}). The periods with an asterisk 
  are the ones used in the linear least square fit (Fig.~\ref{fig:fit-HS2324}).}
\begin{tabular}{cc|ccc}
\hline
\noalign{\smallskip}
$\Pi_i^{\rm O}$ (s) & $\Pi_{i}^{\rm O}$ (s) & $\Pi_{\rm fit}$ (s) & $\delta\Pi$ (s) & $\ell^{\rm O}$ \\
  SEA99             & {\it TESS}    &   &  &     \\
\noalign{\smallskip}
\hline
\noalign{\smallskip}      
1039.02  &           &  	&	 & ? \\
1047.10  &           & 1044.930 &  2.170 & 1 \\ 
         & 1049.877  &  	&	 & ? \\
2005.78  &           &  	&	 & ? \\       
         & 2027.520  &          &	 & ? \\
         & 2029.350* & 2029.350 &  0.000 & 1 \\  
         & 2047.260* & 2045.757 &  1.503 & 1 \\  
2059.97* &           & 2062.164 & $-2.194$ & 1 \\ 
         & 2076.110  &  	&	 & ? \\ 
2078.59* &           & 2078.571 &  0.019 & 1 \\ 
         & 2095.046* & 2094.978 &  0.068 & 1 \\ 
2098.67  &           &  	&	 & ? \\ 
2109.53  & 2110.152* & 2111.385 & $-1.233$ & 1 \\  
         & 2160.970* & 2160.606 &  0.364 & 1 \\ 
2170.49  &           &  	&	 & ? \\ 
         & 2175.290* & 2177.013 & $-1.723$ & 1 \\ 
2194.12  & 2193.420* & 2193.420 &  0.000 & 1 \\ 
         & 2202.990  &  	&	 & ? \\ 
2553.23  &           & 2554.374 & $-1.144$ & 1 \\
2568.86  &           & 2570.781 & $-1.921$ & 1 \\
         & 2682.050  & 2685.630 & $-3.580$ & 1 \\ 
\noalign{\smallskip}
\hline
\end{tabular}
\label{table:HS2324-extended}
\end{table}

\begin{figure} 
  \includegraphics[clip,width=1.0\columnwidth]{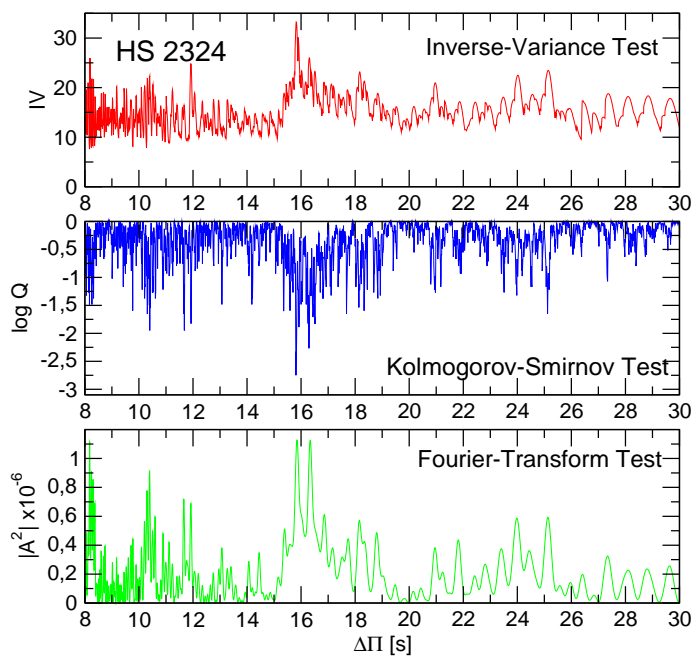}
  \caption{I-V  (upper panel),  K-S  (middle panel),
  and  F-T  (bottom panel)  significance  tests  to  search  for  a
  constant  period  spacing  in   HS~2324.
  The tests are applied to the 21   pulsation periods in
  Table~\ref{table:HS2324-extended}.
  A clear signal of a constant period spacing at $\sim 16$~s
  is  evident. See text for details.}
\label{fig:tests-HS2324} 
\end{figure}

\begin{figure} 
\includegraphics[clip,width=1.0\columnwidth]{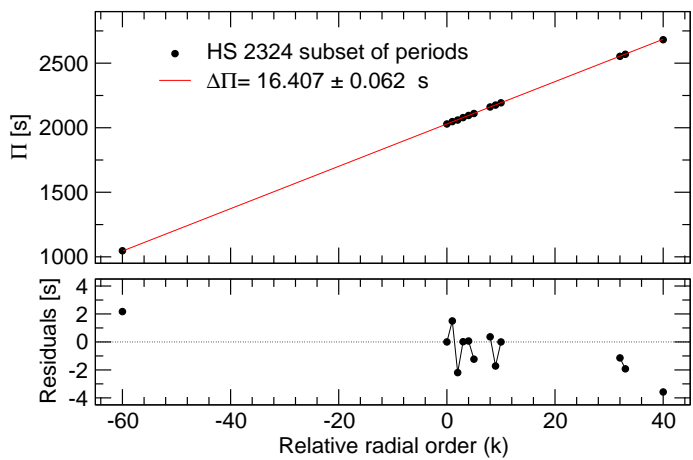}
\caption{Upper panel: linear least-squares fit to the 9 periods of
  HS~2324 marked with asterisk in Table \ref{table:HS2324}. The
  derived dipole ($\ell= 1$) mean period spacing from this fit is
  $\Delta \Pi= 16.407\pm 0.062$~ s. Lower panel: the  residuals of the
  period distribution relative to the mean period spacing,  showing
  observational evidence of mode trapping in the pulsation spectrum of
  HS~2324.  Modes with consecutive radial order are connected with a
  thin black line.}
\label{fig:fit-HS2324} 
\end{figure}

In Fig.~\ref{fig:tests-HS2324} we show the results of applying the
statistical tests to the set of 21 periods of Table
\ref{table:HS2324-extended}. The three tests support the existence of
a mean  period spacing of about $16$~s, which corresponds to our
expectations for  a dipole ($\ell= 1$) sequence\footnote{We can safely rule out 
that this peak corresponds to $\ell= 2$ modes because in that case, 
we should find a peak at $16\ {\rm s}\ \times \sqrt 3\sim 28$~s associated 
to $\ell= 1$ modes, according to 
Eq.  (\ref{eq:1}), which is not observed.}. Assuming that the peak at $\sim 16$~s is 
associated to $\ell= 1$ modes, then if a series of quadrupole ($\ell= 2$) modes 
were present, one  should  find a spacing  of
periods  of $\sim 9$ s. This  is not observed in our analysis 
(see Fig.~\ref{fig:tests-HS2324}).
To determine precisely the period
spacing, we did a linear least-squares fit (plotted in the upper panel
of Fig. \ref{fig:fit-HS2324}) using only nine periods, those marked
with an asterisk in Table~\ref{table:HS2324-extended}.  The reason is
that the other potential $\ell= 1$ periods are much shorter or much
longer and not well constrained, and we risk to assign them a wrong
identification.  We obtain a period spacing $\Delta \Pi= 16.407\pm
0.062$~s and with this value we determine four more periods compatible
with the   $\ell= 1$ sequence, at about 1047, 2553, 2569 and 2682~s.
If we include in the linear least-squares fit also these four
periods, we obtain $\Delta \Pi= 16.371$~s.  We can conclude that
Table~\ref{table:HS2324-extended} contains an incomplete sequence of
13 modes  $\ell = 1$. The remainder periods, on the other hand, can be
associated  with modes with $\ell = 1$ or modes with $\ell = 2$ as
well. In Sect.~\ref{modelling-hs2324} we obtain an estimate of the
stellar mass of HS~2324 on the basis of the period spacing. The lower
panel of Fig. \ref{fig:fit-HS2324} displays the residuals between  the
observed periods and the periods derived from the mean period spacing. 
We note the presence of  several minima in the distribution of residuals, similar
to the case of RX~J2117, which suggests the presence of mode trapping
caused by chemical-composition gradients.

Note that, as discussed in Sect. \ref{obser-hs2324}, a rotational 
splitting of $\sim 4.3\ \mu$Hz cannot be completely 
discarded in the frequency 
spectrum of HS~2324. In such case, we should
have to consider only one component of the two possible incomplete 
triplets constituted by the pairs  $(481.67, 477.316)\ \mu$Hz and 
$(492.769,488.458)\ \mu$Hz when looking for a constant period spacing. 
On the one hand, from the pair $(481.67,477.316)\ \mu$Hz we can adopt 
the frequency $477.316\ \mu$Hz, that corresponds to the period 
2095.046 s, and discard the period 2076.11 s from the analysis. 
This is precisely what we have done in the analysis above. In 
relation to the other pair of frequencies, based on the arguments 
discussed in the Section \ref{obser-hs2324}, we can adopt the 
frequency $492.769\ \mu$Hz that corresponds to the period 2029.35 s\footnote{This implies discarding the period 2047.26 s (frequency 
$488.458\ \mu$Hz) from our analysis, which makes sense since it 
has much lower amplitude than the period  2029.35 s ($492.769\ \mu$Hz)
(see Table \ref{table:HS2324} and Fig. \ref{fig:HS2324-sFT}).}. 
Repeating the calculation of the period spacing, but this time 
neglecting the period of 2047.26 s from the analysis (that is, 
adopting a list of 8 periods), we obtain $\Delta \Pi= 16.421\pm0.120$ s, 
virtually the same period spacing obtained above ($\Delta 
\Pi= 16.407\pm0.062$ s). In summary, if we consider the possible 
existence of two incomplete 
rotational triplets, it does not alter the results in relation to 
the period spacing of HS~2324.

Since the epochs of the two data sets of HS~2324 differ by 22 years,
by comparing the values of the two periods found in both data sets we
can test  their temporal stability.  However, we know that the formal
uncertainties given in Table~2  of \cite{1999A&A...342..745S} are
underestimated as the periods change slightly depending on which
solution is adopted.  By testing different solutions, from the 1997
data we obtain 2109.53$\pm $0.66  and 2194.12$\pm$0.15 with more
realistic uncertainties.  The much larger uncertainty on the first
period is due to the presence of close-by  peaks, not visible in the
{\it TESS} run,  which are probably the cause of the varying amplitude
($\sim$3.3~ppt in 1997, 6.0~ppt in 2019).  In 2019 the periods are
$2110.15 \pm 0.04$ s and $2193.42 \pm 0.06$~s (adopting the formal
uncertainties of the fit).  Only for the second one near 2193~s,
which has a fairly stable amplitude (4.3~ppt in 1997 and 4.0~ppt in
2019, considering also the much redder sensitivity of {\it TESS}),
the period change in time is significant ($4.7\,\sigma$),
corresponding to  $\dot{\Pi}= (-1.0 \pm 0.3) \times 10^{-9}$~s/s.
Even though this value is close to theoretical predictions (see
Sect.~\ref{modelling-hs2324}), the fact that we have only two
measurements and that we know that the periods may have irregular
variations  on different time scales (as we have seen for RX~J2117)
suggests caution.

\subsection{Period spectrum of NGC~6905}
\label{period-spacing-ngc6905}

\begin{figure} 
\includegraphics[clip,width=1.0\columnwidth]{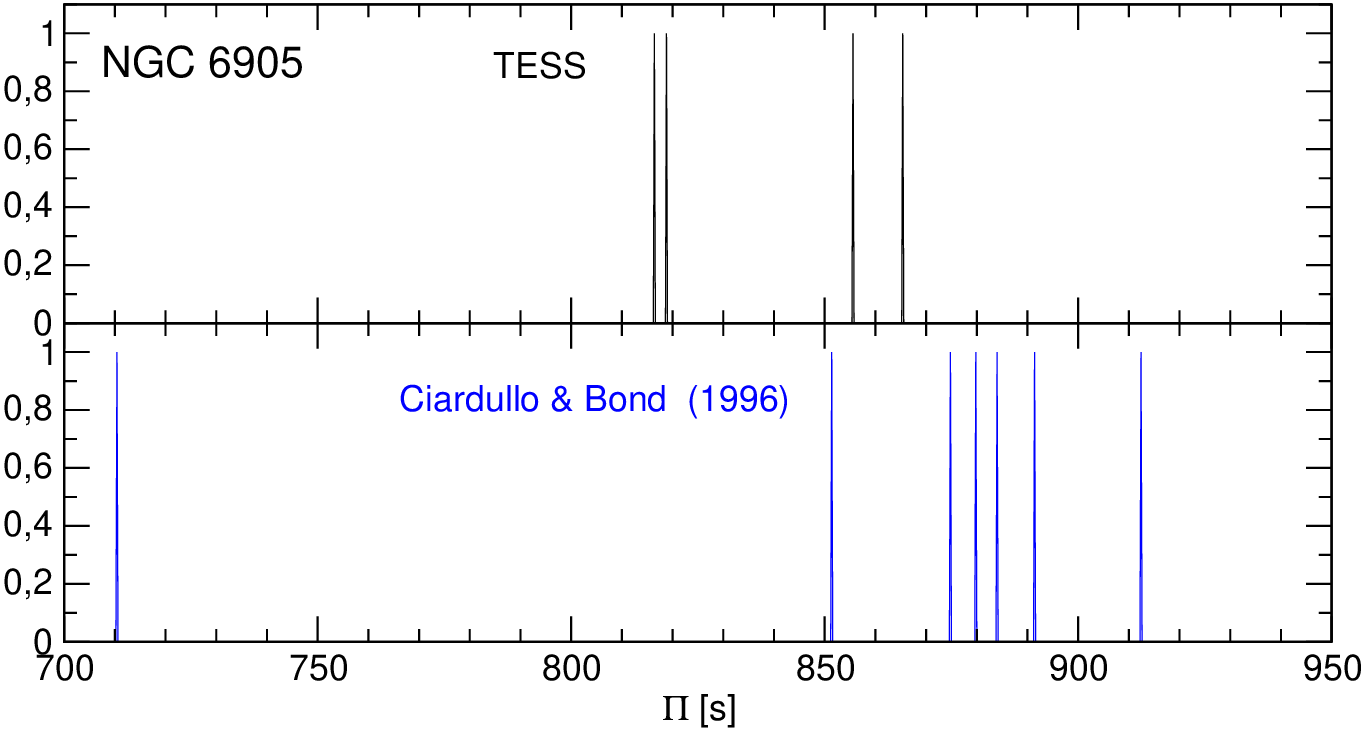}
\caption{Schematic distribution of the pulsation periods of NGC~6905 according 
to {\it TESS} (black lines, upper panel), and according to 
\cite{1996AJ....111.2332C} 
(blue lines, lower panel). The amplitudes have been arbitrarily set to 
one for clarity.}
\label{fig:compara-tess-cb96} 
\end{figure}

As in the cases of RX~J2117 and HS~2324, we find a pulsation spectra
of NGC~6905 quite different as compared with the results of ground-based
observations. In Fig.~\ref{fig:compara-tess-cb96} we schematically
show the 4 periods detected with {\it TESS} (upper panel, black
lines), and the 7 periods detected by \cite{1996AJ....111.2332C}
(lower panel, blue lines), with arbitrary amplitudes set to one to
facilitate visualization. In order to search for a period spacing in
NGC~6905 we enlarged the list of periods by adding the 7 periods
measured by \cite{1996AJ....111.2332C} to the list of periods
collected by {\it TESS}. The extended list of periods to be used in
our analysis, that contains 11 periods, is presented in
Table~\ref{table:NGC6905-extended}. 

\begin{figure} 
\includegraphics[clip,width=1.0\columnwidth]{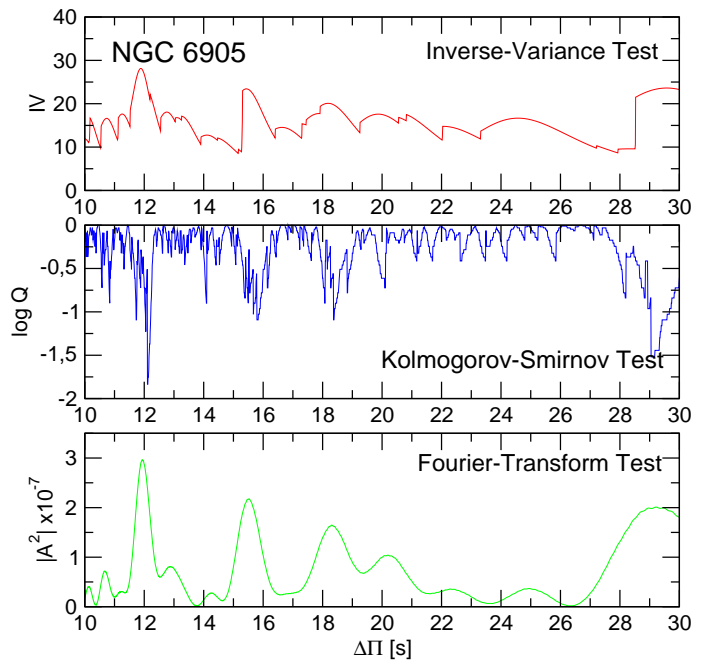}
\caption{I-V  (upper panel),  K-S  (middle panel),
  and  F-T  (bottom panel)  significance  tests  to  search  for  a
  constant  period  spacing  in
NGC~6905.
The tests are applied to the 11 pulsation periods of Table~\ref{table:NGC6905-extended}.
  A clear signal of a constant period spacing at $\sim 12$~s
  is  evident. See text for details.}
\label{fig:tests-NGC6905-extended} 
\end{figure}

We applied the three statistical tests adopting the complete list of
11 periods of Table~\ref{table:NGC6905-extended}, and we did   obtain
a clear indication of a constant period spacing of $\sim 12$~s, as can
be seen in Fig.~\ref{fig:tests-NGC6905-extended}. A priory,  we cannot
know what harmonic degree this apparent constant period spacing
corresponds to.  Being so short, one is tempted to assume that it is a
period spacing of $\ell= 2$ modes,   but in this case, we should find
a period spacing of about 21~s corresponding to $\ell= 1$ modes,
which is absent. The absence of a period spacing at $\sim 21$ s, is not, 
however, a strong reason to discard the possibility that the period
spacing of $\sim 12$~s is associated to a sequence of $\ell= 2$ modes.
Indeed, it could be possible that the $\ell = 1$ modes
(and the associated period spacing) are inhibited for some reason. For
example, it could be that  $\ell= 1$ modes are not excited at the
effective temperature and gravity of NGC~6905, but $\ell= 2$ modes are
unstable. In Sect.~\ref{modelling-ngc6905}, we will consider the 
possibilities that $\Delta \Pi \sim 12$ s is associated to $\ell= 1$ or 
$\ell= 2$ modes when estimating the stellar mass of  NGC~6905.

\begin{figure} 
\includegraphics[clip,width=1.0\columnwidth]{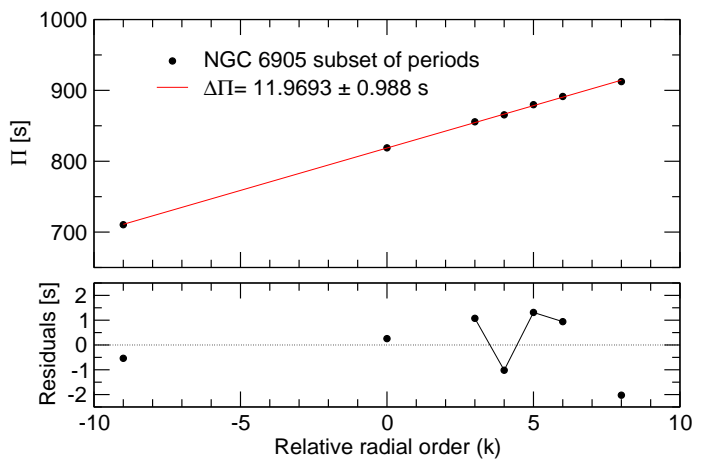}
\caption{Upper panel: linear least-squares fit to the 7 periods of NGC~6905 
marked with asterisk in Table \ref{table:NGC6905-extended}. The derived 
period spacing from this fit is $\Delta \Pi= 11.9693\pm 0.0988$~s. Lower panel: 
residuals of the period distribution relative to the dipole mean period spacing.
Modes with consecutive radial order are connected with a thin black line.}
\label{fig:fit-NGC6905} 
\end{figure}

In order to refine this
period-spacing value, we first performed a linear least-squares fit to
the periods marked  with asterisk in
Table~\ref{table:NGC6905-extended},  except the period at $710.37$~s,
because it is far from the remainder ones and this could affect the
assignment of its relative radial order. We obtain a period spacing
$\Delta \Pi= 11.7769\pm 0.2247$~s. The average value of the residuals
resulting from the difference between the periods observed and those
calculated from the period spacing obtained is $1.395$~s. We repeated
the linear  least-squares fit but this time including the period at
$710.37$~s, and we obtained $\Delta \Pi= 11.9693\pm 0.0988$~s. Note
that this period spacing is slightly longer than that derived
neglecting the  period at $710.37$~s in the linear  least-squares fit,
but the uncertainty is more than two times smaller. In addition,  the
average of the residuals in this case is $1.023$~s, so the fitted
periods match the observed  periods much better than before. For this
reason, we adopt $\Delta \Pi= 11.9693\pm 0.0988$~s   as the period
spacing for NGC~6905.  We show
the periods derived with the fit ($\Pi_{\rm fit}$) and the residuals
($\delta \Pi$) in columns 3 and 4, respectively, of
Table~\ref{table:NGC6905-extended}.  In the last column we show the 
possible identification of the fitted periods with $\ell= 1$ modes, 
although, as stated above, they could be all associated to $\ell= 2$
modes. In the upper panel of
Fig. \ref{fig:fit-NGC6905} we show the fit, whereas in the lower panel
we depict the residuals.


\begin{table}
\centering
\caption{Enlarged list of periods of NGC~6905. Column 1 corresponds to the 
7 periods measured by \cite{1996AJ....111.2332C} (CB96), and column 2 corresponds
to the 4 periods detected by {\it TESS}. Columns 3 to 5 have the 
same meaning as in Table ~\ref{table:RXJ2117-extended}.
The periods with an asterisk are the ones used in the linear least
square fit (Fig. \ref{fig:fit-NGC6905}).}
 \begin{tabular}{cc|ccc}
\hline
\noalign{\smallskip}
$\Pi_i^{\rm O}$ (s) & $\Pi_{i}^{\rm O}$ (s) & $\Pi_{\rm fit}$ (s) & $\delta\Pi$ (s) & $\ell^{\rm O}$ \\
  CB96               &{\it TESS}&          &       \\
\hline
710.37* &             & 710.908 & $-0.538$ &  1 \\
        &    816.337  &         &          &  1 \\    
        &    818.887* & 818.632 &    0.255 &  1 \\ 
851.38  &             &         &          &  ? \\
        &    855.613* & 854.540 &  1.074   &  1 \\ 
        &    865.490* & 866.509 & $-1.109$ &  1 \\ 
874.80  &             &         &          &  ? \\
879.79* &             & 878.479 &  1.311   &  1 \\
884.00  &             &         &          &  ? \\
891.39* &             & 890.448 &  0.942   &  1 \\
912.36* &             & 914.386 & $-2.026$ &  1 \\
\noalign{\smallskip}
\hline
\end{tabular}
\label{table:NGC6905-extended}
\end{table}

\subsection{Period spectrum of NGC1501}
\label{period-spacing-ngc1501}

\begin{figure} 
\includegraphics[clip,width=1.0\columnwidth]{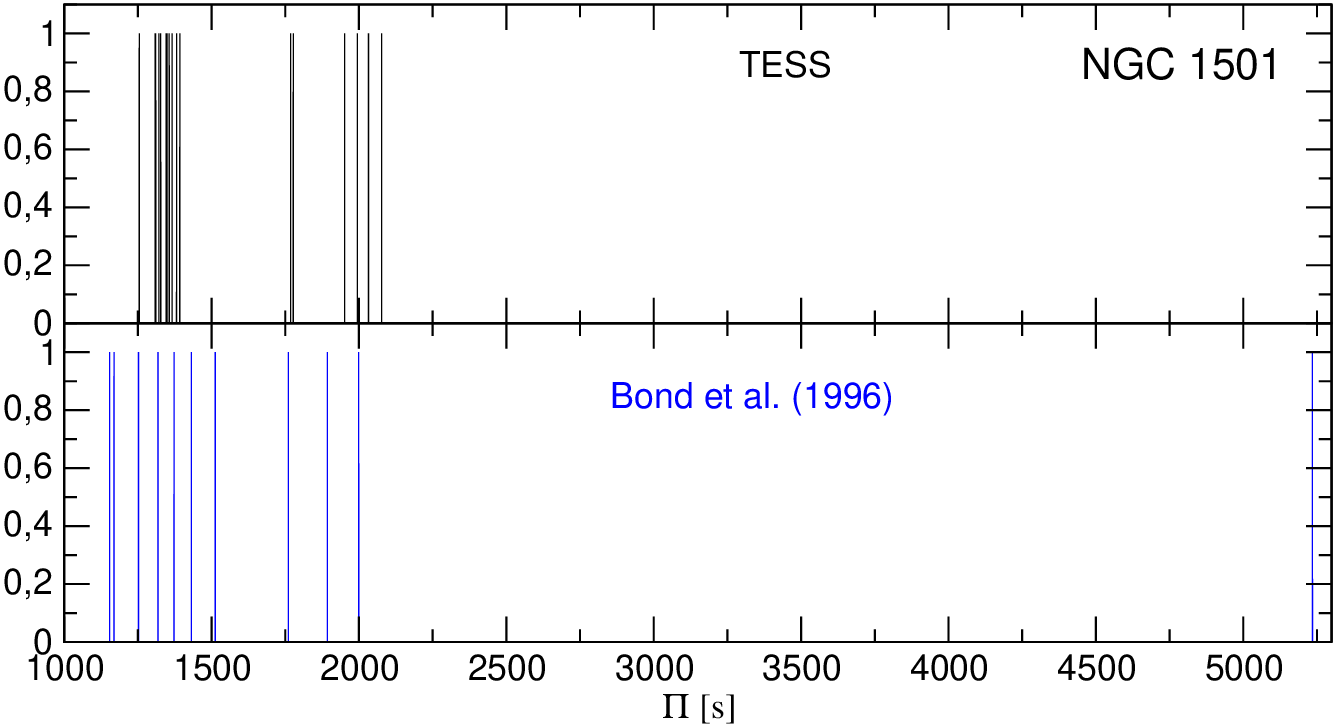}
\caption{Schematic distribution of the pulsation periods of NGC~1501 according 
to {\it TESS} (black lines, upper panel), and according to Table~4 of
\cite{1996AJ....112.2699B} (blue lines, lower panel). The amplitudes 
have been arbitrarily set to one for clarity.}
\label{fig:compara-tess-bea96} 
\end{figure}

As for the targets analyzed before, a comparison of the pulsation
spectrum of NGC~1501 detected by {\it TESS} with that  obtained
through the ground-based monitoring by \cite{1996AJ....112.2699B}
reveals  that the pulsation spectra are markedly different. In
Fig.~\ref{fig:compara-tess-bea96} we schematically show the periods
detected with {\it TESS}  (upper panel, black lines), and the 11
periods detected by \cite{1996AJ....112.2699B} (lower panel, blue
lines) ---specifically, those of their Table~4--- with arbitrary
amplitudes set to one to facilitate visualization. As before, we
consider here the composed pulsation spectrum of NGC~1501, that is,
the list of periods measured by {\it TESS} along with the periods
determined by \cite{1996AJ....112.2699B}.  The extended list of
periods to be used in our analysis is presented in columns 1 and 2 of
Table~\ref{table:NGC1501-extended}.

\begin{figure} 
\includegraphics[clip,width=1.0\columnwidth]{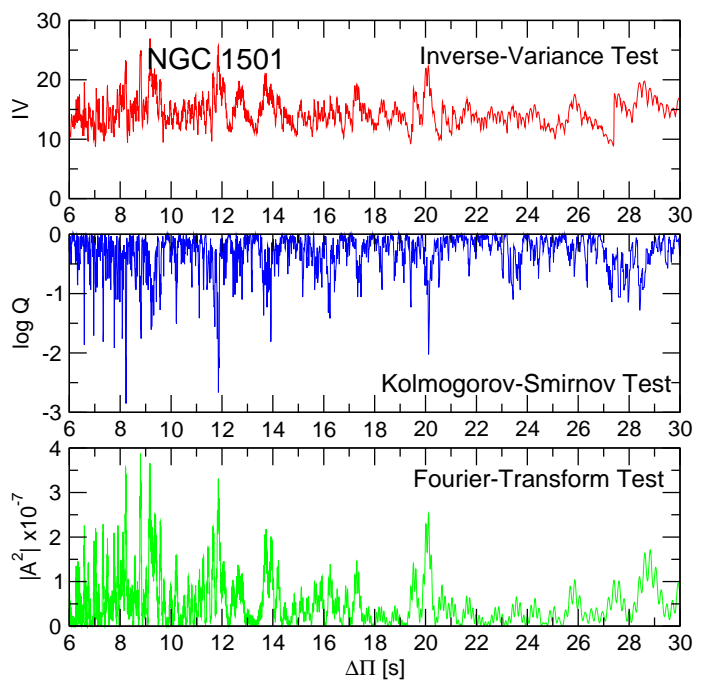}
\caption{I-V  (upper panel),  K-S  (middle panel),
  and  F-T  (bottom panel)  significance  tests  to  search  for  a
  constant  period  spacing  in
NGC~1501.
  The tests are applied to the
  pulsation periods in Table~\ref{table:NGC1501-extended} which are marked with
  an asterisk. The three tests indicate clear signals of two constant
  period spacings of $\sim 11.9$~s and $\sim 20.1$~s.  See text for details.}
\label{fig:tests-NGC1501} 
\end{figure}

In Fig.~\ref{fig:tests-NGC1501} we show the results of applying the
statistical tests to the complete set of 26 periods of
Table~\ref{table:NGC1501-extended}. The three tests support the
existence of two   period spacings of about $11.9$~s and $20.1$~s. The
ratio between the two  values is $20.1/11.9= 1.689$, close to the
expected value of  $\sqrt{3}= 1.732$ if we assume that these are
periods spacings associated to $\ell= 1$ ($\Delta \Pi_{\ell= 1} \sim
20.1$ s)  and $\ell= 2$ ($\Delta \Pi_{\ell= 1} \sim 11.9$ s),
according to Eq.~(\ref{eq:1}). Period spacings due to $\ell= 1$ and
$\ell= 2$  modes in the same object are not common to find in the
context of GW Vir stars,  the only known case so far being the
prototypical variable star PG~1159$-$035 \citep[see,
  e.g.,][]{1991ApJ...378..326W,2008A&A...477..627C}. 

We have alternatively considered  in our analysis the 9 ($m= 0$)
periods of the Table~7 of \cite{1996AJ....112.2699B}, which consist of
a composition of the data from that work plus archival period data on
NGC~1501. However, by putting together those periods with the periods
measured with {\it TESS} in this paper, we do not obtain any clear
pattern of constant period spacing, and therefore, we  do not consider
them here. 

\begin{figure} 
\includegraphics[clip,width=1.0\columnwidth]{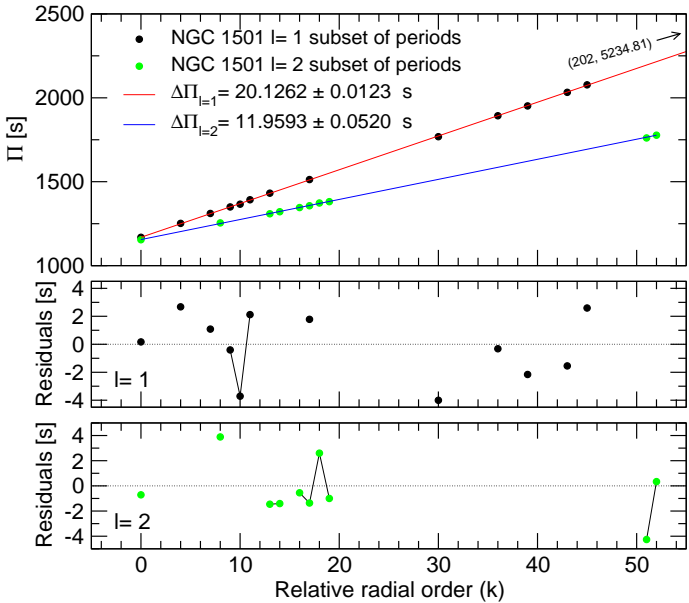}
\caption{Upper panel: the linear least-squares fits to the 14 dipole
  periods  of NGC~1501 marked with an  asterisk and the 9 quadrupole
  periods marked with two asterisks  in Table~\ref{table:NGC1501}. The
  derived period spacings from these fits are $\Delta \Pi_{\ell= 1}=
  20.1262 \pm 0.0123$~s and $\Delta \Pi_{\ell= 2}= 11.9593 \pm
  0.0520$~s, respectively. Middle panel: the residuals of the $\ell=
  1$ period distribution relative to the dipole mean period
  spacing. Lower panel: residuals of the  $\ell= 2$ period
  distribution relative to the quadrupole mean period spacing.  Modes
  with consecutive radial order are connected with a thin black line.}
\label{fig:fit-NGC1501} 
\end{figure}

We perform a linear least-squares fit using the 14 periods marked with
an asterisk  in Table~\ref{table:NGC1501-extended}, which gives a
period spacing  $\Delta \Pi_{\ell= 1}= 20.1262\pm 0.0123$~s associated
with $\ell= 1$ modes. We perform a second  least-squares fit employing
the 9 periods marked with two asterisks  in the same table, which
gives a period spacing  $\Delta \Pi_{\ell= 2}= 11.9593\pm 0.0520$~s
that is probably associated with $\ell= 2$ modes.  With $\Delta
\Pi_{\ell= 2}$, we compute an additional period compatible with the
$\ell= 2$  sequence, at about $1760.7$~s. The ratio between the two
period spacings is  $20.1262/11.9593= 1.683$.  The fits are plotted in
the upper panel of Fig. \ref{fig:fit-NGC1501}, and the residuals  are
depicted in the middle panel ($\ell= 1$) and in the lower panel
($\ell= 2$). 

Note that there are two periods (1318.46~s and 1999.16~s) that do not
fit neither the $\ell= 1$ nor the $\ell= 2$ series of periods with
constant period spacing. For these two periods, we cannot determine
the harmonic degree. These periods either correspond both to $\ell = 1$ or
$\ell = 2$ modes  that are departed from the sequences of equally
spaced periods due to mode trapping effects, or they can be alternatively 
interpreted as $m \neq 0$ components of incomplete rotational multiplets 
---but we have no indication of the presence of rotational multiples in this star. 
Alternatively, these two periods could correspond to $\ell = 3$ modes, although this
is unlikely due to the low chance of being detected by the geometric
cancellation effects \citep{1977AcA....27..203D}. 

In closing this Section, we call the attention about a possible
alternative interpretation of the period spacing derived for NGC~1501. Specifically, 
by examining Fig.~\ref{fig:tests-NGC1501}, we see the presence of a strong peak 
for $\Delta \Pi \sim 8-9$ s. This peak could be associated to the $\ell= 2$ period 
spacing of the star. If this were the case, then we should expect to found a 
peak corresponding to the $\ell= 1$ period spacing at $\Delta \Pi \sim 16$ s. 
However, that peak is not visible in any of the tests. For this reason, 
we rule out this possibility.

In Sect. \ref{modelling-ngc1501} we obtain an estimate of the stellar
mass of NGC~1501 on the basis of the period-spacing values derived in
this section, $\Delta \Pi_{\ell= 1}= 20.1262\pm 0.0123$~s and $\Delta \Pi_{\ell= 2}= 11.9593\pm 0.0520$~s.

\begin{table*}
\centering
\caption{Enlarged list of periods of NGC~1501. Column 1 corresponds to the 
11 periods of Table~4 of \cite{1996AJ....112.2699B} (BEA96), and column 2 corresponds
to the 15 periods detected by {\it TESS} (Table~\ref{table:NGC1501}). 
Columns 3 and 4 (5 and 6) correspond to the fitted periods from the $\ell= 1$ ($\ell= 2$) 
period spacing, and column to 7 gives the ``observed'' harmonic degree.}
\begin{tabular}{cc|ccccc}
\hline
\noalign{\smallskip}
$\Pi^{\rm O}$ (s) & $\Pi^{\rm O}$ (s) & $\Pi_{\rm fit}^{\ell= 1}$ (s) & $\delta \Pi^{\ell= 1}$ (s) & 
$\Pi_{\rm fit}^{\ell= 2}$ (s) & $\delta \Pi^{\ell= 2}$ (s) &$\ell^{\rm O}$ \\
  BEA96           & {\it TESS}        &                     &                  &    &   &         \\
\noalign{\smallskip}
\hline
\noalign{\smallskip}      
1154.36** &            &           &        & 1155.070 & $-0.710$ & 2  \\
1168.90*  &            & 1168.730  & 0.170  &          &        & 1  \\
1251.91*  &            & 1249.235  & 2.675  &          &        & 1  \\
         & 1254.632**  &           &        & 1250.744 & 3.888  & 2  \\
	     & 1309.086**  &           &        & 1310.541 & $-1.455$ & 2  \\
	     & 1310.696*   & 1309.613  & 1.083  &          &        & 1  \\
1318.46  &             &           &        &          &        & ?  \\
         & 1321.095**  &           &        & 1322.500 & $-1.405$ & 2  \\
         & 1345.872**  &           &        & 1346.419 & $-0.547$ & 2  \\
	     & 1349.460*   & 1349.866  & $-0.406$ &          &        & 1  \\
	     & 1356.680**  &           &        & 1358.042 & $-1.362$ & 2  \\
	     & 1366.279*   & 1369.992  & $-3.713$ &          &        & 1  \\
1372.94**  &           &           &        & 1370.337 & 2.603  & 2  \\
         &  1381.295** &           &        & 1382.297 & $-1.002$ & 2  \\
         &  1392.240*  & 1390.118  & 2.122  &          &        & 1  \\
1431.53*  &            & 1430.371  & 1.159  &          &        & 1  \\
1512.66*  &            & 1510.875  & 1.785  &          &        & 1  \\
1760.73  &             &           &        & 1764.994 & $-4.264$ & 2  \\
         &  1768.510*  & 1772.516  & $-4.006$ &          &        & 1  \\
	     &  1777.290** &           &        & 1776.954 &  0.336 & 2  \\
1892.95*  &            & 1893.273  & $-0.323$ &          &        & 1  \\
         &  1951.490*  & 1953.652  & $-2.162$ &          &        & 1  \\
1999.16  &             &           &        &          &        & ?  \\
         & 2032.640*   & 2034.157  & $-1.547$ &          &        & 1  \\
         & 2077.000*   & 2074.409  &  2.591 &          &        & 1  \\
5234.81* &             & 5234.222  &  0.588 &          &        & 1  \\
\noalign{\smallskip}
\hline
\end{tabular}
\label{table:NGC1501-extended}
\end{table*}

\subsection{Period spectrum of NGC~2371}
\label{period-spacing-ngc2371}

\begin{figure} 
\includegraphics[clip,width=1.0\columnwidth]{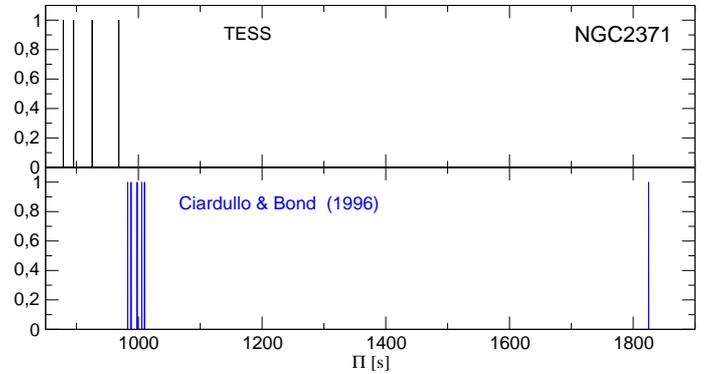}
\caption{Illustrative distribution of the pulsation periods of NGC~2371 according 
to {\it TESS} (black lines, upper panel), and according to 
\cite{1996AJ....111.2332C}, their Table~4 (blue lines, lower panel). The amplitudes 
have been arbitrarily set to one for clarity.}
\label{fig:compara-tess-cb96-ngc2371} 
\end{figure}

NGC~2371 was observed from the ground by
\cite{1996AJ....111.2332C}. An illustrative comparison  between
periods detected from ground-based observations and those detected by
{\it TESS} is shown in Fig.~\ref{fig:compara-tess-cb96-ngc2371}. An
examination of this figure reveals  that both pulsation spectra are
markedly different. In particular, \cite{1996AJ....111.2332C} report
several periods longer than the longest period detected with {\it
  TESS} ($\sim 968$~s).  As before, we adopt an augmented pulsation
spectrum of NGC~2371, composed by the periods  measured by {\it TESS}
along with the periods determined by \cite{1996AJ....111.2332C},
specifically, those of their Table~4 which are considered as secure.
The extended list of periods to be used in our analysis is presented
in Table~\ref{table:NGC2371-extended}.

\begin{figure} 
\includegraphics[clip,width=1.0\columnwidth]{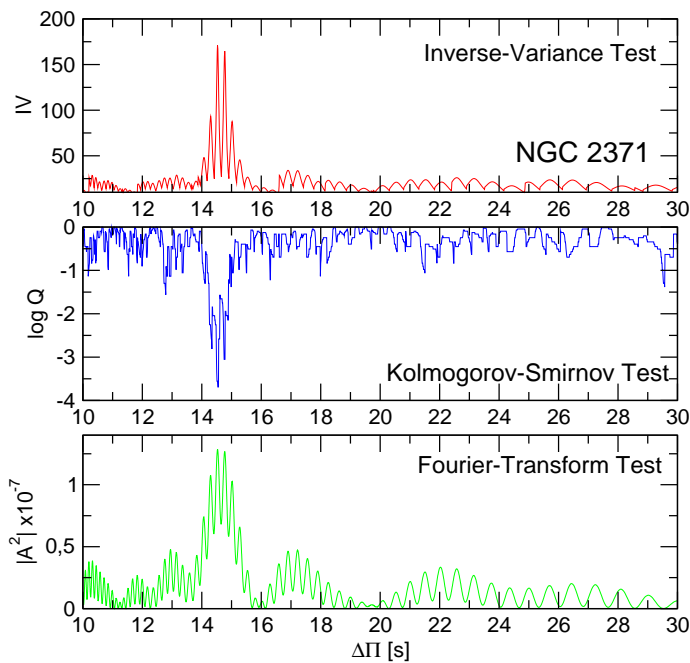}
\caption{I-V  (upper panel),  K-S  (middle panel),
  and  F-T  (bottom panel)  significance  tests  to  search  for  a
  constant  period  spacing  in
NGC~2371.
  The tests are applied to the
  pulsation periods in Table~\ref{table:NGC2371-extended} which are marked with
  an asterisk. The three tests indicate clear signals of a constant
  period spacing of $\sim 14.5$~s.  See text for details.}
\label{fig:tests-NGC2371} 
\end{figure}

\begin{figure} 
\includegraphics[clip,width=1.0\columnwidth]{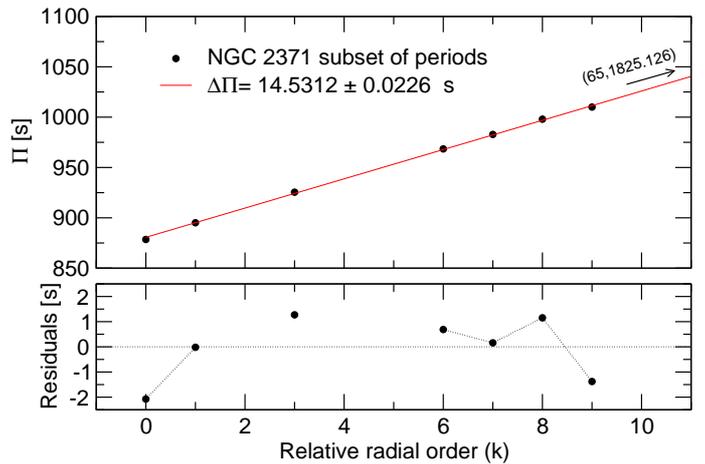}
\caption{Upper panel: linear least-squares fit to the 8 periods of NGC~2371
  marked with asterisk in Table~\ref{table:NGC2371}. Lower panel: the residuals
  between the observed and the fitted periods. The derived period
  spacing from this fit is $\Delta \Pi= 14.5312 \pm 0.0226$~s.}
\label{fig:fit-NGC2371} 
\end{figure}

In Fig.~\ref{fig:tests-NGC2371} we show the results of applying the
statistical tests to the set of periods of
Table~\ref{table:NGC2371-extended} marked with an asterisk. The three
tests support the existence of a mean period spacing of about $14.5$~s
which corresponds to our expectations for a dipole ($\ell= 1$)
sequence.  A linear least-squares fit using the 8 periods marked with
asterisk in Table~\ref{table:NGC2371-extended} gives a period spacing
$\Delta \Pi= 14.5312\pm0.0226$~s.  The fit is plotted in the upper
panel of Fig.~\ref{fig:fit-NGC2371}, and the  residuals are depicted
in the lower panel of the same figure.  In
Sect.~\ref{modelling-ngc2371} we obtain an estimate of the stellar
mass of NGC~2371 on the basis of the period spacing.

\begin{table}
\centering
\caption{Enlarged list of periods of NGC~2371. Column 1 corresponds to the 
6 secure periods of Table~4 of \cite{1996AJ....111.2332C} (CB96), and column 2 
corresponds to the 4 periods detected by {\it TESS} (Table~\ref{table:NGC2371}). 
The  periods  with  an  asterisk  are  the  ones  
used  in  the linear least-square fit (Fig.~\ref{fig:fit-NGC2371}). 
Columns 3 to 5 have the same meaning as in Table~\ref{table:RXJ2117-extended}.}
\begin{tabular}{cc|ccc}
\hline
\noalign{\smallskip}
$\Pi^{\rm O}$ (s) & $\Pi^{\rm O}$ (s) & $\Pi_{\rm fit}$ (s) &  $\delta \Pi$ (s) & $\ell^{\rm O}$\\
  CB96 & {\it TESS} & & & \\
\noalign{\smallskip}
\hline
\noalign{\smallskip}      
         &  878.523* & 880.598 & $-2.075$  & 1 \\
      	 &  895.111* & 895.129 & $-0.018$  & 1 \\
         &  925.465* & 924.192 &  1.273  & 1 \\
         &  968.474* & 967.785 &  0.689  & 1 \\
982.8*    &          & 982.316 &  0.158  & 1 \\
988.2    &           &         &         & ? \\
998.0*    &          & 996.848 & 1.152   & 1 \\
1005.6   &           &         &         & ? \\ 
1010.0*   &          & 1011.379 & $-1.379$ & 1 \\ 
1825.0*   &          & 1825.126 & $-0.126$ & 1 \\ 
\noalign{\smallskip}
\hline
\end{tabular}
\label{table:NGC2371-extended}
\end{table}

\subsection{Period spectrum of K~1$-$16} 
\label{period-spacing-k1-16}

K~1$-$16 was observed from the ground by \cite{1984ApJ...277..211G},
detecting a complex power spectrum with a dominant periodicity of
$\sim 1700$ s. After the delivery  of {\it TESS} data from sectors 14
and 15, the star showed no variability. 
Finally, with the release of observations from the sectors
$16-17$, $19-20$, and $22-26$, a total of six (or seven) periods were
detected. However, the power spectrum is so complex and noisy that the
values of the periods are hard to estimate
(Table~\ref{table:K1-16}). The peak at the frequency 7752~$\mu$Hz corresponds 
to a period of 129 s. This is a very short period as to be excited in 
GW Vir stars by the $\kappa-\gamma$ mechanism, therefore, if this period 
is real, it could correspond to a low-order mode excited by the $\varepsilon$ mechanism acting at the He-burning shell \citep{2009ApJ...701.1008C}. 
Bringing together observations from the
ground and from space, the star exhibits a total of at least 7 periods
which vary in amplitude on weekly time scales. We have not found any
clear regularity in the spacing of periods that can give us any clue to its
stellar mass, as it was the case with the other targets in this
study. Given the complexity and the extremely variable spectrum of
periods of K~1$-$16, and having so few periods available, we are
prevented from carrying out an in-depth asteroseismological analysis
of this star, either with the period spacing or the individual
periods.

\section{Evolutionary models and numerical codes}
\label{models}

\begin{figure} 
\includegraphics[clip,width=1.0\columnwidth]{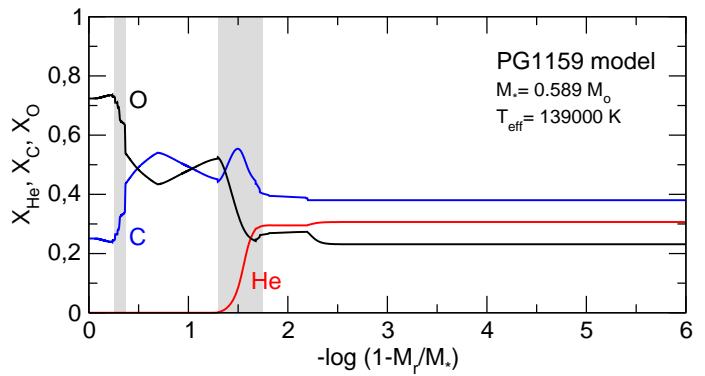}
\caption{The  internal  chemical  profile  of  a template PG~1159  model  
($M_{\star}= 0.589 M_{\sun}$, $T_{\rm eff}= 139\,000$ K)  in  terms  of  the  
outer  fractional  mass.  The locations of the O/C and O/C/He chemical interfaces
are indicated with gray regions.}
\label{fig:xi-dov} 
\end{figure}

The  asteroseismological  analysis  presented  in  this work  relies
on  a  set  of state-of-the-art  stellar  models  that take  into
account the  complete evolution of the PG~1159 progenitor
stars. Specifically, the stellar models were extracted  from the
evolutionary  calculations presented by  \cite{2005A&A...435..631A}
and \cite{2006A&A...454..845M},  who  computed the complete evolution
of model star sequences with initial masses on  the ZAMS in the range
$1 - 3.75\ M_{\sun}$  and assuming  a metallicity of $Z= 0.02$.  All
of the  post-AGB evolutionary sequences  computed with the {\tt
  LPCODE} evolutionary code \citep{2005A&A...435..631A} were followed
through  the  very  late   thermal  pulse  (VLTP)  and  the  resulting
born-again  episode that  give rise  to the  H-deficient, He-,  C- and
O-rich composition characteristic of  PG~1159 stars.  The masses of
the resulting  remnants are  $0.530$, $0.542$, $0.565$, $0.589$,
$0.609$, $0.664$, and  $0.741 \ M_{\sun}$.  In Fig.~\ref{fig:1} the
evolutionary tracks employed  in this  work are shown in the $\log
T_{\rm eff}$ vs. $\log g$ plane.  For details about the input physics
and evolutionary code, and  the   numerical  simulations  performed to
obtain  the  PG~1159 evolutionary sequences  employed here, we refer
the interested reader to the works by \citet{2005A&A...435..631A} and
\citet{2006A&A...454..845M,2007A&A...470..675M,2007MNRAS.380..763M}.
Here, we give a brief description of the chemical structure of our
PG~1159 models. In Fig. \ref{fig:xi-dov} we show the fractional
abundances of the main  chemical species, $^{4}$He, $^{12}$C, and
$^{16}$O, corresponding to a model with $M_{\star}= 0.589 M_{\sun}$
and $T_{\rm eff}= 139\,000$ K.  Clearly visible are the chemical
transition regions of O/C and O/C/He.  The location, thickness, and
steepness  of these chemical interfaces  define the mode-trapping
properties of the models. Generally,  the mode-trapping features in
the pulsation spectrum of these models for periods  shorter than about
$700$ s are induced mostly by the chemical gradient at  the O/C/He
interface, with the O/C chemical transition being much less relevant.
For longer periods, however, it is the core chemical structure in the
O/C interface  that mostly fixes the mode trapping properties
\citep[see, e.g.,][for
  details]{2005A&A...439L..31C,2006A&A...454..863C}.

As mentioned, the star HS~2324 is a hybrid PG~1159 star, i.e.,  it has
H on its surface. The presence of H is not expected to substantially
modify the evolutionary tracks \citep{2019MNRAS.489.1054L} nor the
pulsation properties  of these stars, when compared with the standard
case of PG~1159 stars which lack of H. This is due to the fact that,
at those evolutionary stages, element diffusion is not operative, and
thus vestiges of H left by prior evolution are uniformly distributed
throughout the outer layers of the star,  instead of  forming a pure-H
envelope.  Therefore, as a valid approximation, this star will be
analyzed in this work using  PG~1159 evolutionary models that do not
have H in their surfaces.

We  have computed $\ell=  1, 2$ $g$-mode  adiabatic pulsation  periods in the range  $80-6000$ s  with  the  adiabatic and
nonadiabatic versions of the pulsation code {\tt LP-PUL}
\citep[][]{2006A&A...454..863C,2006A&A...458..259C} and the same
methods  we employed  in the  previous  works of La Plata Stellar
Evolution and Pulsation Research Group\footnote{\tt
  http://fcaglp.fcaglp.unlp.edu.ar/evolgroup/}.  We analyzed about
$4\,000$ PG~1159 models covering a wide range of effective temperatures
($5.4 \gtrsim \log T_{\rm eff} \gtrsim 4.8$), luminosities ($0
\lesssim   \log(L_*/L_{\sun}) \lesssim 4.2$), and  stellar masses
($0.530 \leq M_{\star}/M_{\sun} \leq 0.741$). 

\section{Spectroscopic masses}
\label{spec_mass}

On   the    basis   of   the   evolutionary    tracks   presented   in
Fig.~\ref{fig:1} and the published values of the spectroscopic 
surface gravity and temperature, we derive by interpolation a value of the
spectroscopic   mass of each of the six analyzed stars. In the case of
RX~J2117 we get a stellar mass of $M_{\star}= 0.716 \pm 0.150
\ M_{\sun}$.  As for HS~2324, we get a stellar  mass  of $M_{\star}=
0.532 \pm 0.150 \ M_{\sun}$, while for NGC~6905 and NGC~1501 we derive
$M_{\star}= 0.590 \pm 0.150 \ M_{\sun}$ and $M_{\star}= 0.565 \pm
0.150 \ M_{\sun}$, respectively. We obtain  $M_{\star}= 0.533 \pm
0.150 \ M_{\sun}$ for NGC~2371. Finally,  we derive $M_{\star}= 0.742
\pm 0.150 \ M_{\sun}$ for K~1$-$16.  The uncertainties in the stellar
mass are estimated from the uncertainties in the $T_{\rm eff}$ and
$\log g$ values adopting the extreme values of each parameter when
interpolating between the evolutionary tracks of Fig. \ref{fig:1}.

\section{Asteroseismic modelling}
\label{astero}

The methods we use in this paper to extract information of the stellar
mass and the internal structure of RX~J2117, HS~2324, NGC~2371,
NGC~6905, NGC~1501, NGC~2371,  and K~1$-$16 are the same employed in
our previous works \citep[see][] {2007A&A...461.1095C,
  2007A&A...475..619C,2008A&A...478..869C,2009A&A...499..257C,
  2014MNRAS.442.2278K,2016A&A...589A..40C}. In brief, a way to derive
an estimate of the stellar mass of GW Vir stars is by comparing the
observed period spacing of a target star ($\Delta \Pi$) with the
asymptotic period spacing ($\Delta \Pi_{\ell}^{\rm a}$) computed with
Eq.~(\ref{eq:1}) at the effective temperature of the star \citep[see the pioneer work of ][]{1988IAUS..123..329K}. GW Vir
stars generally do not have all of their pulsation modes in the
asymptotic regime, so there is usually no perfect agreement between
$\Delta \Pi$ and $\Delta \Pi_{\ell}^{\rm a}$.   Therefore, the
derivation of the  stellar  mass  using  the  asymptotic period
spacing  may  not be entirely reliable in pulsating PG~1159 stars that
pulsate with modes characterized by low and intermediate radial
orders, but it gives a good estimate of the stellar mass for stars
pulsating with $g$ modes of high radial order
\citep[see][]{2008A&A...478..175A}. A variation of this approach to
infer the stellar mass of GW Vir stars is to compare $\Delta \Pi$ with
the average of the computed period spacings ($\overline{\Delta
  \Pi_{k}}$). The  average  of  the  computed period  spacings  is
assessed  as $\overline{\Delta \Pi_{k}}= (N-1)^{-1} \sum_k \Delta
\Pi_{k}$, where the "forward" period spacing ($\Delta \Pi_{k}$) is
defined as $\Delta \Pi_{k}= \Pi_{k+1}-\Pi_{k}$ ($k$ being the radial
order) and $N$ is the number of computed periods laying in the range
of the observed periods.  This method is more reliable for the
estimation of the stellar mass of GW Vir stars than that described
above using $\Delta \Pi_{\ell}^{\rm a}$ because, provided that  the
average of  the  computed period  spacings is evaluated at the
appropriate range of periods, the approach is valid for the regimes of
short, intermediate and long periods as well. When the average of the
computed period spacings is taken over a range of periods
characterized by high $k$ values, then the predictions of the present
method become closer to those of the asymptotic period-spacing
approach \citep[][]{2008A&A...478..175A}.  On the other hand, the
present method requires of detailed period computations, at variance
with the method described above, that does  not  involve  pulsational
calculations.  Note that both methods for assessing the stellar mass
rely on the spectroscopic effective temperature, and the results are
unavoidably affected by its associated uncertainty.

Another asteroseismological tool to disentangle the internal structure
of GW Vir stars is to seek theoretical models that best match the
individual pulsation  periods  of  the target star. To measure the
goodness of the match between the theoretical pulsation periods
($\Pi_{\ell,k}$) and the observed individual periods ($\Pi_i^{\rm
  o}$), we follow the same procedure as in our previous works: 

\begin{equation}
  \chi^2(M_{\star}, T_{\rm eff})= \frac{1}{N} \sum_{i= 1}^{N}
      {\rm min}[(\Pi_{\ell,k}-\Pi_i^{\rm o})^2]
\label{eq:3}
\end{equation}  

\noindent where $N$ is the number of observed periods. The observed
periods are shown in Tables~\ref{table:RXJ2117-extended}, \ref{table:HS2324-extended},
\ref{table:NGC6905-extended}, \ref{table:NGC1501-extended}, and 
\ref{table:NGC2371-extended}. In order to find the
stellar model that best replicate the observed periods exhibited by
each target star --- the ``asteroseismological'' model ---, we
evaluate the  function  $\chi^2$ for stellar  masses  $M_{\star}=
0.530, 0.542,  0.565, 0.589, 0.609, 0.664$, and $0.741 M_{\odot}$. For
the  effective  temperature we  employ  a much  finer  grid ($\Delta
T_{\rm eff}= 10-30$ K).  For each target star, the PG~1159 model that
shows the lowest value of $\chi^2$ is adopted as the best-fit
asteroseismological model. 

Below, we employ the tools described above to extract information of
the GW Vir stars considered in this work. 

\subsection{RX~J2117}
\label{modelling-rxj2117}

\begin{figure} 
\includegraphics[clip,width=1.0\columnwidth]{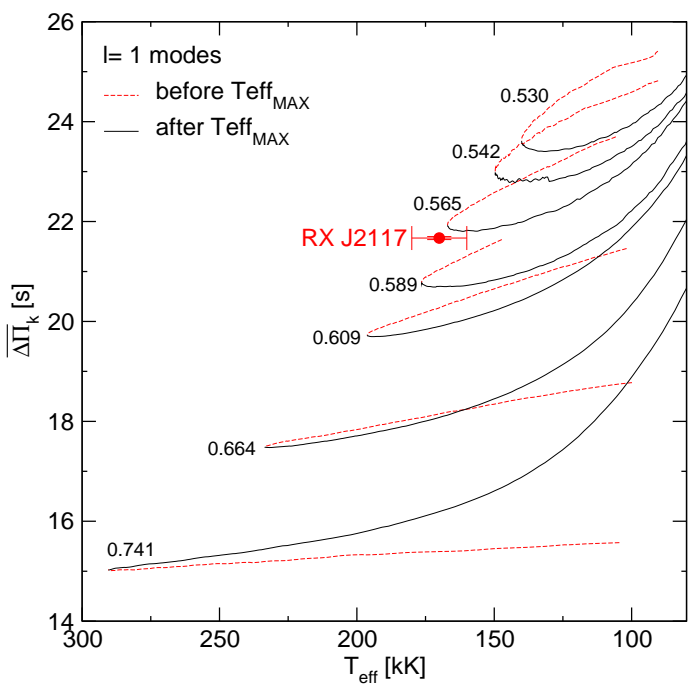}
\caption{Dipole ($\ell= 1$) average of the computed period spacings,
  $\overline{\Delta \Pi_k}$, assessed  in  a  range  of  periods  that
  includes  the  periods  observed  in RX~J2117, shown as solid
  (dashed) curves corresponding to stages before (after) the maximum
  $T_{\rm eff}$ for different stellar masses. The location of RX~J2117
  when we  use the effective temperature derived by
  \cite{1997fbs..conf..217R},  $T_{\rm eff}= 170\,000\pm 10\,000$~K,
  and the period spacing $\Delta \Pi= 21.669\pm 0.030$~s derived in
  Sect.~\ref{period-spacing-rxj2117} is highlighted with a red
  circle. We include the error bars associated to the uncertainties in
  $\overline{\Delta \Pi_k}$ and $T_{\rm eff}$.  The stellar mass
  derived by interpolation is $M_{\star}= 0.569\pm 0.015 M_{\odot}$.}
\label{fig:psp-teff-RXJ2117} 
\end{figure}

\begin{figure*} 
\includegraphics[clip,width= 2.0\columnwidth]{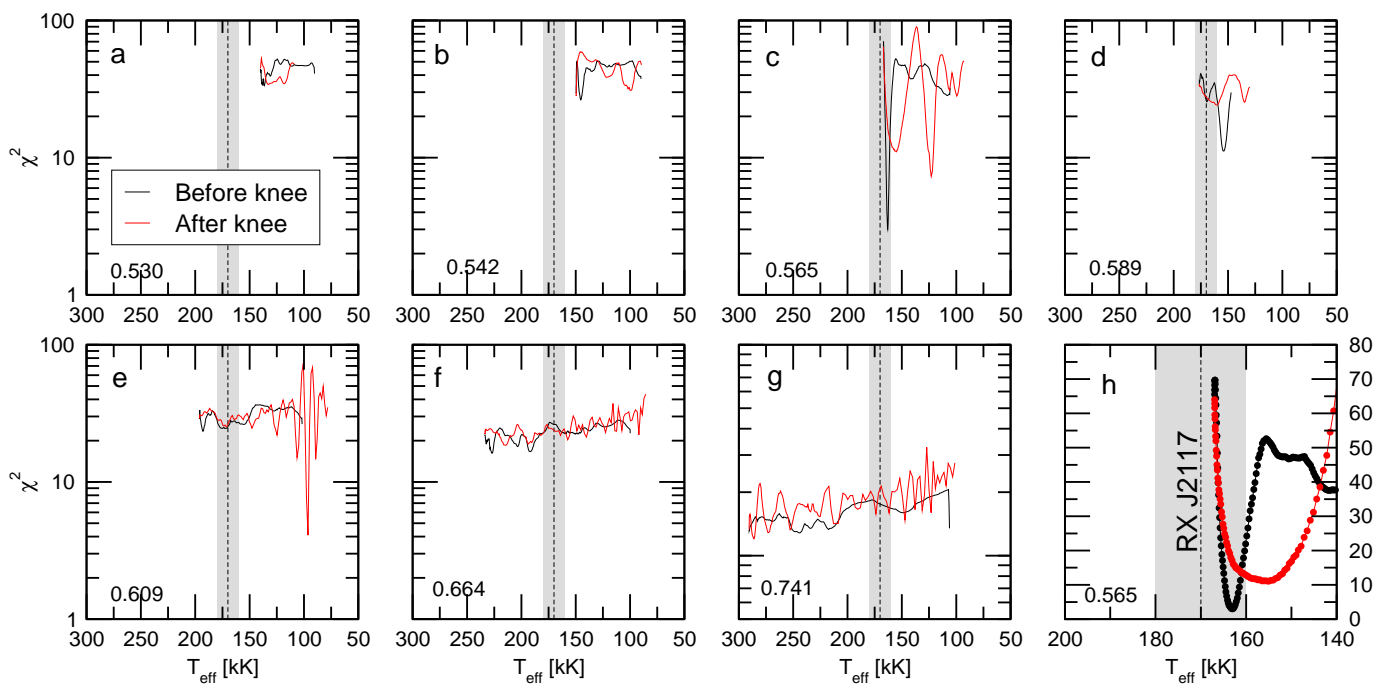}
\caption{The quality  function  of the  period  fit in terms of  the
  effective temperature  for the PG~1159 sequences with different
  stellar masses  (in solar units), indicated  at  the  left-bottom
  corner  of  each  panel.   Black (red) lines correspond to stages
  before (after) the evolutionary knee (see Fig. \ref{fig:1}).   Only
  the periods with $m= 0$ (see Table \ref{table:RXJ2117-extended})
  have been considered. Note  the  strong minimum  in  panel {\bf c},
  corresponding  to $M_{\star}= 0.565 M_{\odot}$.  Panel {\bf h} is a
  zoom of  the region with the  strong minimum seen in panel {\bf c};
  the $y$-axis scale is linear in this case. The vertical dashed line
  is the spectroscopic  $T_{\rm eff}$ of  RX~J2117  (170~kK)  and  the
  gray zone depicts its uncertainties  ($\pm 10$~kK).}
\label{fig:chi2-RXJ2117} 
\end{figure*} 

\begin{table*}
\centering
\caption{Observed and theoretical periods of the asteroseismological
  model for RX~J2117 [$M_{\star}= 0.565 M_{\odot}$, $T_{\rm eff}=
    162\,992$ K, $\log(L_{\star}/L_{\odot})= 3.373$]. Periods are in
  seconds  and rates of period change  (theoretical) are  in units of
  $10^{-12}$ s/s. $\delta \Pi_i= \Pi^{\rm O}_i-\Pi_k$ represents  the
  period differences, $\ell$ the harmonic degree, $k$ the radial
  order, $m$ the azimuthal index.  The last column gives information
  about the pulsational stability/instability   nature  of  the
  modes.}
\begin{tabular}{cc|ccccccc}
\hline
\noalign{\smallskip}
$\Pi_i^{\rm O}$ & $\ell^{\rm O}$ & $\Pi_k$ & $\ell$ & $k$ & $m$ & $\delta \Pi_k$ & $\dot{\Pi}_k$ & 
Unstable\\ 
(s) & & (s) & & & & (s) & ($10^{-11}$ s/s) &  \\
\noalign{\smallskip}
\hline
\noalign{\smallskip}       
 692.267 & 1 & 689.975  & 1 & 30   &   0  & 2.292    & $-2.561$  & no\\ 
 712.975 & 1 & 712.003  & 1 & 31   &   0  & 0.972    & $-4.775$  & no\\
 733.948 & 1 & 732.780  & 1 & 32   &   0  & 1.168    & $-5.217$  & no\\
 757.354 & 1 & 754.864  & 1 & 33   &   0  & 2.490    & $-6.022$  & yes\\
 778.921 & 1 & 777.318  & 1 & 34   &   0  & 1.603    & $-4.702$  & yes\\
 799.495 & 1 & 798.832  & 1 & 35   &   0  & 0.663    & $-2.655.$  & yes\\
 817.375 & 1 & $\hdots$ & 1 & 36   & $-1$ & $\hdots$ & $\hdots$  & $\hdots$\\ 
 821.105 & 1 & 820.614  & 1 & 36   &   0  & 0.491    & $-2.508$  & yes \\      
 824.880 & 1 & $\hdots$ & 1 & 36   & $+1$ & $\hdots$ & $\hdots$  & $\hdots$\\
 843.692 & 1 & 843.310  & 1 & 37   &   0  & 0.382    & $-5.222$  & yes\\ 
 885.736 & 1 & 887.005  & 1 & 39   &   0  & -1.269   & $-7.721$  & yes\\
 902.761 & ? & 902.523  & 2 & 70   &   0  & 0.238    & $-7.666$  & yes\\ 
 907.489 & 1 & 907.453  & 1 & 40   &   0  & 0.036    & $-7.849$  & yes\\
 951.750 & 1 & 953.041  & 1 & 42   &   0  & -1.291   & $-4.485$  & yes\\
 966.785 & 1 & $\hdots$ & 1 & 43   & $-1$ & $\hdots$ & $\hdots$  &  $\hdots$\\
 972.073 & 1 & 974.580  & 1 & 43   &   0  & -2.507   & $-7.938$  & yes\\      
 994.387 & 1 & 994.737  & 1 & 44   &   0  & -0.350   & $-10.487$ & yes\\
1016.467 & 1 & 1016.873 & 1 & 45   &   0  & -0.406   & $-10.678$ & yes\\
1031.978 & 1 & $\hdots$ & 1 & 46   & $-1$ & $\hdots$ & $\hdots$  & $\hdots$\\
1038.120 & 1 & 1039.311 & 1 & 46   &   0  & -1.191   & $-11.222$ & yes\\    
1044.041 & 1 & $\hdots$ & 1 & 46   & $+1$ & $\hdots$ & $\hdots$  & $\hdots$\\  
1058.026 & 1 & 1060.183 & 1 & 47   &   0  & -2.157   & $-8.262$  & yes\\
1103.292 & 1 & 1103.398 & 1 & 49   &   0  & -0.106   & $-12.206$ & yes\\
1124.156 & 1 & 1125.203 & 1 & 50   &   0  & -1.047   & $-14.094$ & yes\\       
1131.200 & 1 & $\hdots$ & 1 & 51   & $+1$ & $\hdots$ & $\hdots$  & $\hdots$\\
1146.346 & 1 & 1146.771 & 1 & 51   &   0  & -0.425   & $-11.240$ & yes\\
1189.956 & 1 & 1188.768 & 1 & 53   &   0  & 1.188    & $-8.643$  & yes\\
1350.870 & ? & 1345.193 & 2 & 105  &   0  & 5.677    & $-9.485$  & no \\
1557.010 & 1 & 1559.013 & 1 & 70   &   0  & -2.003   & $-13.293$ & no \\
1976.060 & ? & $\hdots$ & 1 & 90   & $-1$ & $\hdots$ & $\hdots$  & $\hdots$\\
1997.760 & ? & 1998.018 & 1 & 90   &   0  & -0.258   & $-8.917$  & no  \\
\noalign{\smallskip}
\hline
\end{tabular}
\label{table:RXJ2117-asteroseismic-model}
\end{table*}

\begin{table}
\centering
\caption{The main characteristics of the GW Vir star RX~J2117. The second column  
corresponds to spectroscopic results, whereas the third 
column present results from the asteroseismological model 
of this work.} 
\begin{tabular}{l|cc}
\hline
\hline
Quantity & Spectroscopy &  Asteroseismology \\
         & Astrometry   &   (This work)      \\ 
\hline
$T_{\rm eff}$ [kK]                            & $170 \pm 10^{\rm (a)}$         & $163.0_{-3.7}^{+2.5}$ \\
$M_{\star}$ [$M_{\odot}$]                     & $0.716 \pm 0.15$               & $0.565_{-0.023}^{+0.024}$ \\ 
$\log g$ [cm/s$^2$]                           & $6.0_{-0.2}^{+0.3\rm (a)}$     & $6.62_{-0.07}^{+0.11}$  \\ 
$\log (L_{\star}/L_{\odot})$                  & $3.95 \pm 0.5^{\rm (b)}$       & $3.37 \pm 0.04$ \\  
$\log(R_{\star}/R_{\odot})$                   & $\hdots$                       & $-1.21_{-0.025}^{+0.046}$ \\  
$M_{\rm env}$ [$M_{\odot}$]                   & $\hdots$                       & $0.02$ \\  
$(X_{\rm He},X_{\rm C},X_{\rm O})_{\rm s}$ &   0.39,  0.55,  0.06$^{\rm (a)}$          & 0.39, 0.32,  0.22  \\   
$d$  [pc]                                     & $502.186\pm 12.554^{\rm (c)}$  & $480\pm20$  \\ 
$\pi$ [mas]                                   &  $1.991\pm0.050^{\rm (c)}$     & $2.08\pm0.09$\\ 
\hline
\hline
\end{tabular}
\label{table:modelos-sismo-rxj}

{\footnotesize  References: (a)  \cite{2006PASP..118..183W};  (b) \cite{1993A&A...268..561M}; 
(c) {\it Gaia}; (d) \cite{1993AJ....106.1973A}.}
\end{table}

For this star, we have calculated the average of the computed period spacings for
$\ell= 1$, $\overline{\Delta \Pi_{k}}$,  in terms of the effective
temperature for all the masses considered. The results  are shown in
Fig.~\ref{fig:psp-teff-RXJ2117}, where we depict  $\overline{\Delta
  \Pi_{k}}$ corresponding to evolutionary stages
before the maximum possible  effective temperature, ${T_{\rm
    eff}}_{\rm MAX}$  (that depends on the  stellar mass) with red
dashed lines, and the  phases after that ${T_{\rm eff}}_{\rm MAX}$
(the WD stage itself)  with solid black lines. The location of
RX~J2117 is indicated by a small red  circle with error bars, and
corresponds to the effective temperature of the star  according to
\cite{1997fbs..conf..217R} and the period spacing derived in
Sect.~\ref{obser-rxj2117}. The star has a period-spacing value between
the values of  the curves of $0.565 M_{\odot}$ and $0.589 M_{\odot}$
at the maximum temperature  positions. We perform a linear
interpolation between the  maximum values of effective temperature for
each sequence and obtain $M_{\star}= 0.569 \pm 0.015\,M_\odot$. This
mass value is in good agreement with that inferred  by
\cite{2002A&A...381..122V}, $M_{\star}= 0.56_{-0.04}^{+0.02}
M_{\odot}$,  and \cite{2007A&A...461.1095C}, $M_{\star}=
0.560_{-0.013}^{+0.018}  M_{\odot}$  that also use the period spacing
to infer the stellar mass.  Finally,  we  note  that  our   inferred
stellar  mass  value  of $M_{\star}= 0.569\pm0.015 \,M_{\odot}$  is
in  strong disagreement with  the spectroscopic mass, $M_{\star}=
0.716 \pm 0.150\, M_{\odot}$ (see Sect. \ref{models}). 

Next, we describe our period-to-period fit procedure. The merit
function  $\chi^2(M_{\star}, T_{\rm eff})$  (Eq. \ref{eq:3}) was
evaluated for   stellar  masses $0.530, 0.542, 0.565, 0.589, 0.609,
0.664$, and $0.741\, M_{\odot}$,  and for a very wide interval of the
effective temperatures (depending on the stellar mass) with a very
small step. In our analysis of period-to-period fits, we only
considered the central component ($m= 0$) in the case of multiplets.
Following the results obtained in Sect.~\ref{period-spacing-rxj2117},
for RX~J2117 we assume that a subset of 21 periods are associated to
$\ell= 1$ modes --- those 20 periods marked with an asterisk in
Table~\ref{table:RXJ2117-extended} along with the period at $\sim
1557$ s--- and leave free the assignment as $\ell = 1$ or $\ell = 2$
to the remaining 3 periods. Specifically, we set the value $\ell= 1$
for all the periods except those at $\sim 902$~s,  $\sim 1351$~s, and
$1998$~s\footnote{Note that, on the basis of the analysis made in
  Sect.  \ref{obser-rxj2117} and Sect. \ref{period-spacing-rxj2117},
  we consider the modes  with periods 817.375 s, 824.880 s, 966.785 s,
  1031.978 s, 1044.041 s, 1131.200 s, and 1976.060 s as $m \neq 0$
  components of rotational multiplets,  and thus they are ignored in
  our period-to-period fits.}. We display in
Fig.~\ref{fig:chi2-RXJ2117} our results. We find only one minimum
compatible with the effective temperature of RX~J2117 and its
uncertainties, corresponding to a PG~1159 model  characterized by
$M_{\star}= 0.565\, M_{\odot}$  and $T_{\rm eff}= 162\,992$~K (panel
{\bf c}). There are other minima for other stellar masses, but none of them
have an effective temperature compatible with the $T_{\rm eff}$ of
RX~J2117, and they must be discarded.

We adopt the model characterized by $M_{\star}= 0.565 M_{\odot}$,
$T_{\rm eff}= 162\,992$~K,  and $\log(L_{\star}/L_{\odot})= 3.373$ as
the asteroseismological model for RX~J2117. Note that this model
corresponds to an evolutionary stage just before the star reaches its
maximum effective temperature (${T_{\rm eff}}_{\rm MAX}= 167\,000$
K). Our results are almost identical to those obtained by the analysis
of \cite{2007A&A...461.1095C} (asteroseismological model with
$M_{\star}= 0.565 M_{\odot}$  and $T_{\rm eff}= 163\,418$~K) which was
based on the same PG~1159 stellar models than in the present analysis,
but employing the set of periods measured with ground-based
observations of \cite{2002A&A...381..122V} alone.  In this way, the
incorporation of the new periods detected with {\it TESS} to the
analysis seems to strengthen the validity of the results of
\cite{2007A&A...461.1095C}. In
Table~\ref{table:RXJ2117-asteroseismic-model} we show a  detailed
comparison of the observed periods of RX~J2117 and the theoretical
$m= 0$ periods  of  the  asteroseismological  model. According to our
asteroseismological model, all the $m= 0$ periods exhibited by
RX~J2117 correspond to $\ell= 1$ modes with high radial order $k$,
except two periods that are associated to $\ell= 2$ modes. In order to
quantitatively assess the quality of our period fit, we compute the
average   of   the   absolute   period    differences,
$\overline{\delta \Pi_i}= \left( \sum_{i= 1}^n |\delta \Pi_i|
\right)/n$, where $\delta \Pi_i= (\Pi_{\ell,k}-\Pi_i^{\rm o})$ and $n=
24$,  and the root-mean-square residual, $\sigma= \sqrt{(\sum_{i= 1}^n
  |\delta \Pi_i|^2)/n}= \sqrt{\chi^2}$.  We obtain $\overline{\delta
  \Pi_i}= 1.26$~s and $\sigma= 1.73$~s.  The quality of our fit for
RX~J2117 is slightly worst than that achieved by
\cite{2007A&A...461.1095C} ($\overline{\delta \Pi_i}= 1.08$~s  and
$\sigma= 1.34$~s).  In order to have a global indicator of the
goodness of the period fit that takes into account the number of free
parameters, the number of fitted periods, and the proximity between
the  theoretical and observed periods, we computed the Bayes
Information Criterion \citep[BIC;][]{2000MNRAS.311..636K}: 

\begin{equation}
{\rm BIC}= n_{\rm p} \left(\frac{\log N}{N} \right) + \log \sigma^2,
\end{equation}

\noindent where $n_{\rm p}$ is the number of free parameters of the
models, and $N$ is the number of observed periods. The smaller the
value of BIC, the better the quality of the fit. In our case, $n_{\rm
  p}= 2$ (stellar mass and effective temperature),   $N= 24$, and
$\sigma= 1.73$\,s.  We obtain ${\rm BIC}= 0.59$, which means that our
period fit is excellent.  

We also include in Table~\ref{table:RXJ2117-asteroseismic-model} the
rates of period change ($\dot{\Pi}\equiv d\Pi/dt$) predicted for each
$g$ mode of RX~J2117. Note that all of them are negative
($\dot{\Pi}<0$), implying that the periods are shortening over
time. The rate of  change of periods in WDs and pre-WDs is related
to $\dot{T}$ ($T$ being the temperature at the region of the period
formation) and $\dot{R_{\star}}$ ($R_{\star}$ being the stellar
radius) through the order-of-magnitude expression $(\dot{\Pi}/\Pi) \approx -a\ (\dot{T}/T) + b\ (\dot{R_{\star}}/R_{\star})$
\citep{1983Natur.303..781W}. According  to our asteroseismological
model, the star is heating and contracting before reaching its maximum
temperature (evolutionary knee) and entering its cooling stage. As a
consequence, $\dot{T}>0$ and $\dot{R_{\star}} <0$, and then,
$\dot{\Pi} < 0$. As shown in Fig.~\ref{fig:RXJ2117-period-stability}
of Sect.~\ref{period-spacing-rxj2117}, RX~J2117 exhibits irregular
variations of periods on different time scales. So, a comparison of
our theoretical predictions with the observational trends is not
possible at this time.

Table~\ref{table:RXJ2117-asteroseismic-model} also gives information
about the pulsational stability/instability  nature of the modes
associated with the periods fitted to the observed ones (ninth
column). In particular,  we examine the sign of the computed linear
nonadiabatic growth rates ($\eta_k$).  A positive value of $\eta_k$
means that the mode  is linearly unstable. Interestingly, the interval
of periods corresponding to unstable modes of our asteroseismological
model is almost coincident with the range of excited periods in RX~J2117,
except in the case of the shortest periods (692.267~s, 712.975~s, and
733.948~s) and the longest periods (1350.870~s, 1557.010~s, and
1997.760~s) exhibited by the star, which are not predicted to be
unstable by our theoretical calculations ($\eta_k < 0$). This is
because, for $T_{\rm eff}= 163\,000$~K, the  model sequence of
$M_{\star}= 0.565\, M_{\odot}$ has unstable $\ell= 1$ ($\ell= 2$)
periods in the range $777\ {\rm s}-1181$~s ($416\ {\rm s}-938$ s)
\citep{2006A&A...458..259C}. 

In Table~\ref{table:modelos-sismo-rxj}, we list the main
characteristics of the asteroseismological model for RX~J2117. The
seismological stellar mass ($0.565 M_{\odot}$) is in excellent
agreement with the value derived from the period spacing ($0.569
M_{\odot}$). The average of the dipole ($\ell= 1$) period spacings of
our  asteroseismological model is $\overline{\Delta \Pi}= 22.153$ s
and the asymptotic period spacing is $\Delta \Pi^{\rm a}= 22.064$ s,
in excellent agreement with the  $\ell= 1$ mean period spacing derived
for RX~J2117 in Sect. \ref{period-spacing-rxj2117} ($\Delta \Pi=
21.669 \pm 0.030$~s).  The luminosity of the asteroseismological
model, $\log(L_{\star}/L_{\sun})=  3.37$ is lower than the luminosity
inferred by \cite{1993A&A...268..561M}, $\log(L_{\star}/L_{\sun})=
3.95$, based on the evolutionary tracks of
\cite{1986ApJ...307..659W}. 

The asteroseismological distance can be computed as in
\cite{2007A&A...461.1095C}.   On the basis of the luminosity of the
asteroseismological model ($\log(L_{\star}/L_{\odot})= 3.37\pm0.04$)
and the bolometric correction given by a NLTE model atmosphere with
$T_{\rm eff}= 160$ kK and $\log g= 6.6$ computed with the T\"ubingen
Model Atmosphere Package \citep[$BC= -7.954$; see][for
  details]{2003ASPC..288...31W,2007A&A...461.1095C}, the absolute
magnitude can be assessed as $M_{\rm V}= M_{\rm B}-BC$,  where $M_{\rm
  B}= M_{{\rm B},\odot} - 2.5\ \log(L_{\star}/L_{\odot})$.  We employ
the solar bolometric magnitude $M_{\rm B \odot}= 4.74$
\citep{2000asqu.book.....C}. The seismological distance $d$  is
derived from the relation: $\log d= 15\ [m_{\rm V} - M_{\rm V} +5 -
  A_{\rm V} (d)]$, where we employ the interstellar  extinction law of
\cite{1998A&A...336..137C}.  The  interstellar  absorption $A_{\rm
  V}(d)$  is a nonlinear function of the  distance  and  also  depends
on  the  Galactic  latitude  ($b$).  For  the  equatorial  coordinates
of  RX~J2117  (Epoch  B2000.00, $\alpha= 21^{\rm h} 17^{\rm m} 8.^{\rm
  s}2,\ \delta= +34^{\circ} 12^{'} 27.^{''}58$) the corresponding
Galactic latitude is $b = -10^{\circ} 24^{'} 32.^{''}04$.  We use the
apparent visual magnitude  $m_{\rm V}= 13.16\pm0.01$
\citep{1993A&A...268..561M},  and obtain the seismological distance
and parallax  $d= 480\pm20$ pc and  $\pi= 2.08\pm0.09$ mas,
respectively, being the extinction coefficient  $A_{\rm V}=
0.47\pm0.02$. The uncertainty in the seismological distance comes
mainly from the uncertainty in the luminosity of the
asteroseismological model, which is  admittedly very small
($\Delta{\log(L_{\star}/L_{\odot})= 0.04}$) because this it is solely
attributed to internal errors, i.e., uncertainties typical of the
period-fit procedure. Realistic estimates of these errors (due to
uncertainties in stellar modeling and the pulsation computations) are
probably much higher. A very important check for the validation of the
asteroseismological model for RX~J2117 is the comparison of the
seismological  distance with the distance derived from astrometry. We
have available the estimates  from {\it Gaia}, $d_{\rm G}= 502 \pm 12$
pc and $\pi_{\rm G}= 1.991 \pm 0.05$ mas.  They are in excellent
agreement with the asteroseismological derivations in view of the
uncertainties in both determinations.

We close this section summarizing our findings for RX~J2117. The
seismological stellar mass derived from the period spacing is in
excellent agreement with the stellar mass of the asteroseismological
model --- derived through a fit to the individual periods --- and in
line with what was found by \cite{2002A&A...381..122V} and
\cite{2007A&A...461.1095C}. On the other hand, a  seismological mass
in the range $0.560-0.569 M_{\odot}$ apparently  disagrees  with the
spectroscopic mass, of $0.716\pm0.15 M_{\odot}$, but nevertheless they
are  still compatible each other, given the uncertainties in both
determinations ---particularly in the spectroscopic mass. On the other
hand, the seismological distance derived from the astereoseismological
model is in excellent agreement with the distance measured by {\it
  Gaia}.

\subsection{HS~2324}
\label{modelling-hs2324}

\begin{figure} 
\includegraphics[clip,width=1.0\columnwidth]{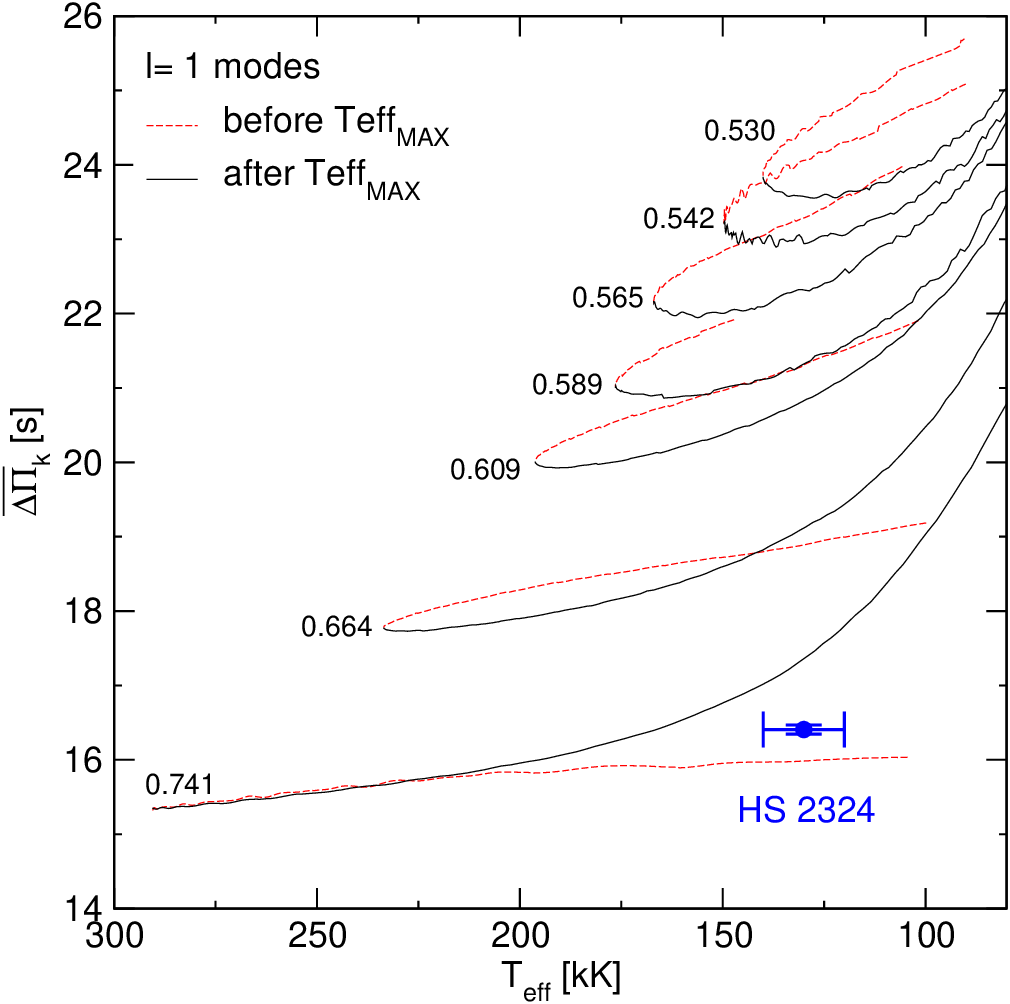}
\caption{Dipole ($\ell= 1$) average of the computed period spacings,
  $\overline{\Delta \Pi_k}$, assessed  in  a  range  of  periods  that
  includes  the  periods  observed  in HS~2324, shown  as solid
  (dashed) curves corresponding to stages before (after) the maximum
  $T_{\rm eff}$ for different stellar masses. The location of HS~2324
  when we  use the effective temperature derived by
  \cite{1996A&A...309..820D}, $T_{\rm eff}= 130\,000 \pm 10\,000$~K,
  and the period spacing $\Delta \Pi= 16.407\pm0.062$~s derived in
  Sect.~\ref{obser-hs2324} is highlighted with a blue circle.  The
  stellar mass derived from linear interpolation  is $M_{\star}=
  0.727\pm 0.017M_{\odot}$ if the star is before the maximum $T_{\rm
    eff}$, and by linear extrapolation, it is $M_{\star}= 0.758 \pm
  0.018 M_{\odot}$ if the star is after the maximum $T_{\rm eff}$.}
\label{fig:psp-teff-HS2324} 
\end{figure} 

\begin{figure*} 
\includegraphics[clip,width= 2.0\columnwidth]{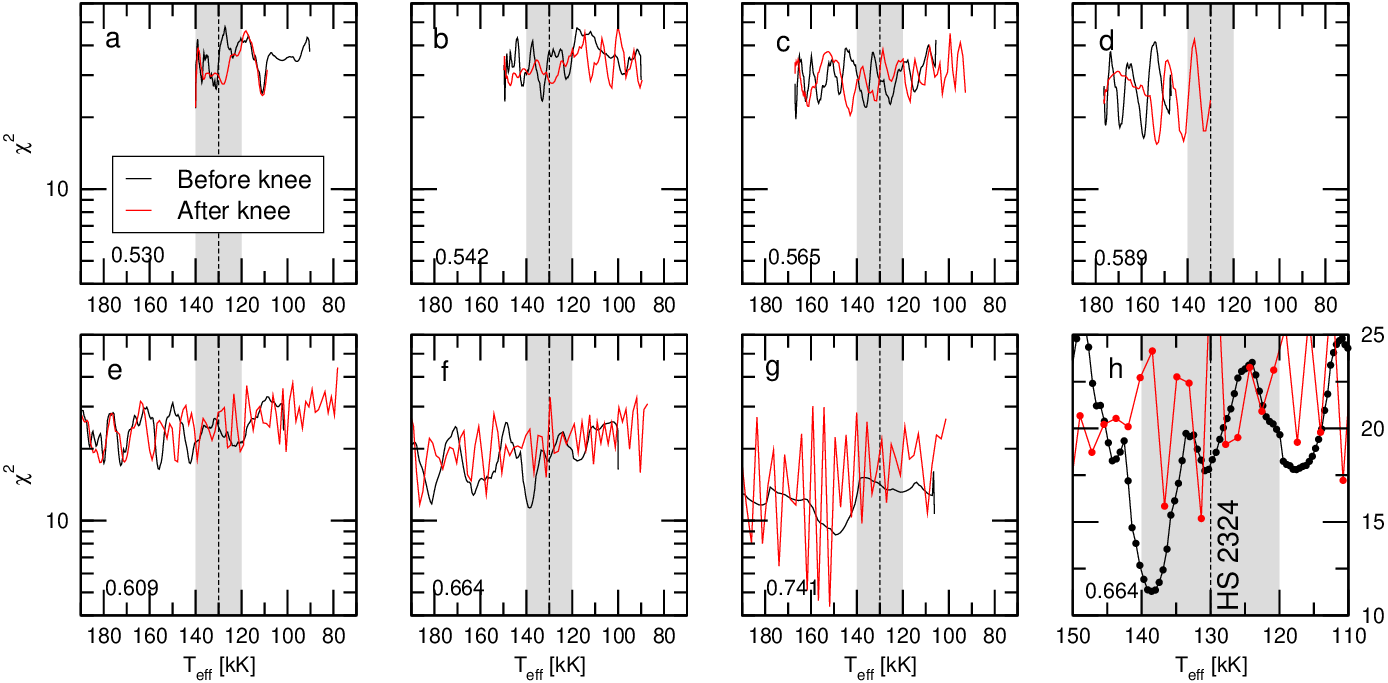}
\caption{Same as Fig. \ref{fig:chi2-RXJ2117}, but for HS~2324. Note
  the minimum  in  panel {\bf f} before the evolutionary knee (black
  curve),   corresponding  to $M_{\star}= 0.664 M_{\odot}$.  Panel
  {\bf h} is a zoom of  the region with the  strong minimum seen in
  panel {\bf f}; the $y$-axis scale is linear in this case. The
  vertical dashed line is the spectroscopic  $T_{\rm eff}$ of  HS~2324
  (130~kK)  and  the  gray zone depicts its quoted uncertainty  ($\pm
  10$~kK).}
\label{fig:chi2-HS2324} 
\end{figure*} 

\begin{table*}
\centering
\caption{Observed and theoretical periods of the asteroseismological
  model for HS~2324 [$M_{\star}= 0.664 M_{\odot}$, $T_{\rm eff}=
    138\,572$~K, $\log(L_{\star}/L_{\odot})= 4.069$].  The columns
  have the same meaning as in Table
  \ref{table:RXJ2117-asteroseismic-model}.}
\begin{tabular}{cc|cccccc}
\hline
\noalign{\smallskip}
$\Pi_i^{\rm O}$ & $\ell^{\rm O}$ & $\Pi_k$ & $\ell$ & $k$ &  $\delta \Pi_k$ & $\dot{\Pi}_k$ & Unstable\\
(s) & & (s) & & &  (s) & ($10^{-11}$ s/s) &  \\
\noalign{\smallskip }
\hline
\noalign{\smallskip}      
1039.020 & ? & 1038.518 & 2 & 101 &  0.502 &  $-48.352$ & no  \\ 
1047.100 & 1 & 1051.555 & 1 &  59 & $-4.455$ &   21.631 & no  \\ 
1049.877 & ? & 1048.216 & 2 & 102 &  1.661 &  $-65.542$ & no  \\
2005.780 & ? & 2006.376 & 1 & 113 & $-0.596$ &  $-18.035$ & yes \\      
2027.520 & ? & 2024.919 & 2 & 197 &  2.601 &  $-43.713$ & yes \\ 
2029.350 & 1 & 2022.871 & 1 & 114 &  6.479 &   $-8.498$ & yes  \\   
2047.260 & 1 & 2041.419 & 1 & 115 &  5.841 &  $-15.661$ & yes  \\   
2059.970 & 1 & 2060.657 & 1 & 116 & $-0.687$ &  $-47.391$ & yes  \\  
2076.110 & ? & 2078.013 & 2 & 202 & $-1.903$ &  $-84.734$ & yes \\  
2078.590 & 1 & 2078.239 & 1 & 117 &  0.351 &  $-93.688$ &  yes \\  
2095.046 & 1 & 2093.885 & 1 & 118 &  1.161 & $-132.285$ &  yes \\  
2098.670 & ? & 2098.804 & 2 & 204 & $-0.134$ &  $-73.882$ & yes \\  
2110.152 & 1 & 2110.330 & 1 & 119 & $-0.178$ & $-137.956$ & yes  \\   
2160.970 & 1 & 2164.402 & 1 & 122 & $-3.432$ &  $-19.931$ & yes  \\  
2170.490 & ? & 2172.624 & 2 & 211 & $-2.134$ &  $-88.824$ & yes \\  
2175.290 & 1 & 2181.536 & 1 & 123 & $-6.246$ &  $-20.287$ & yes \\  
2193.420 & 1 & 2199.885 & 1 & 124 & $-6.465$ &  $-52.300$ & yes \\  
2202.990 & ? & 2203.977 & 2 & 214 & $-0.987$ &  $-55.166$ & yes \\  
2553.230 & 1 & 2553.519 & 1 & 144 & $-0.289$ & $-128.112$ & yes \\        
2568.860 & 1 & 2571.257 & 1 & 145 & $-2.397$ &  $-78.122$ & yes \\         
2682.050 & 1 & 2677.444 & 1 & 151 &  4.606 & $-134.986$ & yes \\  
\noalign{\smallskip}
\hline
\end{tabular}
\label{table:HS2324-asteroseismic-model}
\end{table*}

\begin{table}
\centering
\caption{The main characteristics of the GW Vir star HS~2324. The
  second column   corresponds to spectroscopic results, whereas the
  third  column present results from the asteroseismological model  of
  this work.} 
\begin{tabular}{l|cc}
\hline
\hline
Quantity & Spectroscopy &  Asteroseismology \\
         & Astrometry   &   (This work)      \\ 
\hline
$T_{\rm eff}$ [kK]                            & $130 \pm 10^{\rm (a)}$         & $138.6_{-2.8}^{+3.0}$\\
$M_{\star}$ [$M_{\odot}$]                     & $0.532 \pm 0.150$              & $0.664_{-0.055}^{+0.077}$\\ 
$\log g$ [cm/s$^2$]                           & $6.2\pm0.2^{\rm (a)}$          & $5.72_{-0.10}^{+0.13}$\\ 
$\log (L_{\star}/L_{\odot})$                  & $\hdots$                       & $4.07_{-0.21}^{+0.18}$\\  
$\log(R_{\star}/R_{\odot})$                   & $\hdots$                       & $-0.73_{-0.08}^{+0.06}$\\  
$M_{\rm env}$ [$M_{\odot}$]                   & $\hdots$                       & $0.02$ \\  
$(X_{\rm H},X_{\rm He},X_{\rm C},$            &   0.17,  0.35,  0.42,           & 0.00, 0.47,  0.33,\\   
$X_{\rm O})_{\rm s}$                          &   $0.06^{\rm (a)}$             & 0.13\\
$d$  [pc]                                     & $1448.033\pm 105.266^{\rm (b)}$  & $4379_{-940}^{+1009}$\\ 
$\pi$ [mas]                                   &  $0.691\pm0.0502^{\rm (b)}$ & $0.228_{-0.042}^{+0.063}$\\ 
\hline
\hline
\end{tabular}
\label{table:modelos-sismo-hs2324}

{\footnotesize  References: (a) \cite{2006PASP..118..183W};  (b) {\it Gaia DR2}.}
\end{table}

In Fig. \ref{fig:psp-teff-HS2324} we show $\ell= 1$ $\overline{\Delta
  \Pi_{k}}$ as a function of $T_{\rm eff}$ for all the masses
considered. Note that these curves are slightly different as compared
with those of Fig.~\ref{fig:psp-teff-RXJ2117} that correspond to
RX~J2117. This is because the averages of the computed period spacings  are
calculated considering different ranges of periods for different stars.
The period spacing of HS~2324 is $\Delta \Pi= 16.407\pm0.062$~s, as
derived in Sect.~\ref{obser-hs2324}. In this case, we considered two
possibilities,  that the star is before or after the maximum effective
temperature, that is, the  ``evolutionary knee'' (see
Fig.~\ref{fig:1}). This is because we do not know,  in principle, the
evolutionary stage in which the star is. We find $M_{\star}= 0.727\pm
0.017M_{\odot}$ if the star is before the maximum $T_{\rm eff}$,  and
$M_{\star}= 0.758 \pm 0.018 M_{\odot}$ if the star is after the
maximum $T_{\rm eff}$. These values are much larger than the
spectroscopic  mass derived in Sect.~\ref{models}, of $M_{\star}=
0.532 \pm 0.150 \ M_{\sun}$.

Next, we describe our period-to-period fit analysis for
HS~2324. Again,  the merit function  $\chi^2(M_{\star}, T_{\rm eff})$
(Eq.~\ref{eq:3}) was evaluated  for  all the stellar  masses and
effective temperatures covered by our PG~1159 model  sequences. We
employed the 21 periods of Table~\ref{table:HS2324-extended}.  We
adopted the same approach as for the case of RX~J2117,  that is, we
assumed that there is a subset of 13 periods identified as $\ell= 1$
according to the period spacing derived in
Sect.~\ref{period-spacing-hs2324} (see Table
\ref{table:HS2324-extended}), but the remaining 8 periods  are allowed
to be identified with modes with dipole or quadrupole modes. 

Given the absence of a single global minimum in  the quality function
when evaluated for all the stellar masses and effective temperatures
considered, we were forced to analyze what happens in the range of
effective temperatures  published for HS~2324, that is $120\,000
\lesssim T_{\rm eff} \lesssim 140\,000$~K.  We display in
Fig.~\ref{fig:chi2-HS2324} our results. We note the existence of a
clear  minimum for a model with $M_{\star}= 0.664 M_{\odot}$ and
$T_{\rm eff}= 138\,572$~K. It is  located at the stages previous to
the maximum temperature for this mass.   We adopt this model  as the
asteroseismological model for HS~2324.  In
Table~\ref{table:HS2324-asteroseismic-model} we show a  detailed
comparison  of the observed periods of HS~2324 and the the theoretical
periods  of  the   asteroseismological  model. For this
asteroseismological solution, we have $\overline{\delta \Pi_i}=
2.53$~s and $\sigma= 3.36$~s. The quality of our fit for HS~2324 is
worst than that achieved for RX~J2117 ($\overline{\delta
  \Pi_i}= 1.26$~s  and $\sigma= 1.73$~s).   We also computed the Bayes
Information Criterion. In this case, $N_{\rm p}= 2$ (stellar mass and
effective temperature), $n= 21$, and $\sigma= 3.36$\,s.  We obtain
${\rm BIC}= 1.18$, which means that our fit is not as good as for
RX~J2117 (${\rm BIC}= 0.59$), but still satisfactory.  

According to our  seismological model, the values of the rate of
period change for the $g$ modes of HS~2324 (column 7 of
Table~\ref{table:HS2324-asteroseismic-model}) are $\sim 10-100$ times
greater than for the asteroseismological model of RX~J2117
(Table~\ref{table:RXJ2117-asteroseismic-model}).  This is because the
asteroseismological model of HS~2324 ($M_{\star}= 0.664 M_{\odot}$) is
more massive than the asteroseismological model of RX~J2117
($M_{\star}= 0.565 M_{\odot}$), and the evolution through the PG~1159
stage proceeds considerably faster  for  massive  stars
\citep{2006A&A...454..845M}.  In Sect.~\ref{period-spacing-hs2324} we
called the attention of a possible physical change of the period at
$\sim 2193$~s from 1999 to 2020, with a negative derivative and a
magnitude of $\sim 10^{-9}$~s/s. This value is compatible with the
theoretical rate of change for this period, of  $-0.523 \times
10^{-9}$~s/s (Table~\ref{table:HS2324-asteroseismic-model}).

Column 8 of Table~\ref{table:HS2324-asteroseismic-model} shows that
most of the periods of the asteroseismological model that fit the
observed periods are predicted to be unstable  ($\eta_k > 0$), except
for the three periods shorter than $\sim 2000$~s, for which our
nonadiabatic computations indicate pulsational stability. For $T_{\rm
  eff}= 138\,600$~K, the  model sequence of $M_{\star}= 0.664
M_{\odot}$ has unstable dipole (quadrupole)  periods in the range
$2028\ {\rm s}-465$~s ($1215\ {\rm s}-3095$~s)
\citep{2006A&A...458..259C}\footnote{Note that the nonadiabatic
  pulsation  calculations of \cite{2006A&A...458..259C} correspond to
  PG~1159 models  without H in the envelopes, which is not strictly the
  case of a hybrid GW Vir star, which has some H on its surface. However,
  we do not expect  the results of the pulsation analysis to be
  appreciably modified by  the presence of a trace of H.}.

In Table~\ref{table:modelos-sismo-hs2324}, we list the main
characteristics of the asteroseismological model for HS~2324. The
stellar mass of the seismological model ($0.664 M_{\odot}$) is $\sim
10 \%$ smaller than --- although  still compatible with --- the value
derived from the period spacing ($\sim 0.73 M_{\odot}$), assuming that
the star is before the evolutionary knee. This mass discrepancy is
reflected in  the fact that dipole ($\ell= 1$) mean period spacing of
our  asteroseismological model ($\overline{\Delta \Pi}= 18.81$ s) is
$\sim 14 \%$  longer  than  the $\ell= 1$ mean period spacing derived
for HS~2324 in Sect.~\ref{period-spacing-hs2324} ($\Delta \Pi= 16.407
\pm 0.062$~s).  Given the error ranges of the spectroscopic mass
($0.532 \pm 0.150 \ M_{\sun}$, Sect. \ref{models}) and the
asteroseismic mass determinations, they are compatible  each other.  

We can compute the asteroseismological distance of HS~2324, as we did
for RX~J2117.   We employ the luminosity of the asteroseismological
model of HS~2324, $\log(L_{\star}/L_{\odot})= 4.07^{+0.18}_{-0.21}$, a
bolometric correction of $BC= -7.24$ \citep[extrapolated from the
  value corresponding to PG~1159$-$035 as given
  by][]{1994ApJ...427..415K},   and the interstellar  extinction law
of \cite{1998A&A...336..137C}.  For  the  equatorial  coordinates  of
HS~2324  (Epoch  B2000.00, $\alpha= 23^{\rm h} 27^{\rm m} 15.^{\rm
  s}95,\ \delta= +40^{\circ} 01^{'} 23.^{''}63$) the corresponding
Galactic latitude is $b = -20^{\circ} 3^{'} 0.^{''}86$.  We adopt
$m_{\rm V}= 15.41$ \citep{2011MNRAS.410..899F} and  obtain a
seismological  distance, a parallax, and an extinction coefficient of
$d= 4379_{-940}^{+1009}$ pc, $\pi= 0.228_{-0.042}^{+0.063}$ mas, and
$A_{\rm V}= 0.398$,  respectively. On the other hand, the values
measured by  {\it Gaia} are $d_{\rm G}= 1448 \pm 105$ pc and $\pi_{\rm
  G}= 0.691 \pm 0.050$ mas,   in line with the distance inferred by
\cite{1996A&A...309..820D} ($1500$ pc).  At variance with the case of
RX~J2117, for HS~2324 there is a serious disagreement between our
asteroseismological distance and the distance  obtained with {\it
  Gaia}. This discrepancy must be  largely attributed to the high
luminosity of our seismological model. In this sense,  the uncertainty
of the luminosity of the seismological model ($\Delta
\log(L_{\star}/L_{\odot})\sim 0.2$)  is just formal and it is probably
underrated, since it reflects only internal errors  of the period-fit
processes. Thus, a compatibility  between  the seismological distance
and the  astrometric distance from {\it Gaia} would  be achieved if
we  could employ a more  realistic   estimate  of  the  uncertainties
in the luminosity of  the  asteroseismological  model.

We end this section summarizing our results for HS~2324.  The  stellar
mass  inferred from the  period separation ($\sim 0.73 M_{\odot}$) is
in line  with the mass of the seismological model ($\sim 0.66
M_{\odot}$), and still compatible  with the spectroscopic mass, of
$0.532\pm 0.15 M_{\odot}$, given the large uncertainties in the
parameters with which the spectroscopic mass is determined ---in
particular the $\log g$ value.

\subsection{NGC~6905}
\label{modelling-ngc6905}

\begin{figure} 
\includegraphics[clip,width=1.0\columnwidth]{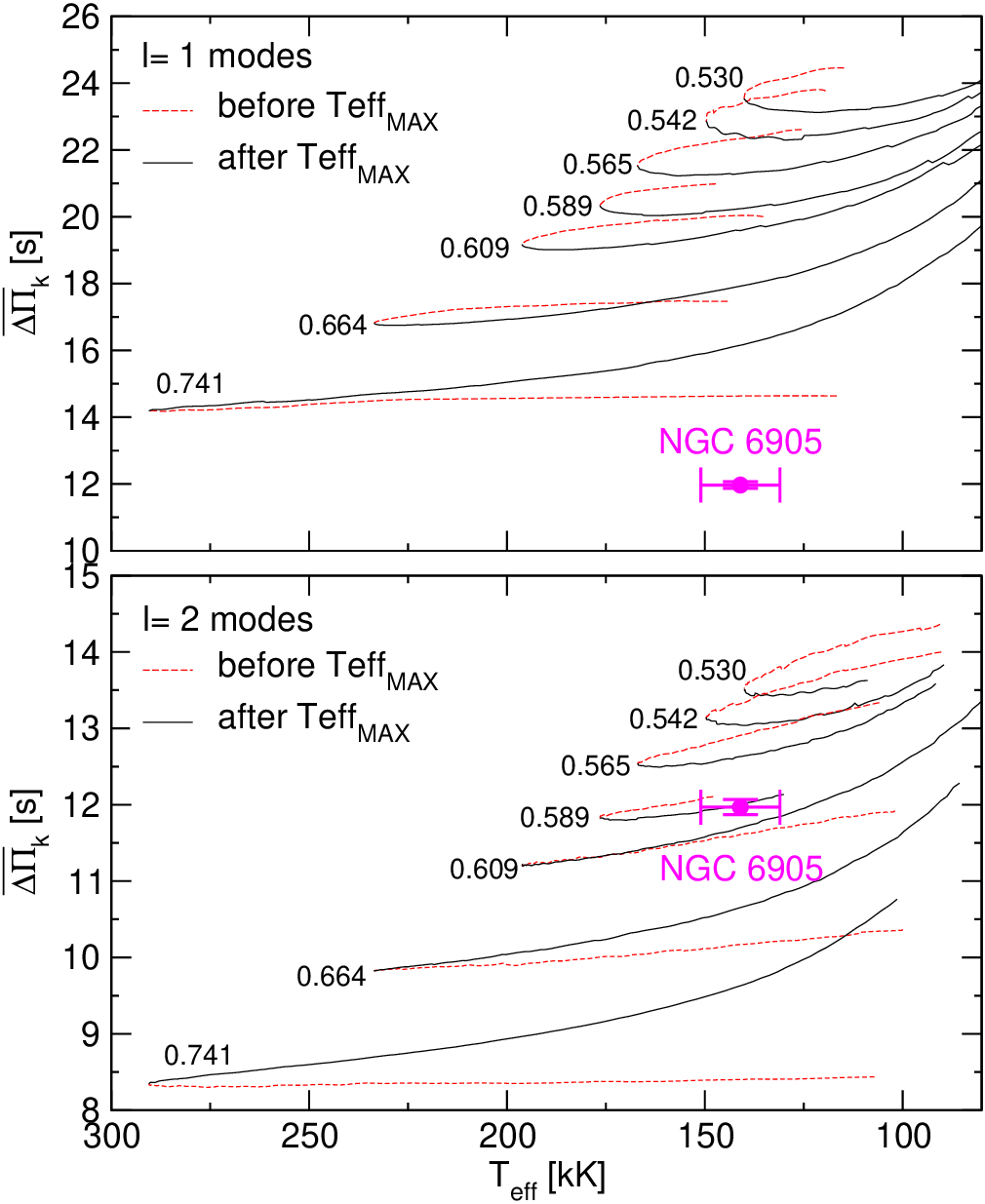}
\caption{Upper panel: dipole ($\ell= 1$) average of the computed period spacings,
  $\overline{\Delta \Pi_k}$, assessed  in  a  range  of  periods  that
  includes  the  periods  observed  in NGC~6905, shown as solid
  (dashed) curves corresponding to stages before (after) the maximum
  $T_{\rm eff}$ for different stellar masses. The location of NGC~6905
  when we  use the effective temperature derived by
  \cite{1996AJ....111.2332C}, $T_{\rm eff}= 141\,000 \pm 10\,000$~K,
  and the period spacing $\Delta \Pi= 11.9693\pm0.0988$~s derived in
  Sect.~\ref{obser-ngc6905} and assumed to be associated to $\ell= 1$ modes, 
  is highlighted with a magenta circle.  The
  stellar mass derived from linear extrapolation is $M_{\star}\sim
  0.818 M_{\odot}$ if the star is before the maximum $T_{\rm eff}$,
  and it is $M_{\star} \sim 0.861 M_{\odot}$ if the star is after the
  maximum $T_{\rm eff}$. Lower panel: same as in upper panel, but for the case
  in which the period spacing is assumed to be associated to $\ell= 2$ modes 
  ($\Delta \Pi= 11.9693\pm0.0988$~s). In this case, the stellar mass
  is $M_{\star}= 0.596 \pm 0.09 M_{\odot}$ and $M_{\star}=
0.590 \pm 0.007 M_{\odot}$ if the star is before or after the
evolutionary knee, respectively.}
\label{fig:psp-teff-NGC6905} 
\end{figure} 

In the upper panel of Fig.~\ref{fig:psp-teff-NGC6905} 
we display the run of the dipole
$\overline{\Delta \Pi_{k}}$ in terms of $T_{\rm eff}$ for all the
masses considered, for the case of NGC~6905.  We assume the period
spacing $\Delta \Pi= 11.9693\pm0.0988$~s to be associated to $\ell= 1$
modes. This period spacing results from considering the periods of
modes detected with {\it TESS}, in addition to the periods from
\cite{1996AJ....111.2332C} (Table \ref{table:NGC6905-extended} of
Sect.~\ref{period-spacing-ngc6905}).  As in the case of HS~2324, for
NGC~6905 we considered two possibilities: the star is before or after
the ``evolutionary knee''. We find $M_{\star}\sim 0.818$ if the star
is before the maximum $T_{\rm eff}$,  and $M_{\star}\sim
0.811M_{\odot}$ if the star is after the maximum $T_{\rm eff}$. Note
that these values are obtained by extrapolation, because the period
spacing of the star, being only about $12$~s, is well below the
average of the computed spacing curves of our models, even the most
massive one ($M_{\star}= 0.741 M_{\sun}$). Therefore, our stellar mass
values are just {\it estimates}. The stellar mass of NGC~6905
according to these estimates is much higher than the spectroscopic
mass of this star, of $M_{\star}= 0.590 \pm 0.150 \ M_{\sun}$, as
derived in Sect.~\ref{models}. 

We also explored the possible situation in which the period spacing
of $\Delta \Pi= 11.9693\pm0.0988$~s correspond to $\ell= 2$ modes. We
then  compared this period spacing with the quadrupole
$\overline{\Delta \Pi_{k}}$ in terms of $T_{\rm eff}$ for all the
masses considered, as shown in the lower panel of 
Fig.~\ref{fig:psp-teff-NGC6905}. In this case, the mass
value is of  $M_{\star}= 0.596 \pm 0.09 M_{\odot}$ and $M_{\star}=
0.590 \pm 0.007 M_{\odot}$ if the star is before or after the
evolutionary knee, respectively. These values are in excellent
agreement with the spectroscopic mass value  ($M_{\star}= 0.59
M_{\sun}$).

We attempted a period-to-period fit for NGC~6905 using the individual
periods of Table \ref{table:NGC6905-extended}. We adopted the same
approach as for the case of RX~J2117 and HS~2324, i.e., we assumed
that there is  a subset of periods identified as $\ell= 1$ according
to the period spacing derived in Sect. \ref{period-spacing-ngc6905}
(see Table \ref{table:NGC6905-extended}),  but the remaining periods
are allowed to be associated with dipole or quadrupole modes  as
well. Note that there are two possibilities with respect to  the
period $\sim 816$ s. Although unlikely, one possibility would be to consider 
that it is a component  $\ell= 1$, $m= \pm 1$ of a rotational triplet, in which
case, we can ignore this period in our  period-fit process and there
are 10 observed periods available.  The other possibility  would be to
consider it as a  $\ell= 2$,  $m = 0$ mode, and therefore  incorporate
it into the analysis. In this case the period-to-period fit is carried
out on a set of 11 periods. The results of our procedure with 10 and
with 11 periods  are very similar, and do not indicate a very clear
asteroseismological model regarding the  uniqueness of the
solutions. Specifically, there are several possible solutions  in
terms of period match within the effective-temperature range of
interest  ($131\,000-151\,000$~K) for NGC~6905. However, most of the
possible solutions  (minima in the quality function) correspond to
stages after the evolutionary knee.  If we accept that  NGC~6905 is
evolving towards the blue, before reaching its maximum  effective
temperature --- as indicated by its position in the $\log T_{\rm
  eff}-\log g$ diagram; see Fig. \ref{fig:1}--- then we do not find
any clear minima in the quality function,  making virtually impossible
to isolate a clear and unambiguous seismological solution\footnote{We
  still could  pick out a possible seismological solution consisting
  of a pre-WD model with a stellar mass $M_{\star}= 0.664 M_{\odot}$
  and an effective temperature  $T_{\rm eff}= 145\,400$ K. However,
  the match between the theoretical and observed periods is very poor,
  with differences of up to $\sim 10$~s, something that is reflected
  by the  high value of the local minimum in the quality function for
  that model ($\chi^2= 16.05$ s$^2$).}. Thus, for NGC~6905 we are
unable to find an asteroseismological model. The harmonic degree of the 
mode with period $\sim 816$~s remains
undetermined, although according to our calculations, the period fits
slightly improve when we consider that  this corresponds to $\ell=
2$. 

We also considered the  possibility that the period
spacing derived in Sect. \ref{period-spacing-ngc6905} corresponds to
modes $\ell= 2$. In this case, in Table \ref{table:NGC6905-extended} the modes 
identified as $\ell= 1$ are now identified as $\ell= 2$. In total, 
we have available 7 quadrupole modes  and 4 modes that can be 
alternatively dipole or quadrupole. We assume that the mode with 
period $\sim 816$~s can correspond to a $\ell= 1$ mode  or a $\ell= 2$ 
mode as well. We have
repeated the period-fit procedure with this identification of modes,
but unfortunately we have not found a satisfactory seismological
solution. In particular, it is very difficult to fit the observed
periods at 816,3 s, 851.4 s,  874.8 s, 884 s with $\ell= 1$ or $\ell=
2$ theoretical periods, if we assume that the rest of the observed
periods correspond to $\ell= 2$ modes. 

\subsection{NGC~1501}
\label{modelling-ngc1501}

\begin{figure} 
\includegraphics[clip,width=1.0\columnwidth]{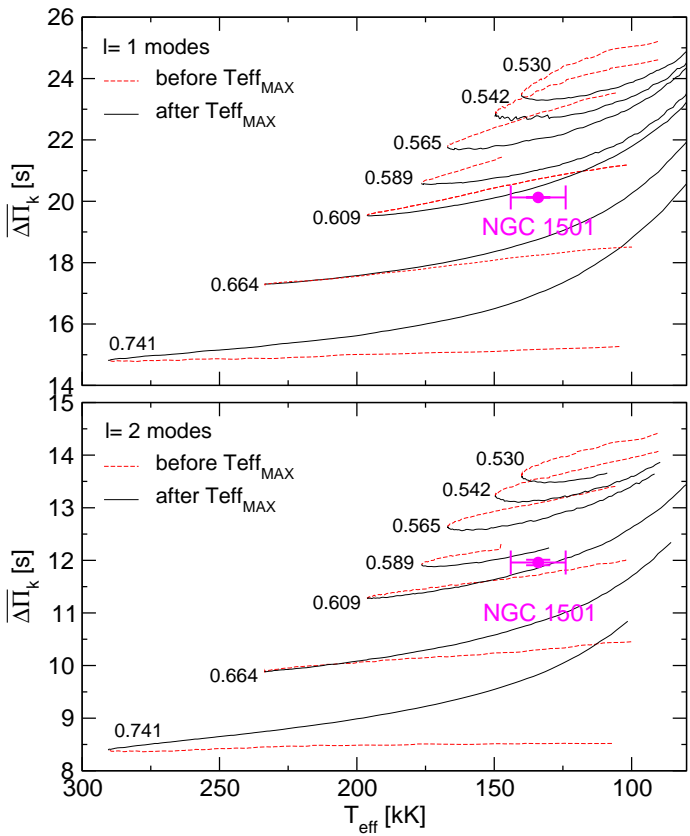}
\caption{Upper panel: dipole ($\ell= 1$) average of the computed
  period spacings, $\overline{\Delta \Pi_k}$, assessed  in  a  range
  of  periods  that includes  the  periods  observed  in NGC~1501,
  shown as solid (dashed) curves corresponding to stages before
  (after) the maximum $T_{\rm eff}$ for different stellar masses. The
  location of NGC~1501 when we  use the effective temperature derived
  by \cite{2006PASP..118..183W},  $T_{\rm eff}= 134\,000 \pm 10\,000$
  K, and the dipole period spacing, $\Delta \Pi_{\ell= 1}= 20.1262\pm
  0.0123$~s (derived in Sect.~\ref{obser-ngc1501}), is highlighted
  with a magenta circle.  The stellar mass derived from linear
  interpolation  is $M_{\star}= 0.621 \pm 0.015 M_{\odot}$ if the star
  is before the maximum $T_{\rm eff}$, and it  is $M_{\star}= 0.624
  \pm 0.015 M_{\odot}$ if the star is after the maximum $T_{\rm
    eff}$. Lower panel: same as in upper panel,  but for the case of
  $\ell= 2$ modes ($\Delta \Pi_{\ell= 2}= 11.9593\pm 0.0520$~s).  In
  this case, the stellar mass is  $M_{\star}= 0.601 \pm 0.015
  M_{\odot}$ and $M_{\star}= 0.606 \pm 0.015 M_{\odot}$, according the
  star is before or after the maximum $T_{\rm eff}$.  }
\label{fig:psp-teff-NGC1501} 
\end{figure} 

\begin{figure*} 
\includegraphics[clip,width= 2.0\columnwidth]{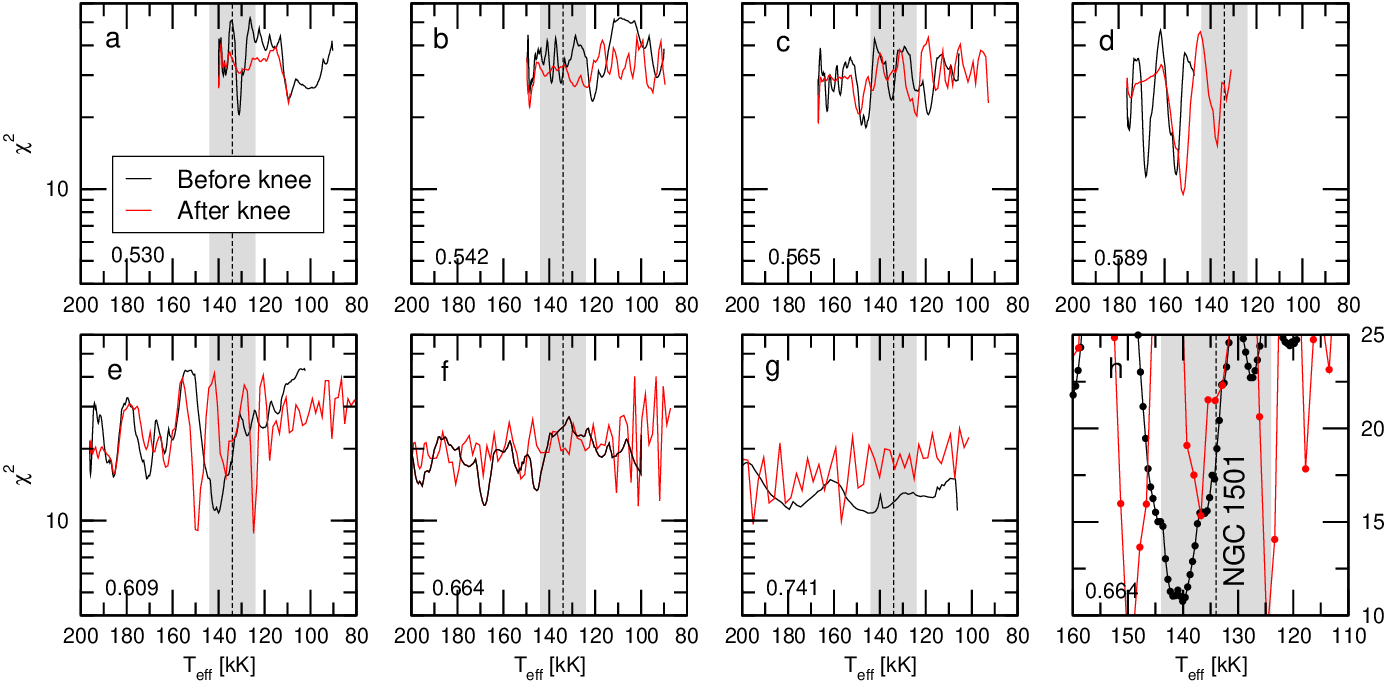}
\caption{Same as Fig.~\ref{fig:chi2-RXJ2117}, but for NGC~1501. Note
  the minimum  in  panel {\bf e} before the evolutionary knee (black
  curve),   corresponding  to $M_{\star}= 0.609 M_{\odot}$.  Panel
  {\bf h} is a zoom of  the region with the  strong minimum seen in
  panel {\bf e}; the $y$-axis scale is linear in this case. The
  vertical dashed line is the spectroscopic  $T_{\rm eff}$ of
  NGC~1501  (134~kK)  and  the  gray zone depicts its uncertainties
  ($\pm 10$~kK).}
\label{fig:chi2-NGC1501} 
\end{figure*} 

We calculated the $\ell= 1$ and $\ell= 2$ averages of the computed
period spacings  for the case of NGC~1501. The curves of
$\overline{\Delta \Pi_{k}}$ in terms of $T_{\rm eff}$ for all the
masses considered are shown in the upper ($\ell= 1$) and lower ($\ell=
2$)  panels of Fig. \ref{fig:psp-teff-NGC1501}.  According to the
analysis of Sect.~\ref{period-spacing-ngc1501}, the dipole period
spacing of NGC~1501 is  $\Delta \Pi_{\ell= 1}= 20.1262\pm 0.0123$~s,
and the quadrupole period spacing is $\Delta \Pi_{\ell= 2}=
11.9593\pm0.0520$~s. As in the case of HS~2324 and NGC~6905,  for
NGC~1501 we considered the two possibilities, i.e., that the object is
before or after the ``evolutionary knee''. From the $\ell= 1$ periods,
we find $M_{\star}= 0.621 \pm 0.015 M_{\odot}$ if the star is before
the evolutionary knee,  and $M_{\star}= 0.624 \pm 0.015 M_{\odot}$ if
the star is after the evolutionary knee. On the other hand, from the
$\ell= 2$ periods we derive $M_{\star}= 0.601 \pm 0.015 M_{\odot}$ and
$M_{\star}= 0.606 \pm 0.015 M_{\odot}$ depending on the star being
before or after the evolutionary knee. These mass values of NGC~1501,
which are in the range $\sim 0.60-0.62 M_{\odot}$, are  $\sim 5-10 \%$
larger than --- but still compatible with --- the spectroscopic mass
of this star ($M_{\star}= 0.565 \pm 0.150 \ M_{\sun}$; see
Sect. \ref{models}).

As for the other target stars analyzed in this study, we have
performed a period-to-period  fit for NGC~1501. We employed the
periods of Table~\ref{table:NGC1501-extended},  except the periods
1318.46~s and 1999.16~s which do not fit neither the $\ell= 1$ nor
the $\ell= 2$ series of periods with constant period spacings. We
assumed that there is a subset of 14 periods corresponding  to $\ell=
1$ modes, and 10 periods identified with $\ell= 2$ modes, according to
the period spacings derived in Sect.~\ref{period-spacing-ngc1501} (see
Table~\ref{table:NGC1501-extended}).    We display in
Fig.~\ref{fig:chi2-NGC1501} our results for this analysis. We note the
existence of a clear minimum for the model with  $M_{\star}= 0.609
M_{\odot}$ and $T_{\rm eff}= 140\,051$~K. It is  located at the stages
previous to the maximum temperature for this mass. The value of the
quality function for this solution is $\chi^2= 10.76$~s$^2$.   We
adopt this model  as the asteroseismological model for NGC~1501. In
Table~\ref{table:NGC1501-asteroseismic-model} we show a  detailed
comparison  of the observed periods of NGC~1501 and the the
theoretical  periods  of  the   asteroseismological  model. For this
asteroseismological solution, we have $\overline{\delta \Pi_i}=
2.81$~s and $\sigma= 3.28$~s. The quality of our fit for NGC~1501 is
worst than that achieved for RX~J2117 ($\overline{\delta
  \Pi_i}= 1.26$~s  and $\sigma= 1.73$~s), but comparable to that
obtained for HS~2324 ($\overline{\delta \Pi_i}= 2.53$~s and $\sigma=
3.36$~s).  We also computed the Bayes Information Criterion. In this
case, $N_{\rm p}= 2$ (stellar mass and effective temperature), $n=
24$, and $\sigma= 3.28$ s.  We obtain ${\rm BIC}= 1.15$, which means
that our fit is not so good as for RX~J2117  (${\rm BIC}= 0.59$), but
very similar to that of HS~2324 (${\rm BIC}= 1.18$), and still
satisfactory.   We have redone the analysis by fixing  the harmonic
degree of the 14 periods labeled as $\ell= 1$ at the outset, but
allowing the other 10 periods to be associated with $\ell = 1$
or $\ell = 2$ modes.  Also, we repeated the above analysis by
incorporating the two periods excluded  into the period fit (1318.46~s
and 1999.16~s), allowing them to be associated to  $\ell= 1$ or $\ell=
2$ modes. In both realizations, we obtained the same
asteroseismological solution than before ($M_{\star}= 0.609 M_{\odot}$
and $T_{\rm eff}= 140\,051$ K), and with similar values of the quality
function.  The fact that we arrive at the same seismological solution
for NGC~1501 implies that the period-fit process automatically assigns
the observed periods that we had originally assumed to be associated
with quadrupole modes (following the strong constraint imposed by the
period spacing; see Table \ref{table:NGC1501-extended}) to theoretical
modes with $\ell= 2$. This adds robustness to the identification of
the harmonic degree of modes made in
Sect. \ref{period-spacing-ngc1501}.

The secular rates of period change for NGC~1501, as predicted by our
seismological model, are shown in column 7 of
Table~\ref{table:NGC1501-asteroseismic-model}. All the values are
negative, reflecting the fact that the star would be evolving towards
high temperatures and rapidly contracting, which shortens the periods
of $g$ modes with time.  The rates of period change are in the range
$(10-160) \times 10^{-11}$ s/s,  being similar to those of the
asteroseismological model of HS~2324 (Table
\ref{table:HS2324-asteroseismic-model}), and two orders of magnitude
higher than those of the asteroseismological model of RX~J2117 (Table
\ref{table:RXJ2117-asteroseismic-model}). 

We include in the column 8 of
Table~\ref{table:NGC1501-asteroseismic-model} the information  about
the stability/instability of the modes of the asteroseismological
model. Strikingly,   all quadrupole modes are predicted to be unstable
($\eta_k > 0$), while most dipole modes are predicted  to be stable
($\eta_k < 0$). This is because  the  model sequence of $M_{\star}=
0.609 M_{\odot}$ has unstable dipole (quadrupole)  periods in the
range $1817\ {\rm s}-4228$~s ($1077\ {\rm s}-3041$~s)
\citep{2006A&A...458..259C}.  Therefore, unlike the previous targets,
the stability calculations of the modes with $\ell = 1$ do not favor
the validity of the seismological model for NGC~1501.  However, we
mention that stability properties of the $g$ modes in these stars
strongly depend on the envelope composition
\citep{2007ApJS..171..219Q}, which is affected by the treatment of
convective boundary mixing during the born again episode. Thus, with a
readjustment of the abundances of O, C and He we do not rule out that
an agreement could be found between the observed $\ell= 1$ periods and
the range of unstable $\ell= 1$ periods of our models, thus
alleviating this discrepancy.  However, a fine tuning of the surface
chemical composition of our models to fit the range of observed
periods is beyond the scope of this paper. This kind of  "nonadiabatic
asteroseismology" of GW Vir stars has been attempted by
\cite{2009JPhCS.172a2077Q}. 

\begin{table*}
\centering
\caption{Observed and theoretical periods of the asteroseismological
  model for NGC~1501  [$M_{\star}= 0.609 M_{\odot}$, $T_{\rm eff}=
    140\,051$ K, $\log(L_{\star}/L_{\odot})= 3.853$]. The columns have
  the same meaning as in
  Table~\ref{table:RXJ2117-asteroseismic-model}.}
\begin{tabular}{cc|cccccc}
\hline
\noalign{\smallskip}
$\Pi_i^{\rm O}$ & $\ell^{\rm O}$ & $\Pi_k$ & $\ell$ & $k$ &  $\delta \Pi_k$ & $\dot{\Pi}_k$ & Unstable\\
(s) & & (s) & & &  (s) & ($10^{-11}$ s/s) &  \\
\noalign{\smallskip }
\hline 
\noalign{\smallskip}      
1154.360 & 2 & 1154.379 & 2 &  98 & $-0.019$ & $-15.441$ & yes \\
1168.900 & 1 & 1167.137 & 1 &  57 & 1.713    & $-10.788$ & no  \\
1251.910 & 1 & 1248.916 & 1 &  61 & 2.994    & $-21.429$ & no  \\
1254.632 & 2 & 1257.656 & 2 & 107 & $-3.024$ & $-15.823$ & yes \\
1309.086 & 2 & 1304.233 & 2 & 111 & 4.853    & $-42.157$ & yes \\
1310.696 & 1 & 1307.716 & 1 &  64 & 2.980    & $-26.766$ & no  \\
1321.095 & 2 & 1315.744 & 2 & 112 & 5.351    & $-47.478$ & yes \\
1345.872 & 2 & 1350.158 & 2 & 115 & $-4.286$ & $-22.660$ & yes \\
1349.460 & 1 & 1348.889 & 1 &  66 & 0.571    & $-16.437$ & no  \\
1356.680 & 2 & 1362.511 & 2 & 116 & $-5.831$ & $-18.491$ & yes \\
1366.279 & 1 & 1367.932 & 1 &  67 & $-1.653$ & $-15.860$ & no  \\
1372.940 & 2 & 1375.144 & 2 & 117 & $-2.204$ & $-18.814$ & yes \\
1381.295 & 2 & 1386.419 & 2 & 118 & $-5.124$ & $-22.368$ & yes \\
1392.240 & 1 & 1388.738 & 1 &  68 & 3.502    & $-14.155$ & no  \\
1431.530 & 1 & 1431.050 & 1 &  70 & 0.480    & $-13.742$ & no \\
1512.660 & 1 & 1511.332 & 1 &  74 & 1.328    & $-35.180$ & no  \\   
1760.730 & 2 & 1759.810 & 2 & 150 & 0.920    & $-47.015$ & yes \\
1768.510 & 1 & 1772.096 & 1 &  87 & $-3.586$ & $-25.077$ & no   \\
1777.290 & 2 & 1771.373 & 2 & 151 &  5.917   & $-41.060$ & yes \\
1892.950 & 1 & 1895.295 & 1 &  93 & $-2.345$ & $-55.804$ & yes \\
1951.490 & 1 & 1953.783 & 1 &  96 & $-2.293$ & $-38.265$ & yes \\
2032.640 & 1 & 2035.792 & 1 & 100 & $-3.152$ & $-40.133$ & yes \\
2077.000 & 1 & 2075.030 & 1 & 102 & 1.970    & $-64.074$ & yes \\
5234.810 & 1 & 5236.174 & 1 & 255 & $-1.364$ & $-156.804$ & no  \\
\noalign{\smallskip}
\hline
\end{tabular}
\label{table:NGC1501-asteroseismic-model}
\end{table*}

We show the main characteristics of the asteroseismological model for
NGC~1501 in Table~\ref{table:modelos-sismo-ngc1501}. The stellar mass
of the seismological model ($0.609 M_{\odot}$) is in excellent
agreement with the mass derived from the period spacing  ($\sim
0.60-0.62 M_{\odot}$), assuming that the star is before the
evolutionary knee. This   agreement in the stellar mass is reflected
in the fact that dipole and quadrupole mean period  spacings of our
asteroseismological model, $\overline{\Delta \Pi}_{\ell= 1}= 20.610$~s
and $\overline{\Delta \Pi}_{\ell= 2}= 11.683$~s, are in excellent
agreement with the $\ell= 1$ and $\ell= 2$ mean period spacings
derived for NGC~1501 in Sect.~\ref{period-spacing-ngc1501}, $\Delta
\Pi_{\ell= 1}= 20.1262 \pm 0.0123$~s and $\Delta \Pi_{\ell= 2}=
11.9593 \pm 0.0520$~s, respectively.  Both the stellar mass of the
seismological model and the mass value derived from the period
spacings are somewhat larger than the spectroscopic mass, $0.565 \pm
0.150 \ M_{\sun}$ (Sect. \ref{models}). The luminosity of the
asteroseismological  model, $\log(L_{\star}/L_{\odot})= 3.85$, is
somewhat larger than that derived by \cite{2004MNRAS.354..558E}
[$\log(L_{\star}/L_{\odot})= 3.70$] in order to provide an ionizing
spectrum that could reproduce the ionization structure implied by the
observed nebular spectrum.

We derive the asteroseismological distance for NGC~1501 as we did for
RX~J2117 and  HS~2324.  We employ the luminosity of the
asteroseismological model of NGC~1501, $\log(L_{\star}/L_{\odot})=
3.85\pm0.02$, a bolometric correction of $BC= -7.46$
\citep[extrapolated from the value corresponding to PG~1159$-$035 as
  given by][]{1994ApJ...427..415K},    and the interstellar
extinction law of \cite{1998A&A...336..137C}.   For  the  equatorial
coordinates  of NGC~1501  (Epoch  B2000.00, $\alpha= 4^{\rm h} 6^{\rm
  m} 59.^{\rm s}40,\ \delta= +60^{\circ} 55^{'} 14.^{''}20$) the
corresponding Galactic latitude is $b = 6^{\circ} 33^{'}
4.^{''}06$. If we adopt $m_{\rm V}= 13.0$ \citep[Montreal White Dwarf
Database;][]{2017ASPC..509....3D},
we obtain a seismological distance, a parallax, and an extinction
coefficient of $d= 824\pm15$ pc, $\pi= 1.22_{-0.03}^{+0.02}$ mas, and
$A_{\rm V}= 0.85$, respectively. On the other hand, the values from
{\it Gaia} are $d_{\rm G}= 1763 \pm  79$ pc and $\pi_{\rm G}= 0.567
\pm 0.025$ mas. The   seismological distance is $\sim 50 \%$ shorter
than that measured by {\it Gaia}.  The  origin of this discrepancy may
reside in the uncertainties related to our seismological model. Again,
as in the case of HS~2314, a better estimate of the uncertainties of
the luminosity of the asteroseismological model probably could
contribute to bringing the asteroseismological distance and parallax
values closer to those derived by {\it Gaia}.

We bring this section to a close by summarizing our results for
NGC~1501.  The  stellar mass  inferred from the  period separations
($\sim 0.60-62 M_{\odot}$) is in line  with the mass of the
seismological model ($\sim 0.61 M_{\odot}$). These values are
somewhat larger than ---but still in agreement with--- the
spectroscopic mass, of $\sim 0.57 M_{\odot}$. 

\begin{table}
\centering
\caption{The main characteristics of the GW Vir star NGC~1501. The second column  
  corresponds to spectroscopic results, whereas the third  column present results
  from the asteroseismological model of this work.} 
\begin{tabular}{l|cc}
\hline
\hline
Quantity & Spectroscopy &  Asteroseismology \\
         & Astrometry   &   (This work)      \\ 
\hline
$T_{\rm eff}$ [kK]                            & $134 \pm 10^{\rm (a)}$         & $140\pm6.5$ \\
$M_{\star}$ [$M_{\odot}$]                     & $0.565 \pm 0.150$              & $0.609_{-0.020}^{+0.055}$ \\ 
$\log g$ [cm/s$^2$]                           & $6.0\pm0.2^{\rm (a)}$         & $5.90_{-0.11}^{+0.14}$  \\ 
$\log (L_{\star}/L_{\odot})$                  & $3.70^{\rm (b)}$                         & $3.85\pm0.02$ \\  
$\log(R_{\star}/R_{\odot})$                   & $\hdots$                       & $-0.84\pm0.05$ \\  
$M_{\rm env}$ [$M_{\odot}$]                   & $\hdots$                       & $0.01$ \\  
$(X_{\rm H},X_{\rm He},X_{\rm C})_{\rm s}$    & 0.50, 0.35, $0.15^{\rm (a)}$   & 0.50, 0.35,  0.10 \\
$d$  [pc]                                     & $1762.801 \pm 78.566^{\rm (c)}$  & $824 \pm 15$  \\ 
$\pi$ [mas]                                   &  $0.567\pm0.025^{\rm (c)}$ & $1.22_{-0.03}^{+0.02}$\\ 
\hline
\hline
\end{tabular}
\label{table:modelos-sismo-ngc1501}

{\footnotesize  References: (a) \cite{1997IAUS..180..114K}, \cite{2006PASP..118..183W}; (b) \cite{2004MNRAS.354..558E};  (c) {\it Gaia}.}
\end{table}

\subsection{NGC~2371}
\label{modelling-ngc2371}

\begin{figure} 
\includegraphics[clip,width=1.0\columnwidth]{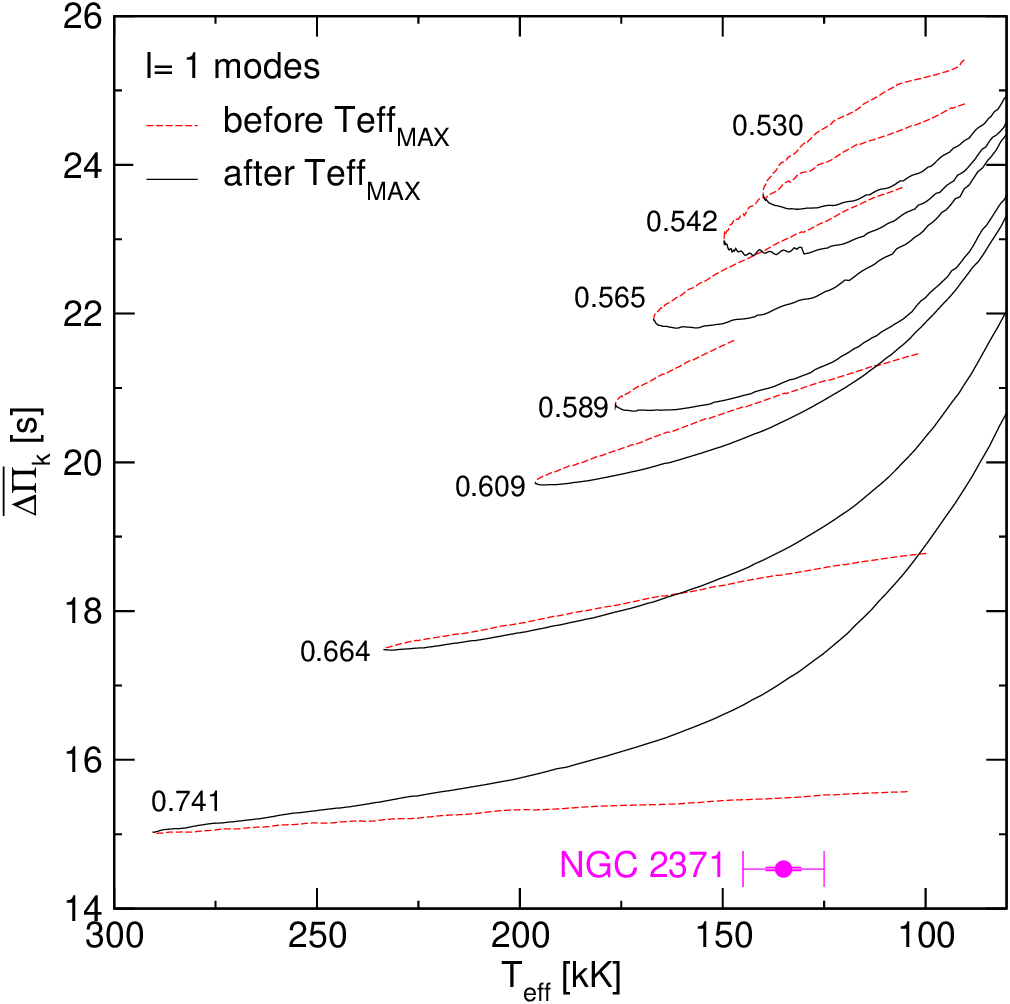}
\caption{Dipole ($\ell= 1$) average of the computed period spacings,
  $\overline{\Delta \Pi_k}$, assessed  in  a  range  of  periods  that
  includes  the  periods  observed  in NGC~2371, shown as solid (dashed) curves
  corresponding to stages before (after) the maximum $T_{\rm eff}$
  for different stellar masses. The location of NGC~2371 when we  use the
  effective temperature derived by \cite{1996AJ....111.2332C}, 
  $T_{\rm eff}= 135\,000 \pm 10\,000$ K, and the period
  spacing $\Delta \Pi= 14.5312\pm0.0226$ s derived in Sect. \ref{obser-ngc2371}, 
  is highlighted with a magenta circle.  The stellar mass derived from linear 
  extrapolation is $M_{\star}\sim 0.760 M_{\odot}$
  if the star is before the maximum $T_{\rm eff}$, and it 
  is $M_{\star} \sim 0.856 M_{\odot}$
  if the star is after the maximum $T_{\rm eff}$.}
\label{fig:psp-teff-NGC2371} 
\end{figure} 

In Fig.~\ref{fig:psp-teff-NGC2371} we display the curves of
$\overline{\Delta \Pi_{k}}$ as function of $T_{\rm eff}$ for all the
masses considered. The period spacing of NGC~2371 is  $\Delta \Pi=
14.5312\pm0.0226$, as derived in the analysis of
Sect. \ref{period-spacing-ngc2371}. As in the case of HS~2324,
NGC~6905,  and  NGC~1501, for NGC~2371 we considered the possibility
that the object is before or after the "evolutionary knee". We find
$M_{\star}= 0.760M_{\odot}$  if the star is before the evolutionary
knee,  and $M_{\star}= 0.856 M_{\odot}$ if the  star is after the
evolutionary knee. As in the case of NGC~6905, these values are
assessed by extrapolation, because the period spacing of the star,
being  only  about  $14.5$~s,  is  well  below  the average  of  the
$\ell= 1$  computed period-spacing curves of our models, even the most
massive one.  Therefore,  our  stellar  mass  values  are  just
estimates. These estimates for the stellar mass of NGC~2371 are much
larger ($\sim 60 \%$) than the spectroscopic mass of this star, of
$M_{\star}= 0.533 \pm 0.150 \ M_{\sun}$ (Sect.~\ref{models}).

\begin{figure*} 
\includegraphics[clip,width= 2.0\columnwidth]{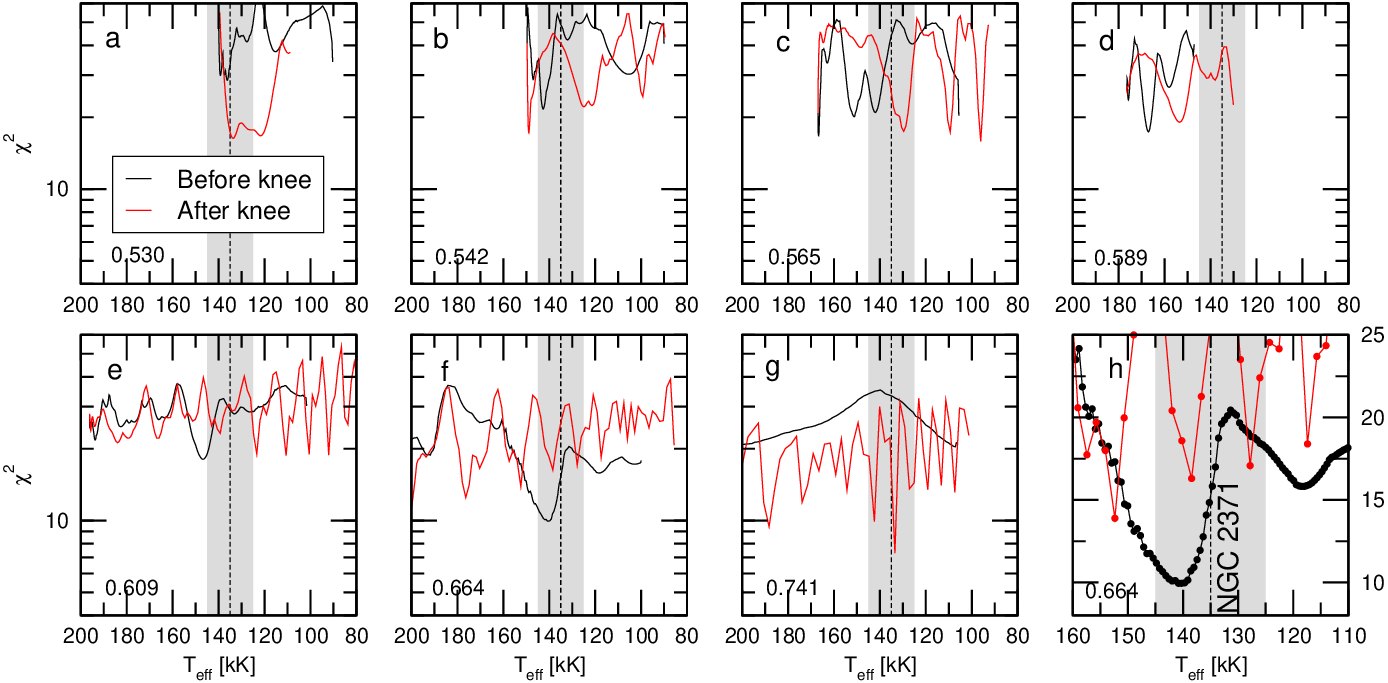}
\caption{Same as Fig. \ref{fig:chi2-RXJ2117}, but for NGC~2371. Note
  the minimum  in  panel {\bf f} before the evolutionary knee (black
  curve),   corresponding  to $M_{\star}= 0.664 M_{\odot}$.  Panel
  {\bf h} is a zoom of  the region with the  strong minimum seen in
  panel {\bf f}; the $y$-axis scale is linear in this case. The
  vertical dashed line is the spectroscopic  $T_{\rm eff}$ of
  NGC~2371  (135~kK)  and  the  gray zone depicts its uncertainties
  ($\pm 10$~kK).}
\label{fig:chi2-NGC2371} 
\end{figure*} 

\begin{table*}
\centering
\caption{Observed and theoretical periods of the asteroseismological
  model for NGC~2371  [$M_{\star}= 0.664 M_{\odot}$, $T_{\rm eff}=
    140\,265$ K, $\log(L_{\star}/L_{\odot})= 4.066$]. The columns have
  the same meaning as in
  Table~\ref{table:RXJ2117-asteroseismic-model}.}
\begin{tabular}{cc|cccccc}
\hline
\noalign{\smallskip}
$\Pi_i^{\rm O}$ & $\ell^{\rm O}$ & $\Pi_k$ & $\ell$ & $k$ &  $\delta \Pi_k$ & $\dot{\Pi}_k$ & Unstable\\
(s) & & (s) & & &  (s) & ($10^{-11}$ s/s) &  \\
\noalign{\smallskip }
\hline
\noalign{\smallskip}      
 878.523 & 1 & 875.320 & 1 & 49   & 3.203    & 1.981     & no \\
 895.111 & 1 & 893.367 & 1 & 50   & 1.744    & 6.166     & no \\ 
 925.465 & 1 & 929.693 & 1 & 52   & $-4.228$ & $-5.008$  & no \\ 
 968.474 & 1 & 964.815 & 1 & 54   & 3.659    & $-47.723$ & no \\ 
 982.800 & 1 & 980.871 & 1 & 55   & 1.929    & $-52.214$ & no \\ 
 988.200 & ? & 986.704 & 2 & 96   & 1.496    & $-1.264$  & no \\ 
 998.000 & 1 & 997.767 & 1 & 56   & 0.233    & $-52.153$ & no \\ 
1005.600 & ? & 1007.777 & 2 & 98  & $-2.177$ & $-10.730$ & no \\ 
1010.000 & 1 & 1015.513 & 1 & 57  & $-5.513$ & $-19.545$ & no \\ 
1825.000 & 1 & 1828.713 & 1 & 103 & $-3.713$ & $-37.306$ & no \\ 
\noalign{\smallskip}
\hline
\end{tabular}
\label{table:NGC2371-asteroseismic-model}
\end{table*}

\begin{table}
\centering
\caption{The main characteristics of the GW Vir star NGC~2371. The second
  column  corresponds to spectroscopic results, whereas the third  column
  present results from the asteroseismological model of this work.} 
\begin{tabular}{l|cc}
\hline
\hline
Quantity & Spectroscopy &  Asteroseismology \\
         & Astrometry   &   (This work)      \\ 
\hline
$T_{\rm eff}$ [kK]                            & $135 \pm 10^{\rm (a)}$         & $140.3_{-5.3}^{+10.6}$\\
$M_{\star}$ [$M_{\odot}$]                     & $0.533 \pm 0.150$              & $0.664_{-0.055}^{+0.077}$\\  
$\log g$ [cm/s$^2$]                           & $6.3\pm0.2^{\rm (a)}$         & $5.74_{-0.12}^{+0.19}$\\ 
$\log (L_{\star}/L_{\odot})$                  & $3.45^{\rm (b)}$                & $4.07_{-0.02}^{+0.01}$\\  
$\log(R_{\star}/R_{\odot})$                   & $-0.98^{\rm (b)}$              & $-0.74_{-0.07}^{+0.04}$\\  
$M_{\rm env}$ [$M_{\odot}$]                   & $\hdots$                       & $0.02$ \\  
$(X_{\rm He},X_{\rm C},X_{\rm O})_{\rm s}$    & 0.54, 0.37, $0.08^{\rm (a)}$  & 0.47, 0.33,  0.13\\   
$d$  [pc]                                     & $1879.138\pm 232.467^{\rm (c)}$ & $1816^{+11}_{-51}$\\ 
$\pi$ [mas]                                   & $0.532\pm0.066^{\rm (c)}$ & $0.554_{-0.007}^{+0.013}$\\ 
\hline
\hline
\end{tabular}
\label{table:modelos-sismo-ngc2371}

{\footnotesize  References: (a) \cite{2004ApJ...609..378H};  (b) \cite{2020MNRAS.496..959G}; (c) {\it Gaia}.}
\end{table}

Below, we describe our period-to-period fit analysis for
NGC~2371. Again,  the merit function  $\chi^2(M_{\star}, T_{\rm eff})$
(Eq.~\ref{eq:3}) was evaluated  for  all the stellar  masses and
effective temperatures covered by our PG~1159 model  sequences. We
employed the 10 periods of Table~\ref{table:NGC2371-extended}.  We
adopted the same approach as for the case of the previous target
stars,  that is, we assumed that there is a subset of 8 periods
identified as $\ell= 1$ according  to the period spacing derived in
Sect.~\ref{period-spacing-ngc2371} (see
Table~\ref{table:NGC2371-extended}), but the remaining 2 periods  are
allowed to be identified with modes with $\ell= 1$ or $\ell= 2$
modes. 

As in the previous cases, given the absence of a single global minimum
in the quality function, we focused on the range of effective
temperatures  compatible with NGC~2371, that is $125\,000 \lesssim
T_{\rm eff} \lesssim 135\, 000$~K.  We display in
Fig.~\ref{fig:chi2-NGC2371} our results. We note the existence of a
clear  minimum for a model with $M_{\star}= 0.664 M_{\odot}$ and
$T_{\rm eff}= 140\,265$~K. It is  located at the stages previous to
the maximum temperature for this mass.   We adopt this model as the
asteroseismological model for NGC~2371.  In
Table~\ref{table:NGC2371-asteroseismic-model} we show a  detailed
comparison  of the observed periods of NGC~2371 and the theoretical
periods of the   asteroseismological model. For this
asteroseismological solution, we  have $\overline{\delta \Pi_i}= 2.79$
s and $\sigma= 3.15$~s. The quality of our fit for NGC~2371 is
 poorer than that achieved for RX~J2117 ($\overline{\delta
  \Pi_i}= 1.26$~s  and $\sigma= 1.73$~s), although comparable to those
of HS~2324 ($\overline{\delta \Pi_i}= 2.53$~s  and $\sigma= 3.36$~s)
and NGC~1501 ($\overline{\delta \Pi_i}= 2.81$~s   and $\sigma=
3.28$~s). We also computed the Bayes Information Criterion. In this
case, $N_{\rm   p}= 2$ (stellar mass and effective temperature), $n=
10$, and $\sigma= 3.15$\,s.  We obtain ${\rm BIC}= 1.20$, which means
that our fit is not so good as for RX~J2117 (${\rm BIC}= 0.59$),
although comparable with those of HS~2324 (${\rm BIC}= 1.18$) and
NGC~1501 (${\rm BIC}= 1.15$).

Most of the rates of period change for NGC~2371 (column 7 of
Table~\ref{table:NGC2371-asteroseismic-model}) are negative, in
concordance with the  fact that the asteroseismological model for this
star is evolving towards higher  effective temperatures and
contracting. The magnitudes of $\dot{\Pi}$ are comparable,  in
average, to those of HS~2324 and NGC~1501. Regarding the pulsational
stability/instability nature of the modes with periods that fit the
periods observed in NGC~2371 (column 9 of Table
\ref{table:NGC2371-asteroseismic-model}), the  nonadiabatic
calculations predict that all the periods (dipole and quadrupole) are
stable,  because at $T_{\rm eff}= 138\,600$~K, the model sequence of
$M_{\star}= 0.664 M_{\odot}$  has unstable $\ell= 1$ ($\ell= 2$)
periods in the range $1962\ {\rm s}-4480$~s ($1190\ {\rm s}-3045$~s)
\citep{2006A&A...458..259C}. Then, our nonadiabatic pulsation results
are in conflict with the observational evidence, similar to the case of
NGC~1501.  Again, we mention that the intervals of unstable periods
are highly dependent on the chemical composition of the outer part of
the PG~1159 stars, which is affected by uncertainties in the treatment
of convective boundary mixing during the born again episode. 

In Table~\ref{table:modelos-sismo-ngc2371}, we list the main
characteristics of the asteroseismological  model for NGC~2371. The
luminosity and radius of the asteroseismological model  ($\log
(L_{\star}/L_{\odot})= 4.07$,  $\log(R_{\star}/R_{\odot})= -0.74$) are
substantially larger than  the values derived by
\cite{2020MNRAS.496..959G} ($\log (L_{\star}/L_{\odot})= 3.45$,
$\log(R_{\star}/R_{\odot})= -0.98$) from stellar atmosphere
models. The stellar mass of the seismological model  ($0.664
M_{\odot}$) is $\sim 14 \%$ lower than the mass inferred from the
period spacing ($\sim 0.76 M_{\odot}$), assuming that the star is
before the evolutionary knee. This mass discrepancy  
is reflected in  the fact that dipole ($\ell= 1$) mean period
spacing of our  asteroseismological model ($\overline{\Delta \Pi}=
18.45$~s) is markedly longer  than  the $\ell= 1$ mean period spacing
for NGC~2371  derived in Sect. \ref{period-spacing-ngc2371} ($\Delta
\Pi= 14.5312 \pm 0.0226$ s).  Both the stellar mass of the
seismological model and the value derived from the period spacing are
in  disagreement with the spectroscopic mass, $0.533 \pm 0.150
\ M_{\sun}$ (Sect. \ref{models}).  This discrepancy between the
spectroscopic mass (based on measured $T_{\rm eff}$ and $\log g$
values) and seismological masses (based on period spacing and
individual periods compared to theoretical expectations) is similar to
that found for HS~2324. 

We employ the luminosity of the asteroseismological model of NGC~2371,
$\log(L_{\star}/L_{\odot})= 4.066^{+0.01}_{-0.02}$, a bolometric
correction of $BC= -7.24$ (the same value used for the  assessment of
the distance to HS~2314) and the interstellar  extinction law of
\cite{1998A&A...336..137C}, to estimate the asteroseismological
distance. For  the  equatorial  coordinates  of NGC~2371  (Epoch
B2000.00, $\alpha= 7^{\rm h} 25^{\rm m} 34.^{\rm s}69,\ \delta=
+29^{\circ} 29^{'} 26.^{''}32$) the corresponding Galactic latitude is
$b = 19^{\circ} 50^{'} 35.^{''}52$. By adopting an  apparent magnitude
of $m_{\rm V}= 13.50$  \citep[Montreal White Dwarf
Database;][]{2017ASPC..509....3D}, we derive $d=
1816_{-51}^{+11}$ pc, $\pi= 0.554_{-0.007}^{+0.013}$ mas, and $A_{\rm
  V}= 0.402$. On the other hand, the values from {\it Gaia} are
$d_{\rm G}= 1879 \pm 232$ pc and $\pi_{\rm G}= 0.532 \pm 0.066$
mas. There is a good agreement between the seismological distance and
parallax and their counterparts measured by {\it Gaia}, given the
uncertainties in both sets of  magnitudes. 

We end this section with a summary of our results for NGC~2371.  The
stellar mass  inferred from the  period separation is larger than the
mass of the seismological model, but they are still in
agreement. However, these values are at variance with the
spectroscopic mass, which is markedly lower. The seismological
distance, on the other hand, is in  good concordance with that
measured by  {\it Gaia}. This suggests that the parameters of the
seismological model are correct, and that the discrepancy in mass
between the seismological inference and the spectroscopic derivation
can come from very uncertain values in the spectroscopic  $T_{\rm
  eff}$ and $\log g$.  

\section{Summary and conclusions}
\label{conclusions}

In this work, we have performed a detailed asteroseismological
analysis of the formerly known GW Vir stars RX~J2117, HS~2324,
NGC~6905,  NGC~1501, NGC~2371, and K~1$-$16 on the basis of the new
{\it TESS} observations  of these stars.  {\it TESS} primary mission
is to search for exoplanets around  bright targets, although the
high-precision photometry capabilities of the mission  are also
allowing us to study low-amplitude stellar variability, including  WD
pulsations \citep[][]{2019A&A...632A..42B,2020A&A...638A..82B}. We studied the
pulsation spectrum of these stars,  and, in one case (RX~J2117), we estimated
the rotation period  on the basis of  frequency-splitting
multiplets. We also determined their  stellar mass and other
structural and  evolutionary properties on the basis of the detailed
PG~1159 evolutionary models of \cite{2005A&A...435..631A} and
\cite{2006A&A...454..845M}. We  have considered adiabatic and
nonadiabatic $g$-mode pulsation  periods  on  PG~1159 evolutionary
models with stellar masses ranging from 0.530 to
$0.741\ M_{\odot}$. These models take into account the complete
evolution of progenitor stars through the thermally pulsing AGB phase
and born-again  episode. 

We estimated a mean period spacing for five of the six studied stars.
To do this, we considered an augmented period spectrum  for  each
star, combining the periods detected in previous works with
ground-based observations  with the more precise periods detected with
{\it TESS}. In this way, we expanded,  when possible, the number of
observed periods, allowing us to arrive at a regime where
asteroseismological methods start to be robust.  The benefit of the
derivation of an underlying period spacing in the observed period
spectrum of GW Vir stars is twofold. On the one hand, it allows to put
strong  constraints on the stellar mass. On the other hand,  it allows
assigning the harmonic degree $\ell$ to a large number of observed
periods,  which simplifies and constrains the subsequent process of
fitting the individual periods  and the derivation of
asteroseismological models. For five of the objects studied, we
constrained the stellar mass by comparing the observed period spacing
with  the average of the computed period spacings. We considered the
cases  where the star is before or after the maximum possible
effective temperature,  although the $\log g$ and $T_{\rm eff}$
spectroscopic values --- and  their comparison with evolutionary
tracks of PG~1159  stars --- strongly suggests that  all the target
stars are evolving before the evolutionary knee in the $T_{\rm eff} -
\log g$  diagram (Fig. \ref{fig:1}). When possible, we also employed
the individual observed periods  to search for a representative
seismological model for each star.  Finally, we derived  seismological
distances whenever this was possible, and compared these estimates
with the  robust astrometric distances measured by {\it Gaia}. 

We  present the main results for each analyzed star below:

\begin{itemize}

\item RX~J2117: The number of secure periods found with the {\it TESS}
  observations  for this star is 15 (see
  Table~\ref{table:RXJ2117}). From two complete rotational triplets
  and three triplets with one lacking component in the {\it TESS}
  frequency spectrum of this  star, we derived a rotation period of
  $P_{\rm rot} \sim 1.04$ days, in agreement with previous works
  \citep{2002A&A...381..122V,2007A&A...461.1095C}. On the basis of a
  subset of 20 periods of  the extended period spectrum (ground-based
  plus {\it TESS} observations;  Table~\ref{table:RXJ2117-extended}),
  we inferred  a $\ell= 1$ constant period spacing of $\Delta \Pi=
  21.669\pm 0.030$~s. This allowed a safe identification of 27 periods
  as dipole modes, with the remainder 4 periods without a secure
  assignation of the value of $\ell$. The comparison of this period
  spacing with the average  of the theoretical period spacings allowed
  us to derive a stellar mass of  $M_{\star}= 0.569\pm0.015
  M_{\odot}$, in perfect agreement with the estimates from
  \cite{2002A&A...381..122V} ($M_{\star}= 0.56 M_{\odot}$),   and
  \cite{2007A&A...461.1095C} ($M_{\star}=0.560 M_{\odot}$)  also using
  the period spacing to derive the stellar mass. We next derived an
  asteroseismological model for this star by means of a
  period-to-period fit  procedure. The magnitude of the rates of
  period change for the asteroseismological model ranges from $\sim
  2.5 \times 10^{-11}$ to $\sim 14.1 \times 10^{-11}$~s/s and 
  all the values are
  negative,  in concordance with the fact that the model is fast
  evolving towards high effective temperatures ($\dot{T_{\rm eff}}
  >0$) and contracting ($\dot{R_{\star}}<0$), a  little before
  reaching the maximum possible temperature. Most of the $g$ modes of
  the asteroseismological  model are predicted to be linearly
  unstable, in concordance with the observations. Two of the periods
  that could not be identified with $\ell= 1$ modes using the constant
  period spacing are identified with $\ell= 2$ modes and the remaining
  two periods with $\ell= 1$ modes,  according to the
  asteroseismological model (Table
  \ref{table:RXJ2117-asteroseismic-model}). The stellar mass of the
  asteroseismological model ($M_{\star}= 0.565\pm0.024 M_{\odot}$) is
  in excellent agreement with the  mass derived from the period
  spacing. Finally,  we derived a seismological distance ($d= 480$~pc)
  in complete consensus with the {\it Gaia} distance ($d_{\rm G}=
  502$~pc), giving robustness to the  asteroseismological model. On
  the other hand, the asteroseismological masses apparently  disagree
  with the spectroscopic mass ($M_{\star}= 0.716 M_{\odot}$), but,
  given the large uncertainties of  this last estimate, we conclude
  that the asteroseismological and spectroscopic determinations  agree with
  each other.\\

\item HS~2324: We detected 12 periods from the {\it TESS} data for
  this star  (Table~\ref{table:HS2324}). At variance with
  \cite{1999A&A...342..745S}, we found no  rotational multiplets with
  a much better frequency resolution of $0.35~\mu$Hz.  This prevented
  us from deriving a rotational period. Using a subset of 9 periods of
  the augmented period spectrum (ground based and {\it TESS} data;
  see Table~\ref{table:HS2324-extended}), we inferred the existence of
  a $\ell= 1$ constant period spacing of $\Delta \Pi= 16.407\pm
  0.062$~s,  which enabled a robust identification of 13 periods as
  $\ell= 1$ modes.   The periods from ground-based observations we
  used in this paper are slightly  different from those in Table~2 of
  \cite{1999A&A...342..745S}. By comparing the derived  dipole period
  spacing with the average of the theoretical period spacings, we
  estimated a stellar mass of  $M_{\star}= 0.727\pm0.017 M_{\odot}$ if
  the star is before the evolutionary knee.  We obtained an
  asteroseismological model for HS~2324 by comparing the individual
  observed and  theoretical periods. Most of the rates of period
  change for the  asteroseismological model are negative, and their
  magnitudes are in the interval  $(\sim 8.5 - 138.0) \times
  10^{-11}$~s/s, which implies that this star must be  evolving much
  faster than RX~J2117. Our stability analysis predicts most of the
  modes of the asteroseismological model to be unstable, in line with
  observations. Most of the observed periods that could not be
  identified with  $\ell= 1$ modes on the basis of the constant period
  spacing are identified with modes  $\ell= 2$ according to the
  asteroseismological model
  (Table~\ref{table:HS2324-asteroseismic-model}).  The stellar mass of
  the asteroseismological model, $M_{\star}= 0.664_{-0.055}^{+0.077}
  M_{\odot}$,  is $\sim 10\%$ lower than the  mass estimated from the
  period spacing, but still compatible with  it. Finally, the
  seismological distance ($d= 4379\pm1000$~pc) is 3 times larger than
  the {\it Gaia} distance ($d_{\rm G}= 1448$~pc), but if we could take
  into account  realistic uncertainties that affect seismological
  distance, probably the disagreement would  not be so great.  The
  asteroseismological masses, on the other hand,   are higher than the
  spectroscopic mass ($M_{\star}= 0.532 M_{\odot}$). However, because
  the large  uncertainties in the spectroscopic mass, both
  spectroscopic and asteroseismological  determinations are compatible
  each other. Additional future observations of HS~2324 will be of great 
  importance in detecting new periods and investigating possible signals 
  of stellar rotation through frequency multiples.
\\

\item NGC~6905: {\it TESS} photometry indicates the presence of only
  4 periods in the pulsation spectrum of this star, as shown in
  Table~\ref{table:NGC6905}. We employed a subset of 7 periods of the 
  enlarged list of
  periods   (see Table~\ref{table:NGC6905-extended}), to uncover the
  existence of a constant period spacing  of $\Delta \Pi= 11.9693\pm
  0.0988$ s, which we can attribute, in principle, to $\ell= 1$ modes.
  This period spacing enabled us to do a secure identification of 7
  periods as dipole modes.  By comparing the inferred $\ell= 1$ period
  spacing  with the average of the theoretical period spacings we
  obtained a stellar mass estimate of  $M_{\star}\sim 0.818 M_{\odot}$
  if the star is before the maximum $T_{\rm eff}$.  This estimate of
  the stellar mass is in strong disagreement with the   spectroscopic mass of
  this star, of $M_{\star}= 0.590 M_{\odot}$.  We also explored the
  possible   situation in which the period spacing
  of $\Delta \Pi= 11.9693\pm0.0988$~s correspond to $\ell= 2$
  modes. In this case, the mass of the star should be $M_{\star}=
  0.596 \pm 0.009 M_{\odot}$ and  $M_{\star}= 0.590 \pm 0.007
  M_{\odot}$ if the star is before or after the evolutionary  knee,
  respectively, in very good agreement with the spectroscopic mass
  value.  Both in the cases where the period spacing corresponds to
  $\ell= 1$ or corresponds to $\ell= 2$ instead, the  results  of
  our period-to-period fit
  procedure did not indicate a clear and unique asteroseismological
  solution. Thus, for NGC~6905, we are unable to find an
  asteroseismological model, which deprives us of the possibility of
  estimating the seismic distance for this star.
\\

\item NGC~1501: We measured 16 periods for this star on the basis of
  {\it TESS}  observations (Table \ref{table:NGC1501}).  
  We considered a subset of 23
  periods of  the enlarged list of periods (i.e., from ground-based
  plus {\it TESS} observations; see Table
  \ref{table:NGC1501-extended}), and we found a $\ell= 1$ constant
  period spacing of  $\Delta \Pi= 20.1262\pm 0.0123$ s and a $\ell= 2$
  constant period spacing of  $\Delta \Pi= 11.9593\pm 0.0520$ s. This
  enabled us to identify 14 periods as $\ell= 1$ modes and 10 periods
  as $\ell= 2$ modes, with the remainder 2 periods without a secure
  assignation of the $\ell$ value. The comparison of these observed
  period spacings  with the average of the theoretical period spacings
  for $\ell= 1$ and $\ell= 2$ allowed  us to infer a stellar mass of
  $M_{\star}= 0.621\pm0.015 M_{\odot}$ and  $M_{\star}= 0.601\pm0.015
  M_{\odot}$, respectively, if the star is before  the maximum $T_{\rm
    eff}$. These values are in close agreement each other.  We derived
  an asteroseismological model for this star on the basis of our usual
  period-to-period fit procedure (Table
  \ref{table:RXJ2117-asteroseismic-model}).  The magnitude of the
  rates of period change for the asteroseismological model ranges from
  $\sim 10 \times 10^{-11}$ to $\sim 160 \times 10^{-11}$ s/s and all the values are
  negative,  as expected on the grounds that the model is is heating
  up and quickly contracting.  Noticeably, all the $\ell= 2$ modes are
  predicted to be unstable by our nonadiabatic calculations,   while  most  $\ell= 1$  modes  are
  predicted  to  be stable.   The stellar mass of the
  asteroseismological model  ($M_{\star}= 0.609_{-0.02}^{+0.06}
  M_{\odot}$) is in perfect concordance with the mass  inferred from
  the period separations, and in agreement with the spectroscopic
  mass, given  its uncertainty.  Finally, we derived a seismological
  distance ($d= 824$~pc)  which is $\sim 50\%$ shorter than the {\it
    Gaia} distance ($d_{\rm G}= 1763$ pc).  The  agreement  between
  these  sets  of  values  could  improve  if  we  could employ   more
  realistic  values  for  the  uncertainties  in the luminosity of
  the  asteroseismological  model.\\

\item NGC~2371: We detected 5 periods from the  {\it TESS} data for
  this star  (Table \ref{table:NGC2371}). Considering a subset of 8
  periods of the enlarged period set  (i.e., those from ground-based
  observations along with {\it TESS} data;  see Table
  \ref{table:NGC2371-extended}), we found a $\ell= 1$ constant  period
  spacing of $\Delta \Pi= 14.5312\pm 0.0226$ s, allowing us to safely
  identify  8 periods with $\ell= 1$ modes, leaving unidentified the
  remaining 2 periods.   The comparison of the derived dipole period
  spacing with the average of the model period spacings results in an
  estimate of the  stellar mass of $M_{\star}\sim 0.760 M_{\odot}$ if
  the star is before the evolutionary knee.  We obtained an
  asteroseismological model for NGC~2371 from a period-to-period fit,
  with most of the rates of period change being negative, and with
  magnitudes  in the range  $(\sim 2 - 52) \times 10^{-11}$
  s/s. Our stability analysis predicts {\it all} of the modes of the
  asteroseismological model to be stable, in conflict with
  observations. The two observed periods that could not be identified
  with  $\ell= 1$ modes on the basis of the constant period spacing
  are identified with $\ell= 2$ modes  according to the
  asteroseismological model (Table
  \ref{table:NGC2371-asteroseismic-model}).  The stellar mass of the
  asteroseismological model, $M_{\star}= 0.664_{-0.055}^{+0.077}
  M_{\odot}$,  is $\sim14 \%$ smaller than the mass inferred from the
  period spacing, assuming that the star  is before the evolutionary
  knee. The asteroseismological distance is  $d= 1816$ pc, in very
  good agreement with the distance measured with {\it Gaia}, of
  $d_{\rm G}= 1880\pm232$ pc. This agreement reinforces the validity
  of the asteroseismological model.  Both the stellar mass of the
  seismological model and the  mass value  derived  from  the  period
  spacing  are  in disagreement with the spectroscopic mass, of
  $M_{\star}= 0.533 \pm 0.150 M_{\odot}$.  This mass discrepancy is
  probably due to the large uncertainty in the spectroscopic $\log g$
  value.  We conclude that, in a similar way than for RX~J2117 and
  HS~2324, these spectroscopic  parameters need to be redetermined for
  NGC~2371.\\
 
\item K~1$-$16: For this star, we have been able to detect only 5 (or
  6) periods (Table \ref{table:K1-16}).  The amplitudes and
  frequencies change significantly in such a way that if we look
  individually  the sectors of {\it TESS}, some modes have zero
  amplitudes, but they have finite amplitudes in  other sectors. Due
  to the complexity and changing nature of the period spectrum of this
  star,  along with the high noise level of its power spectrum, we
  were forced to crudely  estimating the period values, and then added
  a single period detected from the ground.  Considering a composite
  list of periods, we only have 6 periods available, for which we
  found no evidence of any pattern of constant period spacing. Due to
  all these reasons, our  asteroseismological analysis was limited,
  but it revealed the dramatic changes that the pulsations experience
  in this star, as depicted in Fig. \ref{fig:sft1}.

\end{itemize}

The results of this paper demonstrate that the high-quality
observations of {\it TESS}, considered in conjunction with
ground-based observations  (which are usually more uncertain), are
able to provide a  very important input to the asteroseismology of GW
Vir stars, in line with  recent reports for other classes of pulsating
WDs, such as the case of a DBV star
\citep{2019A&A...632A..42B}, pre-ELMV stars
\citep{2020ApJ...888...49W}, warm DA WDs \citep{2020A&A...633A..20A},
and DAV stars \citep{2020A&A...638A..82B}.

The {\it TESS} space mission is demonstrating that it can greatly
contribute to the asteroseismology of WDs and pre-WDs, becoming a worthy successor
of the {\it Kepler} mission, which has  had an excellent performance
at studying pulsating WDs \citep{2020FrASS...7...47C}.  The {\it TESS}
mission, in conjunction with other ongoing space missions such as 
{\it Cheops} \citep{2018A&A...620A.203M} and future space missions like  
{\it PLATO} \citep{2018EPSC...12..969P}, along with new surveys and telescopes
such as the Legacy Survey for Space and Time
\citep{2009arXiv0912.0201L}, which will be operational in the coming
years \citep{2020ApJ...900..139F}, will  probably make stunning
progress in WD asteroseismolgy.

\begin{acknowledgements}
 We  wish  to  acknowledge  the  suggestions  and
 comments of an anonymous referee that strongly improved the original
 version of this work.  This  paper includes data collected with the TESS mission, obtained
 from the MAST data archive at the Space Telescope Science Institute
 (STScI). Funding for the TESS mission is provided by the NASA Explorer
 Program. STScI is operated by the Association of Universities for
 Research in Astronomy, Inc., under NASA con-tract NAS 5–26555.  Part
 of this work was supported by AGENCIA through the Programa de
 Modernizaci\'on Tecnol\'ogica BID 1728/OC-AR, and by the PIP
 112-200801-00940 grant from CONICET. M.U. acknowledges financial
 support from CONICYT Doctorado Nacional in the form of grant number
 No: 21190886. SOK is supported by CNPq-Brazil, CAPES-Brazil and
 FAPERGS-Brazil. KJB is supported by the National Science Foundation 
 under Award No.\ AST-1903828. This research has made use of NASA's Astrophysics
 Data System.
 Financial support from the National Science Centre under projects No.\,UMO-2017/26/E/ST9/00703 and UMO-2017/25/B/ST9/02218 is appreciated.
\end{acknowledgements}

\bibliographystyle{aa}
\bibliography{paper_bibliografia.bib}

\begin{thebibliography}{101}
\expandafter\ifx\csname natexlab\endcsname\relax\def\natexlab#1{#1}\fi

\bibitem[{{Althaus} {et~al.}(2010){Althaus}, {C{\'o}rsico}, {Isern}, \&
  {Garc{\'{\i}}a-Berro}}]{2010A&ARv..18..471A}
{Althaus}, L.~G., {C{\'o}rsico}, A.~H., {Isern}, J., \& {Garc{\'{\i}}a-Berro},
  E. 2010, \aapr, 18, 471

\bibitem[{{Althaus} {et~al.}(2008){Althaus}, {C{\'o}rsico}, {Kepler}, \&
  {Miller Bertolami}}]{2008A&A...478..175A}
{Althaus}, L.~G., {C{\'o}rsico}, A.~H., {Kepler}, S.~O., \& {Miller Bertolami},
  M.~M. 2008, \aap, 478, 175

\bibitem[{{Althaus} {et~al.}(2020){Althaus}, {C{\'o}rsico}, {Uzundag},
  {Vu{\v{c}}kovi{\'c}}, {Baran}, {Bell}, {Camisassa}, {Calcaferro}, {De
  Ger{\'o}nimo}, {Kepler}, \& {Silvotti}}]{2020A&A...633A..20A}
{Althaus}, L.~G., {C{\'o}rsico}, A.~H., {Uzundag}, M., {et~al.} 2020, \aap,
  633, A20

\bibitem[{{Althaus} {et~al.}(2005){Althaus}, {Serenelli}, {Panei},
  {C{\'o}rsico}, {Garc{\'\i}a-Berro}, \& {Sc{\'o}ccola}}]{2005A&A...435..631A}
{Althaus}, L.~G., {Serenelli}, A.~M., {Panei}, J.~A., {et~al.} 2005, \aap, 435,
  631

\bibitem[{{Appleton} {et~al.}(1993){Appleton}, {Kawaler}, \&
  {Eitter}}]{1993AJ....106.1973A}
{Appleton}, P.~N., {Kawaler}, S.~D., \& {Eitter}, J.~J. 1993, \aj, 106, 1973

\bibitem[{{Baran} {et~al.}(2015){Baran}, {Telting}, {N{\'e}meth}, {Bachulski},
  \& {Krzesi{\'n}ski}}]{2015A&A...573A..52B}
{Baran}, A.~S., {Telting}, J.~H., {N{\'e}meth}, P., {Bachulski}, S., \&
  {Krzesi{\'n}ski}, J. 2015, \aap, 573, A52

\bibitem[{Bell(2017)}]{2017PhDT........14C}
Bell, K.~J. 2017, PhD thesis, University of Texas

\bibitem[{{Bell} {et~al.}(2019){Bell}, {C{\'o}rsico}, {Bischoff-Kim},
  {Althaus}, {Bradley}, {Calcaferro}, {Montgomery}, {Uzundag}, {Baran},
  {Bogn{\'a}r}, {Charpinet}, {Ghasemi}, \& {Hermes}}]{2019A&A...632A..42B}
{Bell}, K.~J., {C{\'o}rsico}, A.~H., {Bischoff-Kim}, A., {et~al.} 2019, \aap,
  632, A42

\bibitem[{{Bell} {et~al.}(2017){Bell}, {Hermes}, {Vanderbosch}, {Montgomery},
  {Winget}, {Dennihy}, {Fuchs}, \& {Tremblay}}]{2017ApJ...851...24B}
{Bell}, K.~J., {Hermes}, J.~J., {Vanderbosch}, Z., {et~al.} 2017, \apj, 851, 24

\bibitem[{{Bl{\"o}cker}(2001)}]{2001Ap&SS.275....1B}
{Bl{\"o}cker}, T. 2001, \apss, 275, 1

\bibitem[{{Bogn{\'a}r} {et~al.}(2020){Bogn{\'a}r}, {Kawaler}, {Bell}, {Schrand
  t}, {Baran}, {Bradley}, {Hermes}, {Charpinet}, {Handler}, {Mullally},
  {Murphy}, {Raddi}, {S{\'o}dor}, {Tremblay}, {Uzundag}, \&
  {Zong}}]{2020A&A...638A..82B}
{Bogn{\'a}r}, Z., {Kawaler}, S.~D., {Bell}, K.~J., {et~al.} 2020, \aap, 638,
  A82

\bibitem[{{Bognar} \& {Sodor}(2016)}]{2016IBVS.6184....1B}
{Bognar}, Z. \& {Sodor}, A. 2016, Information Bulletin on Variable Stars, 6184

\bibitem[{{Bond} {et~al.}(1996){Bond}, {Kawaler}, {Ciardullo}, {Stover},
  {Kuroda}, {Ishida}, {Ono}, {Tamura}, {Malasan}, {Yamasaki}, {Hashimoto},
  {Kambe}, {Takeuti}, {Kato}, {Kato}, {Chen}, {Leibowitz}, {Roth}, {Soffner},
  \& {Mitsch}}]{1996AJ....112.2699B}
{Bond}, H.~E., {Kawaler}, S.~D., {Ciardullo}, R., {et~al.} 1996, \aj, 112, 2699

\bibitem[{{Borucki} {et~al.}(2010){Borucki}, {Koch}, {Basri}, {Batalha},
  {Brown}, {Caldwell}, {Caldwell}, {Christensen-Dalsgaard}, {Cochran},
  {DeVore}, {Dunham}, {Dupree}, {Gautier}, {Geary}, {Gilliland}, {Gould},
  {Howell}, {Jenkins}, {Kondo}, {Latham}, {Marcy}, {Meibom}, {Kjeldsen},
  {Lissauer}, {Monet}, {Morrison}, {Sasselov}, {Tarter}, {Boss}, {Brownlee},
  {Owen}, {Buzasi}, {Charbonneau}, {Doyle}, {Fortney}, {Ford}, {Holman},
  {Seager}, {Steffen}, {Welsh}, {Rowe}, {Anderson}, {Buchhave}, {Ciardi},
  {Walkowicz}, {Sherry}, {Horch}, {Isaacson}, {Everett}, {Fischer}, {Torres},
  {Johnson}, {Endl}, {MacQueen}, {Bryson}, {Dotson}, {Haas}, {Kolodziejczak},
  {Van Cleve}, {Chandrasekaran}, {Twicken}, {Quintana}, {Clarke}, {Allen},
  {Li}, {Wu}, {Tenenbaum}, {Verner}, {Bruhweiler}, {Barnes}, \&
  {Prsa}}]{2010Sci...327..977B}
{Borucki}, W.~J., {Koch}, D., {Basri}, G., {et~al.} 2010, Science, 327, 977

\bibitem[{{Bradley} {et~al.}(1993){Bradley}, {Winget}, \&
  {Wood}}]{1993ApJ...406..661B}
{Bradley}, P.~A., {Winget}, D.~E., \& {Wood}, M.~A. 1993, \apj, 406, 661

\bibitem[{{Brassard} {et~al.}(1992){Brassard}, {Fontaine}, {Wesemael}, \&
  {Hansen}}]{1992ApJS...80..369B}
{Brassard}, P., {Fontaine}, G., {Wesemael}, F., \& {Hansen}, C.~J. 1992, \apjs,
  80, 369

\bibitem[{{Calcaferro} {et~al.}(2016){Calcaferro}, {C{\'o}rsico}, \&
  {Althaus}}]{2016A&A...589A..40C}
{Calcaferro}, L.~M., {C{\'o}rsico}, A.~H., \& {Althaus}, L.~G. 2016, \aap, 589,
  A40

\bibitem[{{Chang} {et~al.}(2013){Chang}, {Shih}, {Liu}, {Fan}, {Wu}, {Roques},
  {Doressoundiram}, {Fernand ez}, {Christophe}, \&
  {Dauny}}]{2013A&A...558A..63C}
{Chang}, H.~K., {Shih}, I.~C., {Liu}, C.~Y., {et~al.} 2013, \aap, 558, A63

\bibitem[{{Chen} {et~al.}(1998){Chen}, {Vergely}, {Valette}, \&
  {Carraro}}]{1998A&A...336..137C}
{Chen}, B., {Vergely}, J.~L., {Valette}, B., \& {Carraro}, G. 1998, \aap, 336,
  137

\bibitem[{{Ciardullo} \& {Bond}(1996)}]{1996AJ....111.2332C}
{Ciardullo}, R. \& {Bond}, H.~E. 1996, \aj, 111, 2332

\bibitem[{{C{\'o}rsico}(2020)}]{2020FrASS...7...47C}
{C{\'o}rsico}, A.~H. 2020, Frontiers in Astronomy and Space Sciences, 7, 47

\bibitem[{{C{\'o}rsico} \& {Althaus}(2005)}]{2005A&A...439L..31C}
{C{\'o}rsico}, A.~H. \& {Althaus}, L.~G. 2005, \aap, 439, L31

\bibitem[{{C{\'o}rsico} \& {Althaus}(2006)}]{2006A&A...454..863C}
---. 2006, \aap, 454, 863

\bibitem[{{C{\'o}rsico} {et~al.}(2002){C{\'o}rsico}, {Althaus}, {Benvenuto}, \&
  {Serenelli}}]{2002A&A...387..531C}
{C{\'o}rsico}, A.~H., {Althaus}, L.~G., {Benvenuto}, O.~G., \& {Serenelli},
  A.~M. 2002, \aap, 387, 531

\bibitem[{{C{\'o}rsico} {et~al.}(2008){C{\'o}rsico}, {Althaus}, {Kepler},
  {Costa}, \& {Miller Bertolami}}]{2008A&A...478..869C}
{C{\'o}rsico}, A.~H., {Althaus}, L.~G., {Kepler}, S.~O., {Costa}, J.~E.~S., \&
  {Miller Bertolami}, M.~M. 2008, \aap, 478, 869

\bibitem[{{C{\'o}rsico} {et~al.}(2006){C{\'o}rsico}, {Althaus}, \& {Miller
  Bertolami}}]{2006A&A...458..259C}
{C{\'o}rsico}, A.~H., {Althaus}, L.~G., \& {Miller Bertolami}, M.~M. 2006,
  \aap, 458, 259

\bibitem[{{C{\'o}rsico} {et~al.}(2009{\natexlab{a}}){C{\'o}rsico}, {Althaus},
  {Miller Bertolami}, \& {Garc{\'{\i}}a-Berro}}]{2009A&A...499..257C}
{C{\'o}rsico}, A.~H., {Althaus}, L.~G., {Miller Bertolami}, M.~M., \&
  {Garc{\'{\i}}a-Berro}, E. 2009{\natexlab{a}}, \aap, 499, 257

\bibitem[{{C{\'o}rsico} {et~al.}(2009{\natexlab{b}}){C{\'o}rsico}, {Althaus},
  {Miller Bertolami}, {Gonz{\'a}lez P{\'e}rez}, \&
  {Kepler}}]{2009ApJ...701.1008C}
{C{\'o}rsico}, A.~H., {Althaus}, L.~G., {Miller Bertolami}, M.~M.,
  {Gonz{\'a}lez P{\'e}rez}, J.~M., \& {Kepler}, S.~O. 2009{\natexlab{b}}, \apj,
  701, 1008

\bibitem[{{C{\'o}rsico} {et~al.}(2019){C{\'o}rsico}, {Althaus}, {Miller
  Bertolami}, \& {Kepler}}]{2019A&ARv..27....7C}
{C{\'o}rsico}, A.~H., {Althaus}, L.~G., {Miller Bertolami}, M.~M., \& {Kepler},
  S.~O. 2019, \aapr, 27, 7

\bibitem[{{C{\'o}rsico} {et~al.}(2007{\natexlab{a}}){C{\'o}rsico}, {Althaus},
  {Miller Bertolami}, \& {Werner}}]{2007A&A...461.1095C}
{C{\'o}rsico}, A.~H., {Althaus}, L.~G., {Miller Bertolami}, M.~M., \& {Werner},
  K. 2007{\natexlab{a}}, \aap, 461, 1095

\bibitem[{{C{\'o}rsico} {et~al.}(2007{\natexlab{b}}){C{\'o}rsico}, {Miller
  Bertolami}, {Althaus}, {Vauclair}, \& {Werner}}]{2007A&A...475..619C}
{C{\'o}rsico}, A.~H., {Miller Bertolami}, M.~M., {Althaus}, L.~G., {Vauclair},
  G., \& {Werner}, K. 2007{\natexlab{b}}, \aap, 475, 619

\bibitem[{{Costa} {et~al.}(2008){Costa}, {Kepler}, {Winget}, {O'Brien},
  {Kawaler}, {Costa}, {Giovannini}, {Kanaan}, {Mukadam}, {Mullally}, {Nitta},
  {Proven{\c{c}}al}, {Shipman}, {Wood}, {Ahrens}, {Grauer}, {Kilic}, {Bradley},
  {Sekiguchi}, {Crowe}, {Jiang}, {Sullivan}, {Sullivan}, {Rosen}, {Clemens},
  {Janulis}, {O'Donoghue}, {Ogloza}, {Baran}, {Silvotti}, {Marinoni},
  {Vauclair}, {Dolez}, {Chevreton}, {Dreizler}, {Schuh}, {Deetjen}, {Nagel},
  {Solheim}, {Gonzalez Perez}, {Ulla}, {Barstow}, {Burleigh}, {Good},
  {Metcalfe}, {Kim}, {Lee}, {Sergeev}, {Akan}, {{\c{C}}ak{\i}rl{\i}}, {Paparo},
  {Viraghalmy}, {Ashoka}, {Handler}, {H{\"u}rkal}, {Johannessen}, {Kleinman},
  {Kalytis}, {Krzesinski}, {Klumpe}, {Larrison}, {Lawrence}, {Mei{\v{s}}tas},
  {Martinez}, {Nather}, {Fu}, {Pak{\v{s}}tien{\.{e}}}, {Rosen},
  {Romero-Colmenero}, {Riddle}, {Seetha}, {Silvestri}, {Vu{\v{c}}kovi{\'c}},
  {Warner}, {Zola}, {Althaus}, {C{\'o}rsico}, \&
  {Montgomery}}]{2008A&A...477..627C}
{Costa}, J.~E.~S., {Kepler}, S.~O., {Winget}, D.~E., {et~al.} 2008, \aap, 477,
  627

\bibitem[{{Cox}(2000)}]{2000asqu.book.....C}
{Cox}, A.~N. 2000, {Allen's astrophysical quantities}

\bibitem[{{Dreizler} {et~al.}(1996){Dreizler}, {Werner}, {Heber}, \&
  {Engels}}]{1996A&A...309..820D}
{Dreizler}, S., {Werner}, K., {Heber}, U., \& {Engels}, D. 1996, \aap, 309, 820

\bibitem[{{Dufour} {et~al.}(2017){Dufour}, {Blouin}, {Coutu},
  {Fortin-Archambault}, {Thibeault}, {Bergeron}, \&
  {Fontaine}}]{2017ASPC..509....3D}
{Dufour}, P., {Blouin}, S., {Coutu}, S., {et~al.} 2017, in Astronomical Society
  of the Pacific Conference Series, Vol. 509, 20th European White Dwarf
  Workshop, ed. P.~E. {Tremblay}, B.~{Gaensicke}, \& T.~{Marsh}, 3

\bibitem[{{Dziembowski}(1977)}]{1977AcA....27..203D}
{Dziembowski}, W. 1977, \actaa, 27, 203

\bibitem[{{Eastman} {et~al.}(2010){Eastman}, {Siverd}, \&
  {Gaudi}}]{2010PASP..122..935E}
{Eastman}, J., {Siverd}, R., \& {Gaudi}, B.~S. 2010, \pasp, 122, 935

\bibitem[{{Ercolano} {et~al.}(2004){Ercolano}, {Wesson}, {Zhang}, {Barlow}, {De
  Marco}, {Rauch}, \& {Liu}}]{2004MNRAS.354..558E}
{Ercolano}, B., {Wesson}, R., {Zhang}, Y., {et~al.} 2004, \mnras, 354, 558

\bibitem[{{Faedi} {et~al.}(2011){Faedi}, {West}, {Burleigh}, {Goad}, \&
  {Hebb}}]{2011MNRAS.410..899F}
{Faedi}, F., {West}, R.~G., {Burleigh}, M.~R., {Goad}, M.~R., \& {Hebb}, L.
  2011, \mnras, 410, 899

\bibitem[{{Fantin} {et~al.}(2020){Fantin}, {C{\^o}t{\'e}}, \&
  {McConnachie}}]{2020ApJ...900..139F}
{Fantin}, N.~J., {C{\^o}t{\'e}}, P., \& {McConnachie}, A.~W. 2020, \apj, 900,
  139

\bibitem[{{Feibelman} {et~al.}(1995){Feibelman}, {Kaler}, {Bond}, \&
  {Grauer}}]{1995PASP..107..914F}
{Feibelman}, W.~A., {Kaler}, J.~B., {Bond}, H.~E., \& {Grauer}, A.~D. 1995,
  \pasp, 107, 914

\bibitem[{{Fontaine} \& {Brassard}(2008)}]{2008PASP..120.1043F}
{Fontaine}, G. \& {Brassard}, P. 2008, PASP, 120, 1043

\bibitem[{{Gautschy} {et~al.}(2005){Gautschy}, {Althaus}, \&
  {Saio}}]{2005A&A...438.1013G}
{Gautschy}, A., {Althaus}, L.~G., \& {Saio}, H. 2005, \aap, 438, 1013

\bibitem[{{G{\'o}mez-Gonz{\'a}lez} {et~al.}(2020){G{\'o}mez-Gonz{\'a}lez},
  {Toal{\'a}}, {Guerrero}, {Todt}, {Sabin}, {Ramos-Larios}, \&
  {Mayya}}]{2020MNRAS.496..959G}
{G{\'o}mez-Gonz{\'a}lez}, V.~M.~A., {Toal{\'a}}, J.~A., {Guerrero}, M.~A.,
  {et~al.} 2020, \mnras, 496, 959

\bibitem[{{Grauer} \& {Bond}(1984)}]{1984ApJ...277..211G}
{Grauer}, A.~D. \& {Bond}, H.~E. 1984, \apj, 277, 211

\bibitem[{{Handler} {et~al.}(1997{\natexlab{a}}){Handler}, {Kanaan}, \&
  {Montgomery}}]{1997A&A...326..692H}
{Handler}, G., {Kanaan}, A., \& {Montgomery}, M.~H. 1997{\natexlab{a}}, \aap,
  326, 692

\bibitem[{{Handler} {et~al.}(1997{\natexlab{b}}){Handler}, {Pikall},
  {O'Donoghue}, {Buckley}, {Vauclair}, {Chevreton}, {Giovannini}, {Kepler},
  {Goode}, {Provencal}, {Wood}, {Clemens}, {O'Brien}, {Nather}, {Winget},
  {Kleinman}, {Kanaan}, {Watson}, {Nitta}, {Montgomery}, {Klumpe}, {Bradley},
  {Sullivan}, {Wu}, {Marar}, {Seetha}, {Ashoka}, {Mahra}, {Bhat}, {Babu},
  {Leibowitz}, {Hemar}, {Ibbetson}, {Mashal}, {Meistas}, {Dziembowski},
  {Pamyatnykh}, {Moskalik}, {Zola}, {Pajdosz}, {Krzesinski}, {Solheim}, {Bard},
  {Massacand}, {Breger}, {Gelbmann}, {Paunzen}, \&
  {North}}]{1997MNRAS.286..303H}
{Handler}, G., {Pikall}, H., {O'Donoghue}, D., {et~al.} 1997{\natexlab{b}},
  \mnras, 286, 303

\bibitem[{{Herald} \& {Bianchi}(2004)}]{2004ApJ...609..378H}
{Herald}, J.~E. \& {Bianchi}, L. 2004, \apj, 609, 378

\bibitem[{{Hermes} {et~al.}(2017{\natexlab{a}}){Hermes}, {G{\"a}nsicke},
  {Kawaler}, {Greiss}, {Tremblay}, {Gentile Fusillo}, {Raddi}, {Fanale},
  {Bell}, {Dennihy}, {Fuchs}, {Dunlap}, {Clemens}, {Montgomery}, {Winget},
  {Chote}, {Marsh}, \& {Redfield}}]{2017ApJS..232...23H}
{Hermes}, J.~J., {G{\"a}nsicke}, B.~T., {Kawaler}, S.~D., {et~al.}
  2017{\natexlab{a}}, \apjs, 232, 23

\bibitem[{{Hermes} {et~al.}(2017{\natexlab{b}}){Hermes}, {Kawaler},
  {Bischoff-Kim}, {Provencal}, {Dunlap}, \& {Clemens}}]{2017ApJ...835..277H}
{Hermes}, J.~J., {Kawaler}, S.~D., {Bischoff-Kim}, A., {et~al.}
  2017{\natexlab{b}}, \apj, 835, 277

\bibitem[{{Herwig}(2001)}]{2001ApJ...554L..71H}
{Herwig}, F. 2001, \apjl, 554, L71

\bibitem[{{Howell} {et~al.}(2014){Howell}, {Sobeck}, {Haas}, {Still},
  {Barclay}, {Mullally}, {Troeltzsch}, {Aigrain}, {Bryson}, {Caldwell},
  {Chaplin}, {Cochran}, {Huber}, {Marcy}, {Miglio}, {Najita}, {Smith},
  {Twicken}, \& {Fortney}}]{2014PASP..126..398H}
{Howell}, S.~B., {Sobeck}, C., {Haas}, M., {et~al.} 2014, PASP, 126, 398

\bibitem[{{Jenkins} {et~al.}(2016){Jenkins}, {Twicken}, {McCauliff},
  {Campbell}, {Sanderfer}, {Lung}, {Mansouri-Samani}, {Girouard}, {Tenenbaum},
  {Klaus}, {Smith}, {Caldwell}, {Chacon}, {Henze}, {Heiges}, {Latham},
  {Morgan}, {Swade}, {Rinehart}, \& {Vanderspek}}]{jenkins2016}
{Jenkins}, J.~M., {Twicken}, J.~D., {McCauliff}, S., {et~al.} 2016, in
  \procspie, Vol. 9913, Software and Cyberinfrastructure for Astronomy IV,
  99133E

\bibitem[{{Kawaler}(1988)}]{1988IAUS..123..329K}
{Kawaler}, S.~D. 1988, in IAU Symposium, Vol. 123, Advances in Helio- and
  Asteroseismology, ed. J.~{Christensen-Dalsgaard} \& S.~{Frandsen}, 329

\bibitem[{{Kawaler} \& {Bradley}(1994)}]{1994ApJ...427..415K}
{Kawaler}, S.~D. \& {Bradley}, P.~A. 1994, \apj, 427, 415

\bibitem[{{Kepler}(1993)}]{1993BaltA...2..515K}
{Kepler}, S.~O. 1993, Baltic Astronomy, 2, 515

\bibitem[{{Kepler} {et~al.}(2014){Kepler}, {Fraga}, {Winget}, {Bell},
  {C{\'o}rsico}, \& {Werner}}]{2014MNRAS.442.2278K}
{Kepler}, S.~O., {Fraga}, L., {Winget}, D.~E., {et~al.} 2014, \mnras, 442, 2278

\bibitem[{{Koen} \& {Laney}(2000)}]{2000MNRAS.311..636K}
{Koen}, C. \& {Laney}, D. 2000, \mnras, 311, 636

\bibitem[{{Koesterke}(2001)}]{2001Ap&SS.275...41K}
{Koesterke}, L. 2001, \apss, 275, 41

\bibitem[{{Koesterke} \& {Hamann}(1997)}]{1997IAUS..180..114K}
{Koesterke}, L. \& {Hamann}, W.~R. 1997, in IAU Symposium, Vol. 180, Planetary
  Nebulae, ed. H.~J. {Habing} \& H.~J.~G.~L.~M. {Lamers}, 114

\bibitem[{{Ledoux} \& {Walraven}(1958)}]{1958HDP....51..353L}
{Ledoux}, P. \& {Walraven}, T. 1958, Handbuch der Physik, 51, 353

\bibitem[{{L{\"o}bling} {et~al.}(2019){L{\"o}bling}, {Rauch}, {Miller
  Bertolami}, {Todt}, {Friederich}, {Ziegler}, {Werner}, \&
  {Kruk}}]{2019MNRAS.489.1054L}
{L{\"o}bling}, L., {Rauch}, T., {Miller Bertolami}, M.~M., {et~al.} 2019,
  \mnras, 489, 1054

\bibitem[{{LSST Science Collaboration} {et~al.}(2009){LSST Science
  Collaboration}, {Abell}, {Allison}, {Anderson}, {Andrew}, {Angel}, {Armus},
  {Arnett}, {Asztalos}, {Axelrod}, {Bailey}, {Ballantyne}, {Bankert},
  {Barkhouse}, {Barr}, {Barrientos}, {Barth}, {Bartlett}, {Becker}, {Becla},
  {Beers}, {Bernstein}, {Biswas}, {Blanton}, {Bloom}, {Bochanski}, {Boeshaar},
  {Borne}, {Bradac}, {Brandt}, {Bridge}, {Brown}, {Brunner}, {Bullock},
  {Burgasser}, {Burge}, {Burke}, {Cargile}, {Chand rasekharan}, {Chartas},
  {Chesley}, {Chu}, {Cinabro}, {Claire}, {Claver}, {Clowe}, {Connolly}, {Cook},
  {Cooke}, {Cooray}, {Covey}, {Culliton}, {de Jong}, {de Vries}, {Debattista},
  {Delgado}, {Dell'Antonio}, {Dhital}, {Di Stefano}, {Dickinson}, {Dilday},
  {Djorgovski}, {Dobler}, {Donalek}, {Dubois-Felsmann}, {Durech},
  {Eliasdottir}, {Eracleous}, {Eyer}, {Falco}, {Fan}, {Fassnacht}, {Ferguson},
  {Fernandez}, {Fields}, {Finkbeiner}, {Figueroa}, {Fox}, {Francke}, {Frank},
  {Frieman}, {Fromenteau}, {Furqan}, {Galaz}, {Gal-Yam}, {Garnavich},
  {Gawiser}, {Geary}, {Gee}, {Gibson}, {Gilmore}, {Grace}, {Green}, {Gressler},
  {Grillmair}, {Habib}, {Haggerty}, {Hamuy}, {Harris}, {Hawley}, {Heavens},
  {Hebb}, {Henry}, {Hileman}, {Hilton}, {Hoadley}, {Holberg}, {Holman},
  {Howell}, {Infante}, {Ivezic}, {Jacoby}, {Jain}, {R}, {Jedicke}, {Jee},
  {Garrett Jernigan}, {Jha}, {Johnston}, {Jones}, {Juric}, {Kaasalainen},
  {Styliani}, {Kafka}, {Kahn}, {Kaib}, {Kalirai}, {Kantor}, {Kasliwal},
  {Keeton}, {Kessler}, {Knezevic}, {Kowalski}, {Krabbendam}, {Krughoff},
  {Kulkarni}, {Kuhlman}, {Lacy}, {Lepine}, {Liang}, {Lien}, {Lira}, {Long},
  {Lorenz}, {Lotz}, {Lupton}, {Lutz}, {Macri}, {Mahabal}, {Mandelbaum},
  {Marshall}, {May}, {McGehee}, {Meadows}, {Meert}, {Milani}, {Miller},
  {Miller}, {Mills}, {Minniti}, {Monet}, {Mukadam}, {Nakar}, {Neill}, {Newman},
  {Nikolaev}, {Nordby}, {O'Connor}, {Oguri}, {Oliver}, {Olivier}, {Olsen},
  {Olsen}, {Olszewski}, {Oluseyi}, {Padilla}, {Parker}, {Pepper}, {Peterson},
  {Petry}, {Pinto}, {Pizagno}, {Popescu}, {Prsa}, {Radcka}, {Raddick},
  {Rasmussen}, {Rau}, {Rho}, {Rhoads}, {Richards}, {Ridgway}, {Robertson},
  {Roskar}, {Saha}, {Sarajedini}, {Scannapieco}, {Schalk}, {Schindler},
  {Schmidt}, {Schmidt}, {Schneider}, {Schumacher}, {Scranton}, {Sebag},
  {Seppala}, {Shemmer}, {Simon}, {Sivertz}, {Smith}, {Allyn Smith}, {Smith},
  {Spitz}, {Stanford}, {Stassun}, {Strader}, {Strauss}, {Stubbs}, {Sweeney},
  {Szalay}, {Szkody}, {Takada}, {Thorman}, {Trilling}, {Trimble}, {Tyson}, {Van
  Berg}, {Vand en Berk}, {VanderPlas}, {Verde}, {Vrsnak}, {Walkowicz}, {Wand
  elt}, {Wang}, {Wang}, {Warner}, {Wechsler}, {West}, {Wiecha}, {Williams},
  {Willman}, {Wittman}, {Wolff}, {Wood-Vasey}, {Wozniak}, {Young}, {Zentner},
  \& {Zhan}}]{2009arXiv0912.0201L}
{LSST Science Collaboration}, {Abell}, P.~A., {Allison}, J., {et~al.} 2009,
  arXiv e-prints, arXiv:0912.0201

\bibitem[{{Miller Bertolami} \& {Althaus}(2006)}]{2006A&A...454..845M}
{Miller Bertolami}, M.~M. \& {Althaus}, L.~G. 2006, \aap, 454, 845

\bibitem[{{Miller Bertolami} \&
  {Althaus}(2007{\natexlab{a}})}]{2007A&A...470..675M}
---. 2007{\natexlab{a}}, \aap, 470, 675

\bibitem[{{Miller Bertolami} \&
  {Althaus}(2007{\natexlab{b}})}]{2007MNRAS.380..763M}
---. 2007{\natexlab{b}}, \mnras, 380, 763

\bibitem[{{Miller Bertolami} {et~al.}(2006){Miller Bertolami}, {Althaus},
  {Serenelli}, \& {Panei}}]{2006A&A...449..313M}
{Miller Bertolami}, M.~M., {Althaus}, L.~G., {Serenelli}, A.~M., \& {Panei},
  J.~A. 2006, \aap, 449, 313

\bibitem[{{Motch} {et~al.}(1993){Motch}, {Werner}, \&
  {Pakull}}]{1993A&A...268..561M}
{Motch}, C., {Werner}, K., \& {Pakull}, M.~W. 1993, \aap, 268, 561

\bibitem[{{Moya} {et~al.}(2018){Moya}, {Barcel{\'o} Forteza}, {Bonfanti},
  {Salmon}, {Van Grootel}, \& {Barrado}}]{2018A&A...620A.203M}
{Moya}, A., {Barcel{\'o} Forteza}, S., {Bonfanti}, A., {et~al.} 2018, \aap,
  620, A203

\bibitem[{{Murphy}(2015)}]{2015MNRAS.453.2569M}
{Murphy}, S.~J. 2015, \mnras, 453, 2569

\bibitem[{{Napiwotzki} \& {Schoenberner}(1991)}]{1991A&A...249L..16N}
{Napiwotzki}, R. \& {Schoenberner}, D. 1991, \aap, 249, L16

\bibitem[{{Nather} {et~al.}(1990){Nather}, {Winget}, {Clemens}, {Hansen}, \&
  {Hine}}]{1990ApJ...361..309N}
{Nather}, R.~E., {Winget}, D.~E., {Clemens}, J.~C., {Hansen}, C.~J., \& {Hine},
  B.~P. 1990, \apj, 361, 309

\bibitem[{{O'Donoghue}(1994)}]{1994MNRAS.270..222O}
{O'Donoghue}, D. 1994, \mnras, 270, 222

\bibitem[{{{\O}stensen} {et~al.}(2011){{\O}stensen}, {Bloemen}, {Vu{\v
  c}kovi{\'c}}, {Aerts}, {Oreiro}, {Kinemuchi}, {Still}, \&
  {Koester}}]{2011ApJ...736L..39O}
{{\O}stensen}, R.~H., {Bloemen}, S., {Vu{\v c}kovi{\'c}}, M., {et~al.} 2011,
  \apjl, 736, L39

\bibitem[{{Piotto}(2018)}]{2018EPSC...12..969P}
{Piotto}, G. 2018, in European Planetary Science Congress, EPSC2018--969

\bibitem[{{Provencal} {et~al.}(2009){Provencal}, {Montgomery}, {Kanaan},
  {Shipman}, {Childers}, {Baran}, {Kepler}, {Reed}, {Zhou}, {Eggen}, {Watson},
  {Winget}, {Thompson}, {Riaz}, {Nitta}, {Kleinman}, {Crowe}, {Slivkoff},
  {Sherard}, {Purves}, {Binder}, {Knight}, {Kim}, {Chen}, {Yang}, {Lin}, {Lin},
  {Chen}, {Jiang}, {Sergeev}, {Mkrtichian}, {Andreev}, {Janulis}, {Siwak},
  {Zola}, {Koziel}, {Stachowski}, {Paparo}, {Bognar}, {Handler}, {Lorenz},
  {Steininger}, {Beck}, {Nagel}, {Kusterer}, {Hoffman}, {Reiff}, {Kowalski},
  {Vauclair}, {Charpinet}, {Chevreton}, {Solheim}, {Pakstiene}, {Fraga}, \&
  {Dalessio}}]{2009ApJ...693..564P}
{Provencal}, J.~L., {Montgomery}, M.~H., {Kanaan}, A., {et~al.} 2009, \apj,
  693, 564

\bibitem[{{Quirion} {et~al.}(2007){Quirion}, {Fontaine}, \&
  {Brassard}}]{2007ApJS..171..219Q}
{Quirion}, P.~O., {Fontaine}, G., \& {Brassard}, P. 2007, \apjs, 171, 219

\bibitem[{{Quirion} {et~al.}(2009){Quirion}, {Fontaine}, \&
  {Brassard}}]{2009JPhCS.172a2077Q}
{Quirion}, P.-O., {Fontaine}, G., \& {Brassard}, P. 2009, in Journal of Physics
  Conference Series, Vol. 172, Journal of Physics Conference Series, 012077

\bibitem[{{Rauch} \& {Werner}(1997)}]{1997fbs..conf..217R}
{Rauch}, T. \& {Werner}, K. 1997, in The Third Conference on Faint Blue Stars,
  ed. A.~G.~D. {Philip}, J.~{Liebert}, R.~{Saffer}, \& D.~S. {Hayes}, 217

\bibitem[{{Ricker} {et~al.}(2015){Ricker}, {Winn}, {Vanderspek}, {Latham},
  {Bakos}, {Bean}, {Berta-Thompson}, {Brown}, {Buchhave}, {Butler}, {Butler},
  {Chaplin}, {Charbonneau}, {Christensen-Dalsgaard}, {Clampin}, {Deming},
  {Doty}, {De Lee}, {Dressing}, {Dunham}, {Endl}, {Fressin}, {Ge}, {Henning},
  {Holman}, {Howard}, {Ida}, {Jenkins}, {Jernigan}, {Johnson}, {Kaltenegger},
  {Kawai}, {Kjeldsen}, {Laughlin}, {Levine}, {Lin}, {Lissauer}, {MacQueen},
  {Marcy}, {McCullough}, {Morton}, {Narita}, {Paegert}, {Palle}, {Pepe},
  {Pepper}, {Quirrenbach}, {Rinehart}, {Sasselov}, {Sato}, {Seager},
  {Sozzetti}, {Stassun}, {Sullivan}, {Szentgyorgyi}, {Torres}, {Udry}, \&
  {Villasenor}}]{2015JATIS...1a4003R}
{Ricker}, G.~R., {Winn}, J.~N., {Vanderspek}, R., {et~al.} 2015, Journal of
  Astronomical Telescopes, Instruments, and Systems, 1, 014003

\bibitem[{{Ricker} {et~al.}(2014){Ricker}, {Winn}, {Vanderspek}, {Latham},
  {Bakos}, {Bean}, {Berta-Thompson}, {Brown}, {Buchhave}, {Butler}, {Butler},
  {Chaplin}, {Charbonneau}, {Christensen-Dalsgaard}, {Clampin}, {Deming},
  {Doty}, {De Lee}, {Dressing}, {Dunham}, {Endl}, {Fressin}, {Ge}, {Henning},
  {Holman}, {Howard}, {Ida}, {Jenkins}, {Jernigan}, {Johnson}, {Kaltenegger},
  {Kawai}, {Kjeldsen}, {Laughlin}, {Levine}, {Lin}, {Lissauer}, {MacQueen},
  {Marcy}, {McCullough}, {Morton}, {Narita}, {Paegert}, {Palle}, {Pepe},
  {Pepper}, {Quirrenbach}, {Rinehart}, {Sasselov}, {Sato}, {Seager},
  {Sozzetti}, {Stassun}, {Sullivan}, {Szentgyorgyi}, {Torres}, {Udry}, \&
  {Villasenor}}]{2014SPIE.9143E..20R}
{Ricker}, G.~R., {Winn}, J.~N., {Vanderspek}, R., {et~al.} 2014, in \procspie,
  Vol. 9143, Space Telescopes and Instrumentation 2014: Optical, Infrared, and
  Millimeter Wave, 914320

\bibitem[{{Silvotti}(1996)}]{1996A&A...309L..23S}
{Silvotti}, R. 1996, \aap, 309, L23

\bibitem[{{Silvotti} {et~al.}(1999){Silvotti}, {Dreizler}, {Handler}, \&
  {Jiang}}]{1999A&A...342..745S}
{Silvotti}, R., {Dreizler}, S., {Handler}, G., \& {Jiang}, X.~J. 1999, \aap,
  342, 745

\bibitem[{{Stanghellini} {et~al.}(1991){Stanghellini}, {Cox}, \&
  {Starrfield}}]{1991ApJ...383..766S}
{Stanghellini}, L., {Cox}, A.~N., \& {Starrfield}, S. 1991, \apj, 383, 766

\bibitem[{{Starrfield} {et~al.}(1984){Starrfield}, {Cox}, {Kidman}, \&
  {Pesnell}}]{1984ApJ...281..800S}
{Starrfield}, S., {Cox}, A.~N., {Kidman}, R.~B., \& {Pesnell}, W.~D. 1984,
  \apj, 281, 800

\bibitem[{{Starrfield} {et~al.}(1983){Starrfield}, {Cox}, {Hodson}, \&
  {Pesnell}}]{1983ApJ...268L..27S}
{Starrfield}, S.~G., {Cox}, A.~N., {Hodson}, S.~W., \& {Pesnell}, W.~D. 1983,
  \apjl, 268, L27

\bibitem[{{Tassoul} {et~al.}(1990){Tassoul}, {Fontaine}, \&
  {Winget}}]{1990ApJS...72..335T}
{Tassoul}, M., {Fontaine}, G., \& {Winget}, D.~E. 1990, ApJs, 72, 335

\bibitem[{{Vauclair} {et~al.}(1993){Vauclair}, {Belmonte}, {Pfeiffer},
  {Chevreton}, {Dolez}, {Motch}, {Werner}, \& {Pakull}}]{1993A&A...267L..35V}
{Vauclair}, G., {Belmonte}, J.~A., {Pfeiffer}, B., {et~al.} 1993, \aap, 267,
  L35

\bibitem[{{Vauclair} {et~al.}(2002){Vauclair}, {Moskalik}, {Pfeiffer},
  {Chevreton}, {Dolez}, {Serre}, {Kleinman}, {Barstow}, {Sansom}, {Solheim},
  {Belmonte}, {Kawaler}, {Kepler}, {Kanaan}, {Giovannini}, {Winget}, {Watson},
  {Nather}, {Clemens}, {Provencal}, {Dixson}, {Yanagida}, {Nitta Kleinman},
  {Montgomery}, {Klumpe}, {Bruvold}, {O'Brien}, {Hansen}, {Grauer}, {Bradley},
  {Wood}, {Achilleos}, {Jiang}, {Fu}, {Marar}, {Ashoka}, {Me{\u i}stas},
  {Chernyshev}, {Mazeh}, {Leibowitz}, {Hemar}, {Krzesi{\'n}ski}, {Pajdosz}, \&
  {Zo{\l}a}}]{2002A&A...381..122V}
{Vauclair}, G., {Moskalik}, P., {Pfeiffer}, B., {et~al.} 2002, \aap, 381, 122

\bibitem[{{Wang} {et~al.}(2020){Wang}, {Zhang}, \& {Dai}}]{2020ApJ...888...49W}
{Wang}, K., {Zhang}, X., \& {Dai}, M. 2020, \apj, 888, 49

\bibitem[{{Watson} \& {Werner}(1992)}]{1992IAUC.5603....1W}
{Watson}, T.~K. \& {Werner}, K. 1992, \iaucirc, 5603, 1

\bibitem[{{Werner} {et~al.}(2003){Werner}, {Deetjen}, {Dreizler}, {Nagel},
  {Rauch}, \& {Schuh}}]{2003ASPC..288...31W}
{Werner}, K., {Deetjen}, J.~L., {Dreizler}, S., {et~al.} 2003, in Astronomical
  Society of the Pacific Conference Series, Vol. 288, Stellar Atmosphere
  Modeling, ed. I.~{Hubeny}, D.~{Mihalas}, \& K.~{Werner}, 31

\bibitem[{{Werner} \& {Herwig}(2006)}]{2006PASP..118..183W}
{Werner}, K. \& {Herwig}, F. 2006, PASP, 118, 183

\bibitem[{{Werner} {et~al.}(2007){Werner}, {Rauch}, \&
  {Kruk}}]{2007A&A...474..591W}
{Werner}, K., {Rauch}, T., \& {Kruk}, J.~W. 2007, \aap, 474, 591

\bibitem[{{Werner} {et~al.}(2010){Werner}, {Rauch}, \&
  {Kruk}}]{2010ApJ...719L..32W}
---. 2010, \apjl, 719, L32

\bibitem[{{Winget} {et~al.}(1983){Winget}, {Hansen}, \& {van
  Horn}}]{1983Natur.303..781W}
{Winget}, D.~E., {Hansen}, C.~J., \& {van Horn}, H.~M. 1983, \nat, 303, 781

\bibitem[{{Winget} \& {Kepler}(2008)}]{2008ARA&A..46..157W}
{Winget}, D.~E. \& {Kepler}, S.~O. 2008, \araa, 46, 157

\bibitem[{{Winget} {et~al.}(1991){Winget}, {Nather}, {Clemens}, {Provencal},
  {Kleinman}, {Bradley}, {Wood}, {Claver}, {Frueh}, {Grauer}, {Hine}, {Hansen},
  {Fontaine}, {Achilleos}, {Wickramasinghe}, {Marar}, {Seetha}, {Ashoka},
  {O'Donoghue}, {Warner}, {Kurtz}, {Buckley}, {Brickhill}, {Vauclair}, {Dolez},
  {Chevreton}, {Barstow}, {Solheim}, {Kanaan}, {Kepler}, {Henry}, \&
  {Kawaler}}]{1991ApJ...378..326W}
{Winget}, D.~E., {Nather}, R.~E., {Clemens}, J.~C., {et~al.} 1991, \apj, 378,
  326

\bibitem[{{Wood} \& {Faulkner}(1986)}]{1986ApJ...307..659W}
{Wood}, P.~R. \& {Faulkner}, D.~J. 1986, \apj, 307, 659

\bibitem[{{York} {et~al.}(2000){York}, {Adelman}, {Anderson}, {Anderson},
  {Annis}, {Bahcall}, {Bakken}, {Barkhouser}, {Bastian}, {Berman}, {Boroski},
  {Bracker}, {Briegel}, {Briggs}, {Brinkmann}, {Brunner}, {Burles}, {Carey},
  {Carr}, {Castander}, {Chen}, {Colestock}, {Connolly}, {Crocker}, {Csabai},
  {Czarapata}, {Davis}, {Doi}, {Dombeck}, {Eisenstein}, {Ellman}, {Elms},
  {Evans}, {Fan}, {Federwitz}, {Fiscelli}, {Friedman}, {Frieman}, {Fukugita},
  {Gillespie}, {Gunn}, {Gurbani}, {de Haas}, {Haldeman}, {Harris}, {Hayes},
  {Heckman}, {Hennessy}, {Hindsley}, {Holm}, {Holmgren}, {Huang}, {Hull},
  {Husby}, {Ichikawa}, {Ichikawa}, {Ivezi{\'c}}, {Kent}, {Kim}, {Kinney},
  {Klaene}, {Kleinman}, {Kleinman}, {Knapp}, {Korienek}, {Kron}, {Kunszt},
  {Lamb}, {Lee}, {Leger}, {Limmongkol}, {Lindenmeyer}, {Long}, {Loomis},
  {Loveday}, {Lucinio}, {Lupton}, {MacKinnon}, {Mannery}, {Mantsch}, {Margon},
  {McGehee}, {McKay}, {Meiksin}, {Merelli}, {Monet}, {Munn}, {Narayanan},
  {Nash}, {Neilsen}, {Neswold}, {Newberg}, {Nichol}, {Nicinski}, {Nonino},
  {Okada}, {Okamura}, {Ostriker}, {Owen}, {Pauls}, {Peoples}, {Peterson},
  {Petravick}, {Pier}, {Pope}, {Pordes}, {Prosapio}, {Rechenmacher}, {Quinn},
  {Richards}, {Richmond}, {Rivetta}, {Rockosi}, {Ruthmansdorfer}, {Sandford},
  {Schlegel}, {Schneider}, {Sekiguchi}, {Sergey}, {Shimasaku}, {Siegmund},
  {Smee}, {Smith}, {Snedden}, {Stone}, {Stoughton}, {Strauss}, {Stubbs},
  {SubbaRao}, {Szalay}, {Szapudi}, {Szokoly}, {Thakar}, {Tremonti}, {Tucker},
  {Uomoto}, {Vanden Berk}, {Vogeley}, {Waddell}, {Wang}, {Watanabe},
  {Weinberg}, {Yanny}, {Yasuda}, \& {SDSS Collaboration}}]{2000AJ....120.1579Y}
{York}, D.~G., {Adelman}, J., {Anderson}, Jr., J.~E., {et~al.} 2000, \aj, 120,
  1579

\bibitem[{{Zong} {et~al.}(2018){Zong}, {Charpinet}, {Fu}, {Vauclair}, {Niu}, \&
  {Su}}]{2018ApJ...853...98Z}
{Zong}, W., {Charpinet}, S., {Fu}, J.-N., {et~al.} 2018, \apj, 853, 98

\end{thebibliography}

\end{document}